\documentclass[
 reprint,
 superscriptaddress,
 amsmath,amssymb,
 aps,
 preprintnumbers,
 nofootinbib
]{revtex4-2}
\makeatletter
\def\p@subsection{}
\makeatother

\usepackage{graphicx,amsmath,amssymb,lipsum,bm}

\usepackage[dvipsnames]{xcolor}
\definecolor{xlinkcolor}{rgb}{0.7752941176470588, 0.22078431372549023, 0.2262745098039215}
\definecolor{BrickRed}{rgb}{0.7752941176470588, 0.22078431372549023, 0.2262745098039215}
\definecolor{xlinkcolor}{HTML}{1c1e94}
\usepackage{booktabs} % Include this package in the preamble
\usepackage[normalem]{ulem}
\usepackage[utf8]{inputenc} 
\usepackage{mathtools}
\usepackage{aux_files/aas_macros}
\usepackage{xspace}

\usepackage{ulem}
\usepackage{diagbox}

\usepackage[dvipsnames]{xcolor}
\usepackage{mathrsfs}

\usepackage[colorlinks=true,citecolor=xlinkcolor,linkcolor=xlinkcolor,urlcolor=xlinkcolor, backref=false,pdfborder={0 0 0}]{hyperref}

\newcommand{\beqa}{\begin{eqnarray}}
\newcommand{\eeqa}{\end{eqnarray}}

\newcommand{\be}{\begin{equation}}
\newcommand{\ee}{\end{equation}}
\newcommand{\beq}{\begin{equation}}
\newcommand{\eeq}{\end{equation}}

\newcommand{\KK}{{\cal K}}

\renewcommand\k{{\bm k}}
\newcommand\q{\bm{q}}

\newcommand\GG{\Gamma_3}

\newcommand{\bseq}{\begin{subequations}}
\newcommand{\eseq}{\end{subequations}}

\renewcommand{\ln}{\mathop{\rm ln}\nolimits}

\def\ltsima{$\; \buildrel < \over \sim \;$\xspace}
\def\gtsima{$\; \buildrel > \over \sim \;$\xspace}
\def\simlt{\lower.5ex\hbox{\ltsima}}
\def\simgt{\lower.5ex\hbox{\gtsima}}

\newcommand{\hMpc}{\, h\mathrm{Mpc}^{-1}\, }

\newcommand{\lya}{Ly-$\alpha$\xspace}
\newcommand{\Lya}{Ly-$\alpha$\xspace}

\newcommand{\td}{\delta}

\newcommand{\vk}{\mathbf{k}}

\newcommand{\vq}{\mathbf{q}}
\newcommand{\vp}{\mathbf{p}}
\newcommand{\vx}{\mathbf{x}}

\def\gsim{\raise0.3ex\hbox{$\;>$\kern-0.75em\raise-1.1ex\hbox{$\sim\;$}}}
\def\lsim{\raise0.3ex\hbox{$\;<$\kern-0.75em\raise-1.1ex\hbox{$\sim\;$}}}

\def\beqn#1{\begin{equation}\label{#1}}
\def\eeqn{\end{equation}}

\def\beqa#1{\begin{eqnarray}\label{#1}}
\def\eeqa{\end{eqnarray}}

\def\hMpc{\, h{\text{Mpc}}^{-1}\xspace}

\def\Z2{$\mathcal{Z_2}$}

\newcommand {\ignore}[1]{}

\renewcommand{\arraystretch}{1.3} 

\DeclareRobustCommand{\ion}[2]{%
\relax\ifmmode
\ifx\testbx\f@series
{\mathbf{#1\,\mathsc{#2}}}\else
{\mathrm{#1\,\mathsc{#2}}}\fi
\else\textup{#1\,{\mdseries\textsc{#2}}}%
\fi}
\newcommand{\MIT}{Center for Theoretical Physics -- a Leinweber Institute, Massachusetts Institute of Technology, Cambridge, MA 02139, USA}
\newcommand{\IAIFI}{The NSF AI Institute for Artificial Intelligence and Fundamental Interactions, Cambridge, MA 02139, USA}

\newcommand{\KavliMIT}{Department of Physics and Kavli Institute for Astrophysics and Space Research, Massachusetts Institute of Technology, Cambridge, MA 02139, USA
}

\renewcommand{\vp}{\bm{p}}

\renewcommand{\GG}{\mathcal{G}}
\renewcommand{\KK}{\mathcal{K}}
\newcommand{\TT}{\mathcal{T}}
\newcommand{\cs}{\,,\quad}
\newcommand{\bpsi}{\bm\psi}

\newcommand{\zz}[1]{(#1)_{zz}}
\newcommand{\etaE}{\eta^{\rm E}}
\renewcommand{\td}[1]{\widetilde{#1}}

\begin{document}

\preprint{MIT-CTP/6150}

\title{Simulation-Based Priors and BAO Shifts from the Lyman--$\alpha$ Forest at the Field Level}

\author{Divij Sharma}
\email{d\_sharma@mit.edu}
\affiliation{\KavliMIT}
\affiliation{\IAIFI}

\author{Roger de Belsunce}
\affiliation{\MIT}
\affiliation{\IAIFI}

\author{Mikhail M. Ivanov}
\email{ivanov99@mit.edu}
\affiliation{\MIT}
\affiliation{\IAIFI}

\date{\today}

\begin{abstract}
We present an analytic forward model for the Lyman-$\alpha$ (\lya) forest flux field based on cubic order large-scale structure perturbation theory 
(effective field theory) calculations. 
We validate our cubic model on two sets of hydrodynamic simulations spanning $2 \leq z \leq 4$: the small-volume CAMELS suite, which spans a wide range of cosmological and astrophysical parameters, 
and the large-volume, high-resolution Sherwood simulations. 
We find qualitatively similar results for both at the power-spectrum level. The perturbative model reproduces the three-dimensional flux power spectrum of the simulations to $\sim 5\%$ down to $k \simeq 1$--$1.5\,h\,{\rm Mpc}^{-1}$, depending on the redshift. 
This improves compared to previous quadratic perturbative forward models by up to 40\%. 
Exploiting cosmic-variance cancellation
in the field-level approach, we carry out precision measurements of the \Lya forest bias parameters from the CAMELS simulation suite. 
We use these measurements to analytically predict the shifts in the Baryon Acoustic Oscillations (BAO) scaling parameters induced by non-linearities in the three-dimensional \Lya forest clustering as a function of cosmological parameters. We find sub-percent shifts in both the radial and transverse directions with a noticeable dependence on cosmology and redshift. Finally, from the field-level fits we construct simulation-based priors
that can be used for cosmological analyses of the \Lya forest with the currently observing Dark Energy Spectroscopic Instrument (DESI).
In addition, our cubic model captures the connected four-point function (trispectrum) of the forest, which opens up novel opportunities for 
accurate covariance matrix estimation
for DESI and its successor, DESI-II. 
\end{abstract}

\maketitle

\section{Introduction}\label{sec:intro}
The Lyman-$\alpha$ (\lya) forest provides a unique window onto the high-redshift Universe. 
Neutral hydrogen in the intergalactic medium (IGM) imprints a series of absorption features on the spectra of distant quasars. Fluctuations in the transmitted flux trace the underlying matter distribution~\cite{McQuinn:2015icp}, allowing the \lya forest to probe large-scale structure (LSS) over $2\lesssim z\lesssim5$ and on scales down to Mpc and below that are inaccessible to current galaxy surveys~\cite{Ivanov:2023yla}.

The \lya forest has already yielded a wealth of cosmological information and can be split into small high-resolution samples observed with, e.g., the High Resolution Echelle Spectrometer \citep[HIRES;][]{Vogt:1994,OMeara:2021} and the UV-Visual Echelle Spectrograph \citep[UVES;][]{Dekker:2000,Murphy:2019} and large medium-resolution samples observed with the ongoing Dark Energy Spectroscopic Instrument (DESI;~\cite{DESI:2025zpo, DESI:2025zgx}) and its predecessors~\cite {SDSS:2004aee, BOSS:2013rpr, eBOSS:2018qyj}.  The one-dimensional flux power spectrum, $P_{\rm 1D}$, measures fluctuations along individual quasar sightlines and is therefore sensitive to the matter distribution on small scales, yielding constraints on the amplitude and shape of the matter power spectrum, as well as fundamental physics including the sum of neutrino masses and non-standard dark matter scenarios~\cite{Viel:2006kd, Palanque-Delabrouille:2015pga, Garny:2020rom, Viel:2013fqw, Irsic:2017ixq, Irsic:2017yje}. Complementarily, correlations between different sightlines measure the three-dimensional clustering of the \lya forest and probe the matter distribution on large scales. In particular,  measurements of the baryon acoustic oscillation (BAO) feature provided constraints on the cosmic expansion history during matter domination~\cite{McDonald:2006qs,BOSS:2013ola, BOSS:2012gof, eBOSS:2020tmo,DESI:2024lzq, DESI:2025zpo}, and it has been used to measure the broadband shape of the large-scale three-dimensional \Lya correlation function~\citep{BOSS:2013ola, Cuceu:2021hlk, Cuceu:2022wbd, Gordon:2023dua, Cuceu:2025nvl}. Recent advances in measuring the three-dimensional flux power spectrum, $P_{\rm 3D}$, pave the way to extracting additional information from its full shape~\cite{deBelsunce:2025bqc}, similar to galaxy clustering analyses \cite{Chudaykin:2025aux}.

The \lya forest is highly complementary to current galaxy-clustering surveys. It samples a substantially larger cosmological volume, much of which will remain beyond the reach of galaxy surveys until Stage-V spectroscopic observations~\cite{Spec-S5:2025uom}. Its high redshift also places the observed modes closer to the quasi-linear regime, extending the range of scales on which perturbative techniques can be applied. Moreover, hydrodynamical simulations resolve the IGM down to $\sim$kpc scales, allowing perturbative descriptions of the forest to be calibrated with high precision~\cite{Bolton:2016bfs, CAMELS_presentation,  Chabanier:2024knr, Bird:2023evb}. Finally, the shot-noise contribution to the \lya forest decreases towards higher redshift, in agreement with scaling-Universe expectations, providing information complementary to that available from galaxy or quasar clustering alone~\cite{deBelsunce:2025bqc}. Realizing this potential requires accurate theoretical models that relate the underlying dark matter field to the observed \lya forest flux. In this work, we use the effective field theory (EFT) framework~\cite{Ivanov:2023yla,deBelsunce:2024rvv,deBelsunce:2025edy,deBelsunce:2025bqc} which provides a systematic description of the large-scale \lya forest field by including all contributions allowed by the symmetries of the tracer~\cite{McDonald:2009dh,Baumann:2010tm,Carrasco:2013mua,Ivanov:2023yla, deBelsunce:2025edy}. 

A key science driver of the currently observing DESI survey is to measure the expansion of the Universe through the measurement of the BAO feature. However, non-linearities in the matter clustering shift and damp the BAO -- leading to biases in current linear-theory-based \Lya forest analyses~\cite{Chen:2024tfp, DESI:2025zpo, DESI:2025zgx, deBelsunce:2024rvv}. Crucially, the EFT framework allows us to either consistently model these effects and unbias the BAO feature~\cite{Hadzhiyska:2025cvk} or, by determining the values of the bias parameters, quantify the magnitude of the shifts in the BAO scaling parameters to establish a systematic error budget~\cite{deBelsunce:2024rvv}. An alternative is to perform BAO reconstruction similar to galaxy clustering analyses~\cite{Hadzhiyska:2026thl}. 

In this work, we provide a systematic error budget for DESI \Lya forest analyses by extending the field-level fitting formalism of Refs.~\cite{deBelsunce:2025bqc,deBelsunce:2026tks} to cubic order for the \lya forest. Fitting the simulated flux field directly to the perturbative model exploits the shared phases between the simulation and model fields, leading to cosmic-variance cancellation and precise measurements of the EFT bias parameters. We apply this approach to two complementary suites of hydrodynamical simulations: the CAMELS simulations~\cite{CAMELS_DR1, CAMELS_DR2, CAMELS_DR3}, which span a broad range of cosmological and astrophysical parameters, and the large-volume, high-resolution Sherwood simulations~\cite{Bolton:2016bfs}. This allows us both to test the validity of the forward model and to study the dependence of the nonlinear \lya bias parameters on the underlying physical parameters.

The same bias parameter measurements provide the ingredients for enabling (and improving) full-shape measurements of the \Lya forest using simulation-based priors (SBPs;~\cite{Ivanov:2024xgb,Ivanov:2024hgq}). The large number of EFT parameters required to model the \Lya forest can be constrained using ensembles of numerical simulations to set priors on nuisance parameters. For galaxy clustering, correlated SBPs have already shown promise in improving the constraining power of EFT analyses while retaining their perturbative analytic control~\cite{Ivanov:2024xgb, Chudaykin:2026nls, Chen:2025jnr, Ivanov:2024hgq}. We extend this procedure to the \Lya forest by using the distribution of bias parameters obtained from CAMELS. These can be readily combined with full-shape analyses of the \lya forest. Together, the cubic-order field-level model, the resulting BAO-shift predictions, and the simulation-based priors provide a route towards cosmological analyses of both $P_{\rm 1D}$ and $P_{\rm 3D}$ from DESI \lya forest data.

This paper is organized as follows: In Sec.~\ref{sec:model}, we develop the cubic-order field-level EFT model of the \lya forest. Sec.~\ref{sec:sims} describes the CAMELS and Sherwood hydrodynamical simulations. In Sec.~\ref{sec:methods}, we introduce the field-level transfer-function fits, the extraction of the EFT bias parameters, and the physics causing the BAO shift. We present the performance of the forward model and the resulting bias-parameter measurements in Sec.~\ref{sec:results}, including their implications for nonlinear BAO shifts and simulation-based priors. We summarize and conclude in Sec.~\ref{sec:conclusions}.

\section{Field-level forward model}\label{sec:model}

We model the \lya flux fluctuation $\delta_F \equiv F/\bar F - 1$, where $F$ is the transmitted flux fraction, with the EFT~\cite{McDonald:2009dh, Carrasco:2012cv, Baumann:2010tm, Desjacques:BiasReview, Ivanov:2023yla}, extending the quadratic field-level model of Refs.~\cite{deBelsunce:2025bqc, deBelsunce:2026tks} to cubic order in the linear density $\delta_1$. The model is derived in Sec.~\ref{sec:Lmodel}, where an interested reader can find the details.  Here we limit ourselves to the executive summary.

\textbf{Lagrangian forward model.}
 As for galaxies, the flux field can be considered a weighted Lagrangian field mapped to redshift space~\cite{Schmittfull:2020trd},
\begin{align}
    \delta_F^s(\bm k) &= \int d^3q\,\big[1 + w(\bm q)\big]\,e^{-i\bm k\cdot[\bm q + \bm\psi^s(\bm q)]} \,,
    \label{eq:lya_forward_model_rsd}\\
    \bm\psi^s &= \bm\psi + f\,(\hat{\bm z}\cdot\dot{\bm\psi})\,\hat{\bm z} = \sum_n R^{[n]}\bm\psi^{(n)} \,,
    \nonumber
\end{align}
where $\bm\psi = \sum_n\bm\psi^{(n)}$ is the LPT displacement, $\dot{\bm\psi} = d\bm\psi/d\ln D_+$, $f = d\ln D_+/d\ln a$, $\hat{\bm z}$ is the line of sight (LOS), and $R^{[n]}_{ij} = \delta_{ij} + nf\,\hat z_i\hat z_j$.
Fundamentally, the weight $w(\q)$ is the most general local function of the distortion tensors $\partial_i\psi^{(n)}_j$ allowed by the symmetries of the forest, i.e.\ rotations about the LOS. Through cubic order, it has 22 independent terms,
which matches the number of Eulerian
Lyman-$\alpha$
bias operators at
the third order.

To ease comparison with
the literature, 
we equivalently rewrite 
our model 
in the basis of operators from Refs.~\cite{Desjacques:2018pfv, Ivanov:2023yla}.
This amounts to a simple rotation in the operator space. 
Using $O_L$ to denote the Lagrangian seed of each
Eulerian operator (e.g.\ $\delta_L = \delta_1$, $\eta_L = -f\eta + \dots$  etc.), the Lagrangian bias model can be written as
\begin{equation}
\begin{aligned}
    w ={}& \sum_O b^{\rm L}_O\,O_L(\bm q) \,,\\
    \{O\} ={}& \big\{\delta_1,\ \eta\big\}\cup \big\{\delta_1^2,\ \mathcal G_2,\ \delta_1\eta,\ \eta^2,\ (KK)_\parallel,\ \Pi^{[2]}_\parallel\big\}\\
    &\cup \big\{\delta_1^3,\ \delta_1\mathcal G_2,\ \mathcal G_3,\ \Gamma_3,\ \eta\delta_1^2,\ \eta^2\delta_1,\ \eta^3,\\
    &\qquad \delta_1(KK)_\parallel,\ \eta(KK)_\parallel,\ \delta_1\Pi^{[2]}_\parallel,\ \eta\Pi^{[2]}_\parallel,\\
    &\qquad \eta\mathcal G_2,\ (K\Pi^{[2]})_\parallel,\ \Pi^{[3]}_\parallel\big\} \,.
\end{aligned}
\label{eq:lya_cubic_bias_expansion}
\end{equation}
All operators are built from the linear tidal tensor $\Pi_{ij} = \partial_i\partial_j\nabla^{-2}\delta_1$ and its LOS projections ($\mathcal O_\parallel \equiv \hat z^i\hat z^j\partial_i\partial_j\nabla^{-2}\mathcal O$). In particular
\begin{equation}
\begin{aligned}
    &\eta \equiv \hat z^i\hat z^j\Pi_{ij} \,,\qquad
    K_{ij} = \Pi_{ij} - \tfrac13\delta_{ij}\delta_1 \,,\\
    &\mathcal G_2 = \Pi_{ij}\Pi_{ji} - \delta_1^2 \,,\qquad
    (KK)_\parallel = \hat z^iK_{ik}K_{kj}\hat z^j \,,
\end{aligned}
\end{equation}
and $\Pi^{[2]}$, $\Pi^{[3]}$ are the members of the convective hierarchy of Refs.~\cite{Mirbabayi:2014zca,Desjacques:BiasReview}. $\mathcal G_3$, $\Gamma_3$ and the remaining operators have the standard definitions given in Sec.~\ref{sec:Lmodel}. The linear LOS velocity gradient is $\partial_\parallel v_\parallel/(aH) = -f\eta$, so at linear order $\delta_F^s = (b_1 - b_\eta f\mu^2)\,\delta_1$ with $b_1 = 1 + b^{\rm L}_\delta$ and $b_\eta = b^{\rm L}_\eta - 1$~\cite{McDonald:1999dt, Seljak:2012tp}. The operators $(KKK)_\parallel$ and $\eta K^2$ of Ref.~\cite{Desjacques:2018pfv} are not independent: by the Cayley--Hamilton theorem and the identity $K^2 = \mathcal G_2 + \tfrac23\delta_1^2$ they are combinations of the operators above, so we ignore these operators in what follows.

\textbf{Shifted operators.}
 We keep the linear redshift-space displacement $R^{[1]}\bm\psi^{(1)}$ in the exponent of our Lagrangian model in order to resum the large-scale bulk flows, which are not small even when the density is perturbative~\cite{Schmittfull:2018yuk, Modi:2019qbt}. The remainder of $\bm\psi^s$ is expanded to cubic order. Every operator then enters as a shifted field
\begin{equation}
    \widetilde O(\bm k) \equiv \int d^3q\,O(\bm q)\,e^{-i\bm k\cdot[\bm q + R^{[1]}\bm\psi^{(1)}(\bm q)]} \,,
    \label{eq:shifted_operator}
\end{equation}
and $\widetilde 1 \equiv \delta_Z$ is the redshift-space Zel'dovich density. Two features of the expansion fix the form of the linear operators:
First, the shifted velocity field is not an independent operator. Mass conservation of the Zel'dovich map gives the exact identity
\begin{equation}
\begin{aligned}
    f\,\widetilde\eta ={}& \delta_Z - \widetilde\delta_1 - \tfrac12\widetilde{\mathcal G}_2 - f\,\widetilde{(KK)}_\parallel\\
    &+ \tfrac f3\,\widetilde{\delta_1\eta} + \tfrac f9\,\widetilde{\delta_1^2} - \tfrac{1+f}{3}\,\widetilde{\mathcal G}_3 \,,
\end{aligned}
\label{eq:eta_identity}
\end{equation}
so $\eta$ is traded for $\delta_Z$ as in the forward model of~\cite{deBelsunce:2025bqc}.
Second, the higher-order displacements generate shift operators,
\begin{equation}
\begin{aligned}
    &S_3 = \bm\psi^{(2)}\cdot\nabla\delta_1 \,,\qquad
    S_3^\delta = R^{[2]}_{ij}\psi^{(2)}_j\partial_i\delta_1 \,,\\
    &S_3^\eta = f R^{[2]}_{ij}\psi^{(2)}_j\partial_i\eta \,,\qquad
    \Psi_2^\eta = \tfrac{14}{3}\,\psi^{(2)}_\parallel\partial_\parallel\eta \,,\\
    &\mathcal G_2^\parallel = \hat z^i\hat z^j\partial_i\partial_j\nabla^{-2}\mathcal G_2 \,,
\end{aligned}
\end{equation}
whose coefficients are not free: they are fixed by the equivalence principle in terms of the two linear bias parameters, together with the transverse part of $\bm\psi^{(3)}$, whose derivative along the LOS does not vanish. Since their coefficients are fixed by the linear terms, it is natural to combine these cubic operators with their linear counterparts:
\begin{align}
    \delta_1^{\prime} &\equiv \widetilde\delta_1 - \widetilde S_3^\delta \,,\nonumber\\
    \eta^{\prime} &\equiv \eta_{\rm new} - \widetilde S_3^\delta - 2\,\widetilde S_3^\eta + \tfrac{3f^2}{7}\,\widetilde\Psi_2^\eta + f\mu^2\widetilde S_3 \,,\label{eq:eta_new}\\
    \eta_{\rm new} &\equiv \delta_Z - \tfrac{3f}{7}\,\widetilde{\mathcal G}_2^\parallel \,.\nonumber
\end{align}
At quadratic order only $\eta_{\rm new}$ is needed~\cite{deBelsunce:2026tks}. $\delta_1^{\prime}$ and $\eta^{\prime}$ are the cubic completions.

With these definitions, the redshift-space flux field depends on 22 operators,
\be 
\begin{aligned}
    \delta_F^s ={}& \beta_1\,\delta_1^{\prime} + \beta_\eta\,\eta^{\prime}
    + \sum_{O\,\in\,\rm quadratic}\beta_O\,\widetilde O\\
    &+ \sum_{O\,\in\,\rm cubic}\beta_O\,\tilde{O} + \epsilon \,,
\end{aligned}
\label{eq:field_level_model}
\ee
where the sums run over the quadratic and cubic operators of eq.~\eqref{eq:lya_cubic_bias_expansion}, and $\epsilon$ is the stochastic part, uncorrelated with all operators.  The coefficients $\beta_O$ are linear functions of the Eulerian bias parameters of the one-loop CLASS-PT model of Ref.~\cite{Ivanov:2023yla} (Sec.~\ref{sec:Lmodel}). The cubic operators are renormalized so that they carry no UV-divergent overlap with $\delta_1$. 

\textbf{Transfer functions.}
In the fit, we promote the above $\beta_O$ to free transfer functions $\beta_a(k,\mu)$, which absorb higher-order corrections.
To remove the degeneracies in our basis and make the transfer function measurements more stable,  we orthogonalize the operators in each $(k,\mu)$ bin with the Gram--Schmidt procedure, $\langle\widetilde O_a^\perp\widetilde O_b^{\perp*}\rangle \propto \delta_{ab}$:
\begin{equation}
    \delta^s_{F,~\rm model}(\bm k) = \sum_a\beta_a(k,\mu)\,\widetilde O_a^\perp(\bm k) + \epsilon(\bm k) \,,
    \label{eq:field_level_transfer_model}
\end{equation}
with the operator list
\begin{equation}
\begin{aligned}
    \{O_a\} = \{&\delta_1^{\prime},\ \eta^{\prime},\ \delta_1^2,\ \mathcal G_2,\ \eta^2,\ \delta_1\eta,\ (KK)_\parallel,\\
    &\delta_1^3,\ \delta_1\mathcal G_2,\ \mathcal G_3,\ \Gamma_3,\ \delta_1\Pi^{[2]}_\parallel,\ \eta\Pi^{[2]}_\parallel,\\
    &\Pi^{[3]}_\parallel,\ \eta\mathcal G_2,\ \eta\delta_1^2,\ \delta_1(KK)_\parallel,\ \eta^2\delta_1,\\
    &\eta(KK)_\parallel,\ \eta^3,\ \mathcal K_3\} \,.
\end{aligned}
\label{eq:final_operator_basis}
\end{equation}
Two operators of eq.~\eqref{eq:lya_cubic_bias_expansion} are treated specially in the model.  $(K\Pi^{[2]})_\parallel$ is replaced by $ \mathcal K_3$ defined as~\cite{Schmittfull:2020trd}
\begin{equation}
    \mathcal K_3 = \hat z^i\,\Pi_{ik}\,\partial_k\partial_j\nabla^{-2}\mathcal G_2\,\hat z^j \,,
\end{equation}
which spans the same space modulo other local cubic operators in the basis. $\Pi^{[2]}_\parallel$ is not fitted separately, because at quadratic order it is fully degenerate with $(KK)_\parallel$, $\delta_1\eta$, $\delta_1^2$ and $\mu^2\mathcal G_2$
once the bias parameters 
are promoted to the transfer functions~\cite{deBelsunce:2025bqc}.

For comparison, we also use the linear and quadratic models of Ref.~\cite{deBelsunce:2025bqc}, with
\begin{equation}
\begin{aligned}
    \{O_a\}_{\rm lin} ={}& \{\widetilde\delta_1,\ \widetilde\eta\} \,,\\
    \{O_a\}_{\rm quad} ={}& \{\widetilde\delta_1,\ \eta_{\rm new},\ \widetilde{\delta_1^2},\ \widetilde{\mathcal G}_2,\ \widetilde{\eta^2},\ \widetilde{\delta_1\eta},\\
    &\ \widetilde{(KK)}_\parallel,\ \delta_1^3\} \,,
\end{aligned}
\end{equation}
respectively; $\delta_1^3$ is included in the quadratic model for direct comparison with Refs.~\cite{deBelsunce:2025bqc, deBelsunce:2026tks}.

Given a ground truth
Lyman-$\alpha$ field
$\delta_F^s$, the optimal
best-fit transfer function in each bin
$\k$ follows from the minimization of the functional $\langle|\delta^s_{F,~\rm sim}(\k)-\delta^s_{F,~\rm model}(\k)|^2\rangle $,
which  yields:
\begin{equation}
    \beta_a(k,\mu)
    =
    \frac{
        \left\langle
       \delta^s_{F,~\rm sim}(\bm k)
        \widetilde{\mathcal O}_a^{\perp *}(\bm k)
        \right\rangle'
    }{
       D(\k)
    }\,,
    \label{eq:beta_measured}
\end{equation}
where $D(\k)\equiv \left\langle
        |\widetilde{\mathcal O}_a^\perp|^2
        \right\rangle'$ and  the primes denote the Dirac delta-functions
stripped off. In terms of the original, non-orthogonalized fields, the transfer functions are defined through the Cholesky factorization
\begin{equation}
\begin{aligned}
    \beta_a(k,\mu)
=
    \frac{1}{
        D(\k)
    }
    \sum_{A,B}
        M_{aB}\,
        \beta_A\,
        \left\langle
        \widetilde{\mathcal O}_A
        \widetilde{\mathcal O}_B^*
        \right\rangle' \,,
\end{aligned}
\label{eq:beta_gram_schmidt}
\end{equation}
where $M$ is cross-power matrix of the non-orthogonalized shifted fields, which is the usual Gram matrix, 
\begin{equation}
    M_{ab}(k)
    =
    \left\langle
        \widetilde{\mathcal O}_a(\bm k)
        \widetilde{\mathcal O}_b^*(\bm k)
    \right\rangle'.
    \label{eq:operator_gram}
\end{equation}
The residual between the theory and the truth captures the stochasticity of the Lyman-$\alpha$  field. Its power spectrum is called the error power spectrum, 
\be 
P_{\rm err}(k)=\langle |\epsilon(\k)|^2\rangle'\,.
\ee 
Note that the field-level model we use is formally equivalent to the two-loop
EFT expansion at the level of the flux power spectrum. The reason is that the transfer functions, by construction, capture 
all correlations between the full deterministic 
field and the EFT fields through the cubic order~\cite{Baldauf:2015zga}.
Therefore, the linear transfer functions 
capture $\langle \delta_1 \delta_5\rangle$, the quadratic ones
account for $\langle \delta_2 \delta_4\rangle$, while the cubic ones
capture $\langle \delta_3 \delta_3\rangle$ correlations. 
This exhaust all two-loop diagrams~(see e.g.~\cite{Carrasco:2013mua,Ivanov:2026zos}). The first power spectrum correction
not captured by our model is 
$\langle \delta_4 \delta_4\rangle$,
which is a three-loop diagram.

\section{Simulation data}\label{sec:sims}

We validate our field-level model on two sets of hydrodynamical simulations of the \lya forest: CAMELS-SB35 \cite{CAMELS_presentation} and Sherwood \cite{Bolton:2016bfs}. 

For testing our field-level model over a broad range of cosmologies and astrophysics, we employ the second generation of the Cosmology and Astrophysics with MachinE Learning Simulations (CAMELS) project \cite{CAMELS_presentation, CAMELS_DR1, CAMELS_DR3}. We use the SB35 set of the IllustrisTNG suite \cite{Pillepich:2017jle}, consisting of 1024 hydrodynamical simulations. These simulations are run in $(50h^{-1}{\rm Mpc})^3$ boxes with $512^3$ dark matter and gas particles, and sample a 35-dimensional parameter space of 5 cosmological and 30 astrophysical parameters. The 5 cosmological parameters are $[{\Omega_m,\Omega_b,h,n_s,\sigma_8}]$, and the astrophysical parameters vary aspects of the IllustrisTNG subgrid model \cite{Pillepich:2017jle, Weinberger:2016ueb}. In particular, the second generation suite extends the astrophysical parameter space of previous CAMELS simulations to include variations in amplitude and timing of the UV background, which are relevant for \lya observables. The 35-dimensional space is sampled in a Sobol sequence for uniform coverage, and simulations have different random seeds for their initial conditions. For this work, we analyze the \lya flux fields at redshifts $z \in [2,3,4]$. For each simulation and redshift, we consider three LOS directions corresponding to the three Cartesian axes of the simulation box. Unless otherwise stated, all results presented below are first averaged over these three LOS directions before computing statistics over the simulation ensemble.

In this work, we use simulation SB35-637 as a representative example for direct comparisons between the linear, quadratic, and cubic-order forward models. Within the full suite, this simulation lies closest to the Planck best-fit cosmology \cite{Planck:2013pxb}, with $(\Omega_m,\Omega_b,h,n_s,A_s) =(0.2998,\,0.0497,\,0.6606,\,0.9462,\,2.13\times10^{-9})$, and we therefore refer to it as our near-Planck simulation throughout the following sections. To test the robustness
of our model across different cosmologies, we also use simulation SB35-465, which lies far from Planck, with $(\Omega_m,\Omega_b,h,n_s,A_s)
=(0.1588,\,0.0366,\,0.5734,\,1.0277,\,2.13\times10^{-9})$. We refer to this as our far-from-Planck simulation throughout the following sections.

In addition to CAMELS, we use a higher-resolution simulation from the Sherwood suite \cite{Bolton:2016bfs}. We use two series of Sherwood snapshots. First, one at redshift $z=2.8$ in a $(160\,h^{-1}{\rm Mpc})^3$ comoving volume with $2048^3$ dark matter and gas particles each. Second, a series of four snapshots at redshifts $z=2.0,\, 2.4,\, 2.8,\, 3.2$ in a $(80\,h^{-1}{\rm Mpc})^3$ comoving volume with $1024^3$ dark matter and gas particles each. The $L=160\,h^{-1}{\rm Mpc}$ box will be used for our key Sherwood results.
The initial conditions are generated using the \texttt{N-GenIC} code \cite{Springel:2005mi}. This simulation adopts a fiducial cosmology based on the Planck 2013 best-fit values \cite{Planck:2013pxb}, with $(\Omega_m,\Omega_b,h,\sigma_8,n_s) = (0.308,\,0.0482,\,0.678,\,0.829,\,0.961)$. 
The advantage of using Sherwood is access to a substantially larger volume and higher resolution to calibrate our field-level techniques with access to more quasi-linear modes. 

\section{Methods}\label{sec:methods}
In this section, we describe how we apply our cubic-order \lya forest flux forward model to the hydrodynamic simulations introduced in Sec.~\ref{sec:sims}. We first explain how our measured transfer functions are fit and matched to the perturbative model to infer bias coefficients and counterterms in Sec. \ref{sec:fitting}. We then describe how we use these coefficients to reconstruct the
three-dimensional and one-dimensional \lya flux
power spectra in Secs. \ref{sec:Pff_reconstruction} and \ref{sec:P1D}, respectively, providing an independent
validation of the inferred parameters.
Next, we explain how the distribution of EFT coefficients across the CAMELS-SB35 suite can be
used to construct simulation-based priors in Sec. \ref{sec:SBP_methods}. Finally, we describe how the nonlinear bias coefficients
are used to predict the BAO shifts in Sec. \ref{sec:bao_shift}. 

\subsection{Transfer function fits}\label{sec:fitting}

Our forward model, as given in Eqs. \ref{eq:field_level_transfer_model} and \ref{eq:final_operator_basis}, benefits from cosmic variance cancellation since we use the same initial conditions for the forward model as for the simulation while measuring the transfer functions \cite{Schmittfull:2018yuk, Schmittfull:2020trd, deBelsunce:2025bqc, deBelsunce:2026tks}. This enables us to get tight constraints on bias parameters from the transfer functions without cosmic variance. Hence, field-level measurements of EFT parameters provide an optimal path for constructing simulation-based priors \cite{Ivanov:2024xgb} for cosmological analyses.

As elaborated in Ref. \cite{deBelsunce:2026tks}, the key idea is that on sufficiently large scales, the transfer functions, $\beta_a(k,\mu)$, are fixed by the scale-independent coefficients of the bias expansion, $b_a$. 
To extract the underlying fundamental 
bias parameters, we forward model the orthogonalization procedure and fit the transfer functions simultaneously.

The errors on transfer functions can be estimated analytically~\cite{Schmittfull:2018yuk,Kokron:2026vgd}. For the orthogonalized fields they are
uncorrelated and given by the error power spectrum of the model,
\begin{equation}
{\rm Var}\big(\beta^\perp_i\big)=\frac{P_{\rm err}(k,\mu)}{N(k,\mu)\,D_i(k,\mu)}\,,
\end{equation}
where $i$  is the bin index, and $N$ is the number of Fourier modes in the bin.
Having the theory prediction then allows us to form a Gaussian likelihood over the 
transfer functions in the $(k,\mu)$ bins.

The theory prediction for the same
quantities is obtained in two steps. First, the EFT expansion fixes the
coefficient of every operator in terms of the $22$ fundamental scale-independent bias parameters
$b_j$. Second, these coefficients are projected onto the orthogonalized operators with
the measured Gram matrix, which reproduces the orthogonalization applied to the data.
This way we forward model the transfer functions using the standard EFT expansion 
\eqref{eq:field_level_model}.

The operator $\Pi^{[2]}_\parallel$, which is not among the $21$ measured transfer
functions, enters through its projection on the other operators. Its bias parameter is
constrained only by the angular dependence of the transfer functions. Each transfer
function additionally receives a counterterm $k^2(c_0+c_1\mu^2+c_2\mu^4)$ that absorbs
higher-order corrections. Those of the linear operator are the counterterms of the
one-loop power spectrum, and the remaining ones are nuisance parameters.

The resulting model is linear in the bias parameters and counterterms, so the Gaussian
likelihood of all transfer functions in all bins with $k_{\min}\leq k\leq k_{\max}$ is
maximized in closed form by a single weighted least-squares solve, which also yields the
parameter covariance marginalized over the counterterms. For the Sherwood box and
$k_{\max}=1\,h\,{\rm Mpc}^{-1}$ this amounts to $1428$ data points for $85$ parameters.
We choose $k_{\max}$ as the largest value for which bias parameter
shifts by less than $3\sigma$ relative to a fit at low $k_{\max}$. The fitted bias
parameters and linear-operator counterterms are then used in the one-loop power spectrum
model reconstruction and 1D power spectrum fits (see below). Technical details of the procedure are given in App.~\ref{app:fit}.

\subsection{Power spectrum reconstruction}\label{sec:Pff_reconstruction}
Once we have the Eulerian bias parameters from the transfer functions, we can use them for a perturbative prediction to reconstruct the \lya three-dimensional power spectrum. This offers an additional test of the field-level model and our bias extraction and also allows us to extend our \lya modeling to larger scales \cite{deBelsunce:2026tks}.

We reconstruct the power spectrum using the one-loop EFT model \cite{Ivanov:2023yla} in \texttt{CLASS-PT} \cite{Chudaykin:2020aoj},
\begin{equation}
\begin{aligned}
    P_F(k,\mu)
    ={}&
    P_F^{\rm tree}(k,\mu)
    + P_F^{\rm 1-loop}(k,\mu)
    \\
    &+ P_F^{\rm ct}(k,\mu)
    + P_F^{\rm st}(k,\mu).
\end{aligned}
\label{eq:PF_EFT}
\end{equation}
The tree-level contribution is \cite{Ivanov:2023yla}
\begin{equation}
    P_F^{\rm tree}(k,\mu)
    =
    K_1^2(\bm k)P_{\rm lin}(k),
    K_1(\bm k)
    =
    b_1-b_\eta f\mu^2,
    \label{eq:PF_tree}
\end{equation}
The one-loop contribution is given by \cite{Ivanov:2023yla}
\begin{align}
    P_F^{\rm 1-loop}(k,\mu)
    ={}&
    2\int_{\bm q}
    K_2^2(\bm q,\bm k-\bm q)
    P_{\rm lin}(q)
    P_{\rm lin}(|\bm k-\bm q|)
    \nonumber\\
    &+
    6K_1(\bm k)P_{\rm lin}(k)
    \int_{\bm q}
    K_3(\bm k,-\bm q,\bm q)
    P_{\rm lin}(q),
    \label{eq:PF_oneloop}
\end{align}
Here, $\bm k$ is the Fourier wavevector, $k=|\bm k|$ its magnitude, and $\mu=k_\parallel/k=\hat{\bm k}\cdot\hat{\bm z}$ the cosine
of the angle between the wavevector and the line of sight $\hat{\bm z}$. The variable $\bm q$ denotes the internal momentum integrated over in the loop contributions, with $\int_{\bm q}\equiv
\int {\rm d}^3q/(2\pi)^3$.

The redshift-space kernels $K_2$, $K_3$ [as given in eq. (3.19) of \cite{Ivanov:2023yla}] contain the quadratic and cubic bias parameters, which we set to our best-fit values obtained from the transfer functions. 
The counter terms are parametrized by \cite{Ivanov:2023yla}
\begin{equation}
    P_F^{\rm ct}(k,\mu)
    =
    -2
    \left(
        c_0+c_2\mu^2+c_4\mu^4
    \right)
    K_1(\bm k)k^2P_{\rm lin}(k).
    \label{eq:PF_counterterms}
\end{equation}
 And, finally, the stochastic contribution is given by
\begin{equation}
P_F^{\rm st}(k,\mu)
=
P_{\rm shot}
+
\frac{k^2}{k_{\rm NL}^2}
\left(
a_0+a_2\mu^2+a_4\mu^4
\right),
\label{eq:PF_stochastic}
\end{equation}
where $k_{\rm NL}$ is the non-linear scale which we use as a normalization  ($k_{\rm NL}\simeq 5~\hMpc$ for $z=3$), $P_{\rm shot}$ is the large-scale constant contribution, and the remaining terms capture the leading scale- and angle-dependent corrections \cite{deBelsunce:2024rvv, Hadzhiyska:2025cvk}. 

The above expression can be used as a model
for $P_{\rm err}$, which indeed captures its scale and angle-dependence quite well. 
Note, however, that 
$P_{\rm shot},a_{0,2,4}$
are counterterms in EFT, which have
both the finite parts, responsible for the 
error power spectrum, and ``infinite''
parts that subtract the errors of the $P_{22}$-like loop integrals~\cite{Ivanov:2026nlf} 
The ``infinite'' parts are numerically rather small for the $\Lambda$CDM power
spectrum at the redshifts of interest.
However, since the ``finite'' parts of
$P_{\rm shot},a_{0,2,4}$ are small too,
which implies that we cannot use the 
$P_{\rm err}$-based measurements
of $P_{\rm shot},a_{0,2,4}$
in our power spectrum predictions.
Because of that, we treat $P_{\rm shot},a_{0,2,4}$ as free parameters
in the $P_{FF}$ fits, while all other parameters, i.e. bias and counterterms, 
are fixed to their field-level transfer 
function-based values.

\subsection{1D \texorpdfstring{\lya}{Lyα} power spectrum}\label{sec:P1D}
The one-dimensional \lya flux power spectrum can be obtained by integrating the three-dimensional power spectrum over transverse modes \cite{Ivanov:2023yla, McDonald:2001fe},
\begin{equation}
    P_{\rm 1D}(k_\parallel)
    =
    \frac{1}{2\pi}
    \int_{k_\parallel}^{\infty}
    {\rm d}k\, k\,
    P_F(k,k_\parallel).
    \label{eq:P1D}
\end{equation}
This integral, however, is done over modes that lie outside perturbative control and are not captured by our forward model. Hence, following the EFT formalism, the integral is evaluated only up to a finite cutoff $k_{\parallel,~\rm max}$. 
In practice, we use $k_{\parallel,~\rm max}=k_{\rm max}$ for the integral over the loop contributions
and a fixed $k_{\parallel,~\rm max}=20~\hMpc$ for the integral over the 
tree-level part, which is appropriate given
a weak sensitivity of the 1D integrals to $P_F^{\rm tree}$ and a strong sensitivity to $P_F^{\rm 1-loop}$. This choice is simply a particular renormalization scheme.

Since the theory should not depend on the arbitrary cutoff scale, suitable counterterms are added to remove the cutoff dependence. Specifically, following \cite{Ivanov:2024jtl, Garny:2020rom, Ivanov:2023yla}, we parametrize the counterterm contributions as the following,\footnote{We ignore the logarithmic corrections~\cite{Ivanov:2023yla}. These are found to be negligibly small both for Sherwood and CAMELS. }
\begin{equation}
    P_{\rm 1D}^{\rm ctr}(k_\parallel)
    =
    C_0
    + C_2 k_\parallel^2
    + C_4 k_\parallel^4 ,
    \label{eq:P1D_counterterms}
\end{equation}
such that the full one-dimensional model is \cite{Ivanov:2023yla}
\begin{equation}
    P_{\rm 1D}^{\rm model}(k_\parallel)
    =
    P_{\rm 1D}(k_\parallel;k_{\parallel~\rm max})
    + P_{\rm 1D}^{\rm ctr}(k_\parallel).
    \label{eq:P1D_model}
\end{equation}
Here, $C_0, C_2, C_4$ absorb the sensitivity of the model to short-scale physics not explicitly captured by our forward model.

\subsection{Constructing simulation-based priors}\label{sec:SBP_methods}
The field-level measurements of the EFT bias coefficients can be used to construct simulation-based priors for cosmological analyses of the \lya forest \cite{Ivanov:2024xgb}.
Cosmic variance cancellation in the field-level fits allows us to measure these parameters for individual simulations, and the CAMELS suite provides a broad sampling of cosmological and astrophysical parameter space.

For each simulation and redshift, we extract
the vector of EFT coefficients,
\begin{equation}
\bm b_i(z)
=
\left(
b_{1,i},b_{\eta,i},b_{2,i},
b_{\mathcal G_2,i},\ldots
\right),
\end{equation}
where $i$ labels the CAMELS simulation. The ensemble of these measurements characterizes the joint distribution of the coefficients over the parameter space sampled by the simulation suite. 

We use this ensemble to capture correlations and degeneracies among EFT parameters. These correlations are important for cosmological constraints, see e.g. \cite{Ivanov:2024xgb}.

\subsection{Nonlinear BAO shifts}\label{sec:bao_shift}
The bias coefficients measured using our model can also be used to predict the nonlinear shift of the BAO feature in the \lya forest. Nonlinear structure formation modifies the BAO signal through long-wavelength displacements broadening the acoustic feature and mode coupling shifting the inferred BAO scale. For the \lya forest, the shift depends on the nonlinear LOS-dependent bias coefficients \cite{deBelsunce:2024rvv, Chen:2024tfp, Hadzhiyska:2025cvk}.

We follow the treatment of Ref. \cite{deBelsunce:2024rvv}. Conventionally, BAO information is described by the radial and transverse dilation parameters, 
\begin{equation}
    \alpha_\parallel
    =
    \frac{H^{\rm fid}(z)\,r_d^{\rm tem}}
         {H(z)\,r_d},
    \qquad
    \alpha_\perp
    =
    \frac{D_A(z)\,r_d^{\rm tem}}
         {D_A^{\rm fid}(z)\,r_d},
\label{eq:Alphas_def}       
\end{equation}
where $H(z)$ and $D_A(z)$ are the Hubble parameter and angular
diameter distance, respectively, and $r_d$ is the sound horizon at
the drag epoch. Equivalently, this information can be recast using the isotropic and Alcock--Paczynski combinations
\begin{equation}
    \alpha_{\rm iso}
    =
    \left(\frac{\alpha_\parallel}{\alpha_\perp^2}\right)^{1/3},
    \qquad
    \alpha_{\rm AP}
    =
    \frac{\alpha_\parallel}{\alpha_\perp}.
\label{eq:isoAP_def}
\end{equation}
A shift in the physical BAO feature away from linear theory used in conventional analyses therefore produces corresponding shifts
$\Delta\alpha_\parallel$ and $\Delta\alpha_\perp$ in the inferred
distance scales.

At next-to-leading order, the origin of this effect comes from the quadratic-quadratic contribution to the one-loop power spectrum,
\begin{equation}
    P^{(22)}_F(\bm{k})
    =
    2
    \int_{\bm{p}}
    \left[
        K_2(\bm{p},\bm{k}-\bm{p})
    \right]^2
    P_{\rm lin}(p)
    P_{\rm lin}(|\bm{k}-\bm{p}|),
    \label{eq:bao_p22}
\end{equation}
where $K_2$ is the quadratic \lya kernel as defined in Ref.~\cite{Ivanov:2023yla}.
The contribution responsible for shifting the BAO arises from the coupling between modes much longer than the acoustic scale and modes carrying the BAO feature. The relevant quadratic operators include a
long-wavelength displacement acting on the density,
$\bm{\Psi}\cdot\nabla\delta$, and local nonlinear clustering,
$\delta^2$. Their cross-correlation generates a contribution of the
form
\begin{equation}
    \left\langle
        \bm{\Psi}(\bm{x})\,\delta(\bm{0})
    \right\rangle
    \cdot\nabla\xi_{\rm w}(\bm{x}),
\end{equation}
where $\xi_{\rm w}$ denotes the oscillatory part of the power spectrum. Because the displacement is correlated with the large-scale density,
\begin{equation}
    \left\langle
        \bm{\Psi}(\bm{x})\,\delta(\bm{0})
    \right\rangle
    \simeq
    -\frac{\bm{x}}{3}\,
    \left\langle\delta^2\right\rangle_x ,
\end{equation}
this coupling coherently dilates or contracts the acoustic feature.
Only modes with wavelengths larger than the BAO scale contribute to this effect. Its amplitude is therefore proportional to
the mean-square long-wavelength density fluctuation on the acoustic
scale, which we denote by $\sigma_d^2$
\cite{deBelsunce:2024rvv}. The complete analytic form of the resulting RSD nonlinear \lya forest BAO shift can be found in Appendix A of \cite{deBelsunce:2024rvv}, first derived in the context of galaxy clustering in Ref.~\cite{Chen:2024tfp}. Importantly, the BAO shift is not given by the linear-order models alone and requires the nonlinear bias values \cite{deBelsunce:2024rvv}, where our work can be especially useful. 

For each set of nonlinear bias fits, we therefore evaluate the corresponding shifts in the radial and transverse BAO shift parameters,  $\Delta\alpha_\parallel$ and $\Delta\alpha_\perp$. The resulting shifts depend on the linear \lya biases as well as the quadratic coefficients $\{ b_2,\, b_{\mathcal G_2},\, b_{\delta\eta},\, b_{\eta^2},\, b_{(KK)_\parallel},\, b_{\Pi_\parallel^{[2]}}\}$, and on the amplitude of long-wavelength density fluctuations on the
BAO scale. We refer the reader to Ref. \cite{deBelsunce:2024rvv} for
the complete expression and details of its projection onto the BAO shift parameters.

\section{Results}\label{sec:results}

In this section, we present the results of applying our analytic forward model, introduced in Sec.~\ref{sec:model}, to the Sherwood and CAMELS-SB35 simulations, introduced in Sec.~\ref{sec:sims}. We use the larger volume and higher resolution of the Sherwood simulation to establish the accuracy of our perturbative description before extending the analysis to the CAMELS-SB35 suite. The CAMELS suite then allows us to demonstrate our model performance across a broad range of cosmological and astrophysical parameters.

We begin by validating different statistics of our model against the true simulation statistics. First, we validate our model at the power spectrum level
against simulations and compare the performance against linear- and quadratic-order models in Sec. \ref{sec:2ptResults}. 
We then present performance at the level of phases in Sec. \ref{sec:rccResults}, followed by performance directly at the field level through configuration-space flux maps. 

After validating model performance, we move towards results using extracted EFT coefficients. We present the measured transfer functions in Sec. \ref{sec:TFandBiasResults}. We then present results on bias extraction from transfer functions and their joint distribution across the CAMELS-SB35 suite in Sec. \ref{sec:SBP}.
Using nonlinear biases, we present our power spectrum reconstruction results in Sec. \ref{sec:PffReconstructionResults}.
In Sec. \ref{sec:BAOshiftResults}, we present our BAO shift predictions on the CAMELS suite. 

\subsection{Power spectrum validation}\label{sec:2ptResults}

\begin{figure*}
    \centering
    \includegraphics[width=\textwidth]{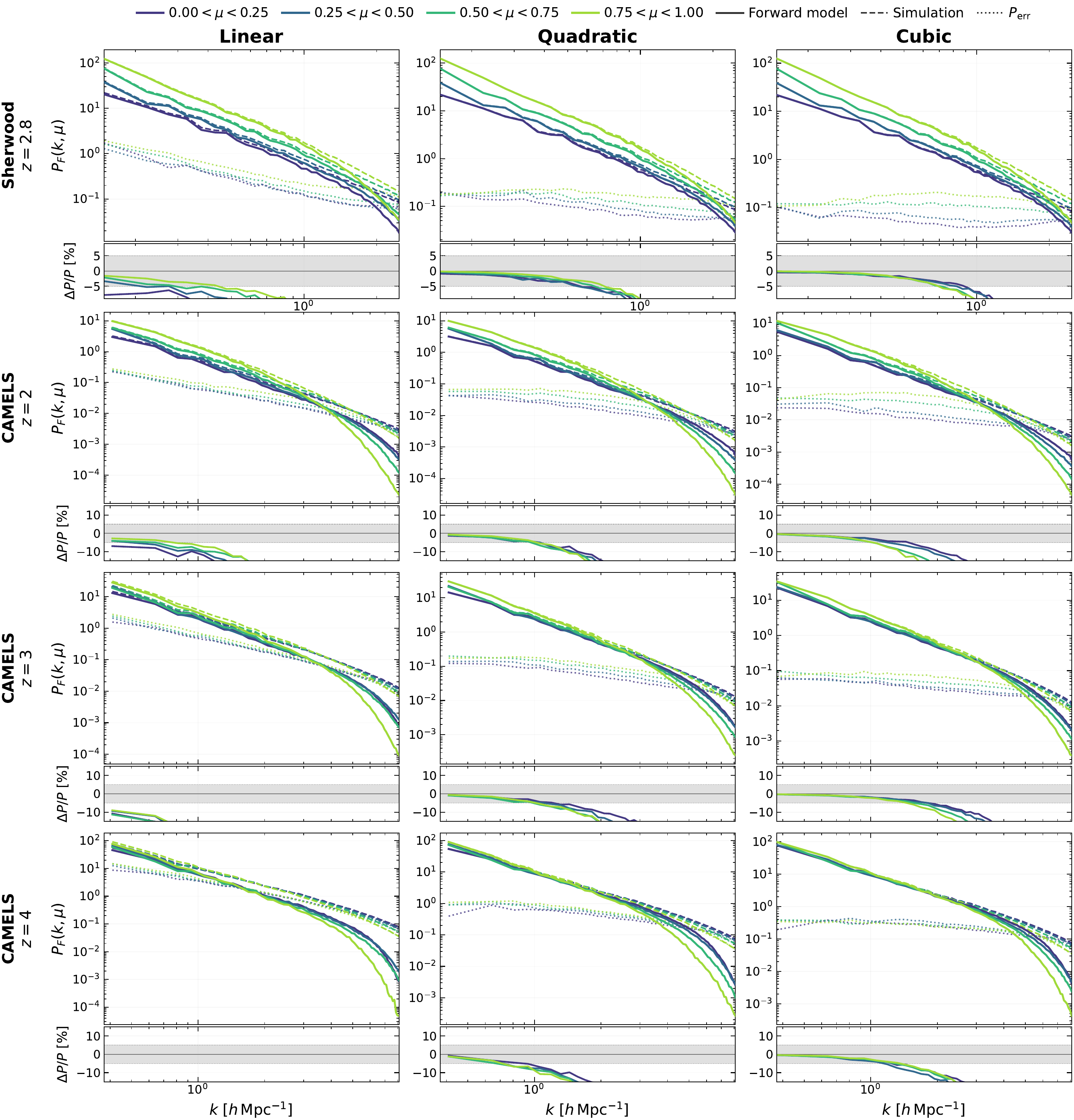}
    \caption{\textbf{Comparison of the three-dimensional flux power spectrum performance of linear, quadratic, and cubic EFT models.}
    The left, middle, and right columns show the linear-, quadratic-, and cubic-order predictions, respectively, while the four rows correspond to the higher-resolution Sherwood simulation at $z=2.8$ and the near-Planck CAMELS simulation at $z=2,3,4$, respectively. In each panel, the four colors denote the LOS angular bins $0<\mu<0.25$, $0.25<\mu<0.50$, $0.50<\mu<0.75$, and $0.75<\mu<1$, where $\mu=k_\parallel/k$, in purple, blue, green, and light green, respectively. For CAMELS, the measurements are averaged over the three LOS directions, $x,y,z$, while the Sherwood results use the $z$-direction as the LOS. In the upper panels, the solid lines show the forward model predictions and the dashed lines show the simulated flux power spectra. The dotted lines show the corresponding error power spectrum $P_{\rm err}(k,\mu)\equiv\langle|\delta_F^{\rm sim}-\delta_F^{\rm model}|^2\rangle$. The lower panels show the percent difference $100[P_F^{\rm model}/P_F^{\rm sim}-1]$. The gray shaded region highlights agreement within $\pm 5\%$. The cubic model reproduces the simulated flux power spectrum to approximately $5\%$ up to $k\simeq 1-1.5\,h\,{\rm Mpc}^{-1}$, with improved performance relative to the linear and quadratic models.
    }
    \label{fig:linear_quadratic_cubic_comparison}
\end{figure*}

We first validate the three-dimensional flux power spectrum predicted by our forward model and present the improvement from linear to quadratic and cubic order. Fig. \ref{fig:linear_quadratic_cubic_comparison} compares these three models against the Sherwood simulation at $z=2.8$ and the near-Planck CAMELS simulation at $z=2,3,4$. The columns correspond to increasing perturbative order, while the rows correspond to the different simulations and redshifts. Each panel shows four angular wedges, allowing us to see the accuracy of the model as a function of scale, line-of-sight orientation, and redshift.

The upper panels of Fig. \ref{fig:linear_quadratic_cubic_comparison} show the measured power spectra, $P_F^{\rm sim}(k,\mu)$, of the simulation together with our model prediction, $P_F^{\rm model}(k,\mu)$, and the field-level error power spectrum, $P_{\rm err}(k,\mu)$. The latter quantifies the power in the stochastic component of our forward model \cite{deBelsunce:2026tks},
\begin{equation}
P_{\rm err}(k,\mu)
\equiv
\left\langle
\left|
\delta^s_{F,\rm sim}(\bm k)
-
\delta^s_{F,\rm model}(\bm k)
\right|^2
\right\rangle',
\label{eq:Perr}
\end{equation}
Here, the subscript $\mathrm{model}$ denotes the corresponding linear-, quadratic-, or cubic-order prediction.
The lower panels show the fractional difference between the model and the simulation power spectra, with the gray band indicating the $\pm 5\%$ region.

The improvement with perturbative order is clear in the fractional power-spectrum residuals, the field-level error power, and the range of scales over which the models remain accurate. The failure of the linear model is striking, already on the largest scales accessible in the box,  with residuals exceeding $5\%$ at $k \simeq 0.2-0.5 \,h\,{\rm Mpc}^{-1}$ for Sherwood and at even smaller wavenumbers for CAMELS, particularly at $z=3,4$. Moreover, its error power spectrum $P_{\rm err}(k,\mu)$ exhibits a pronounced scale dependence. The onset of this scale dependence is indicative of the breakdown of the linear-order perturbative description. 
Including the quadratic operators absorbs part of these missing contributions into the deterministic model, substantially reducing both the power-spectrum mismatch and the amplitude and scale dependence of the error power \cite{deBelsunce:2025bqc, deBelsunce:2026tks}. 

The quadratic model extends the approximate $5\%$ agreement to $k \simeq 0.8-1 \,h\,{\rm Mpc}^{-1}$ for Sherwood and CAMELS at all redshifts and $\mu$ bins.
The cubic model further improves the agreement, yielding a smaller and more weakly scale-dependent error power spectrum and extending the range of accurately reproduced wavenumbers to $k \simeq 1-1.5 \,h\,{\rm Mpc}^{-1}$ for Sherwood and CAMELS across redshifts and $\mu$ bins. 
Following baseline expectations, the behavior of the error power spectrum improves with increasing redshift and the residual contribution becomes increasingly important at higher $k$, consistent with the deviations seen in the power spectrum residuals in the lower panel. This agreement at the level of the power spectrum demonstrates that our EFT approach successfully captures the two-point statistics of the nonlinear \lya flux field across a wide range of cosmological and astrophysical hydrodynamical simulations. 

\subsection{Cross-correlation coefficient}\label{sec:rccResults}

\begin{figure*}
    \centering
    \includegraphics[width=\textwidth]{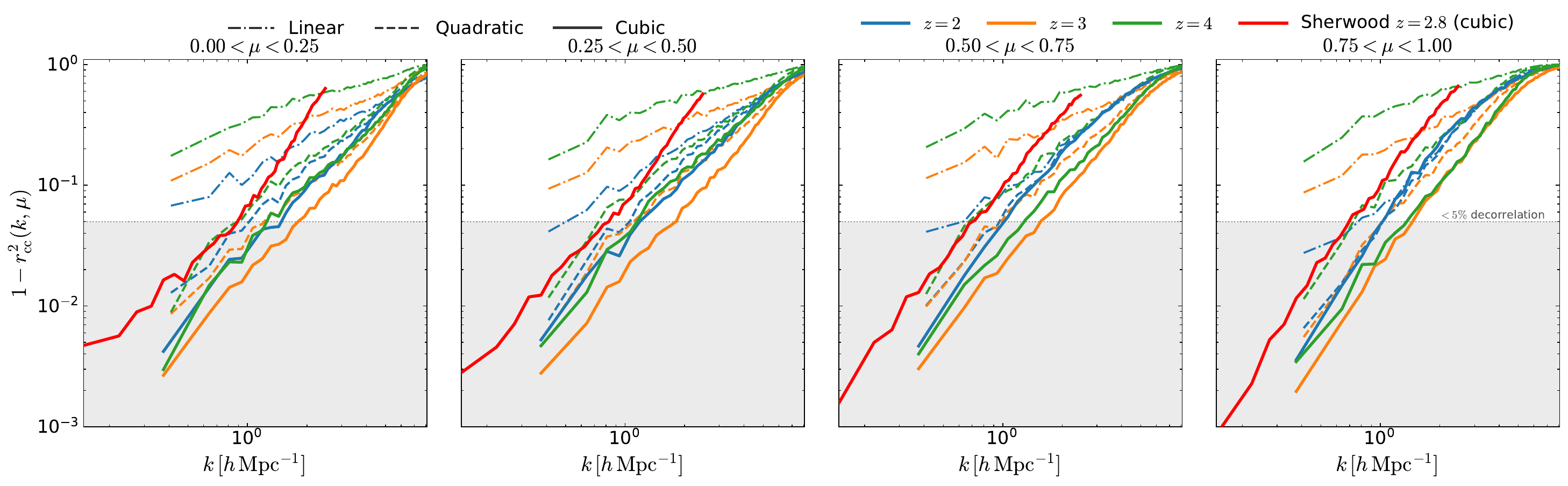}
    \caption{\textbf{Our cubic forward model preserves the Fourier-phase information down to $k\simeq 1\,h\,{\rm Mpc}^{-1}$.} We show the field-level decorrelation, $1-r_{\rm cc}^2(k,\mu)$, where $r_{\rm cc}$ is the cross-correlation coefficient between the reconstructed and simulated \lya flux fields, as a function of wavenumber in four line-of-sight angle bins. The blue, orange, and green curves correspond to the near-Planck CAMELS simulation, respectively, while the red curve shows the cubic reconstruction of the Sherwood simulation at $z=2.8$.
    Dash-dotted, dashed, and solid lines denote the linear, quadratic, and cubic CAMELS forward models, respectively. The Sherwood results are shown only at cubic order. The gray band highlights decorrelation below $5\%$.
    }
    \label{fig:rcc}
\end{figure*}

A crucial hurdle for the field-level procedure is that it needs to match at the level of the amplitudes and phases and not only for the summary statistics. Having validated our model's ability to match power spectra in the previous section, we now turn to this next level of validation. To study the phase-level correlation between our model and the true field, Fig. \ref{fig:rcc} shows their Fourier-space decorrelation \cite{deBelsunce:2026tks}, $1-r_{\rm cc}^2$, where the cross-correlation coefficient is defined as 
\begin{equation}
r_{\rm cc}(k,\mu)
=
\frac{
P_{\rm cross}(k,\mu)
}{
\sqrt{
P_F^{\rm model}(k,\mu)
P_F^{\rm sim}(k,\mu)
}
}.
\label{eq:rcc_def}
\end{equation}
In the above equation, $P_{\rm cross}(k, \mu)=\left\langle \delta^s_{F,\rm model}(\bm k) \delta^{s*}_{F,\rm sim}(\bm k) \right\rangle'$ is the cross spectrum between the model prediction and the simulated flux field. The cross-correlation is a proxy for how spatially correlated the densities are in Fourier space and the cosmological information that can be extracted from \lya modeling. 

In Fig.~\ref{fig:rcc} we first consider the Sherwood simulation at $z=2.8$ (red curves). The cubic reconstruction exhibits low decorrelation on large scales, with the mismatch gradually increasing toward smaller scales. The decorrelation remains at the few-percent level up to $k \sim 1\,h\,{\rm Mpc}^{-1}$, highlighting that our model can accurately recover Fourier-phase information of the Sherwood flux field. 

Additionally, we compare the decorrelation of the linear, quadratic, and cubic models for the CAMELS simulations at $z=2,3,4$. As in the power spectrum comparison, successive nonlinear operators improve agreement with the true hydrodynamic flux field. While all three models remain correlated with CAMELS on large scales, their performance increasingly diverges toward smaller scales, where the cubic model shows the most accurate phase agreement. Similar to Sherwood, the cubic model maintains decorrelation at the few-percent level to $k \sim 1.5 h\,{\rm Mpc}^{-1}$. 

\begin{figure*}
    \centering
    \includegraphics[width=\textwidth]{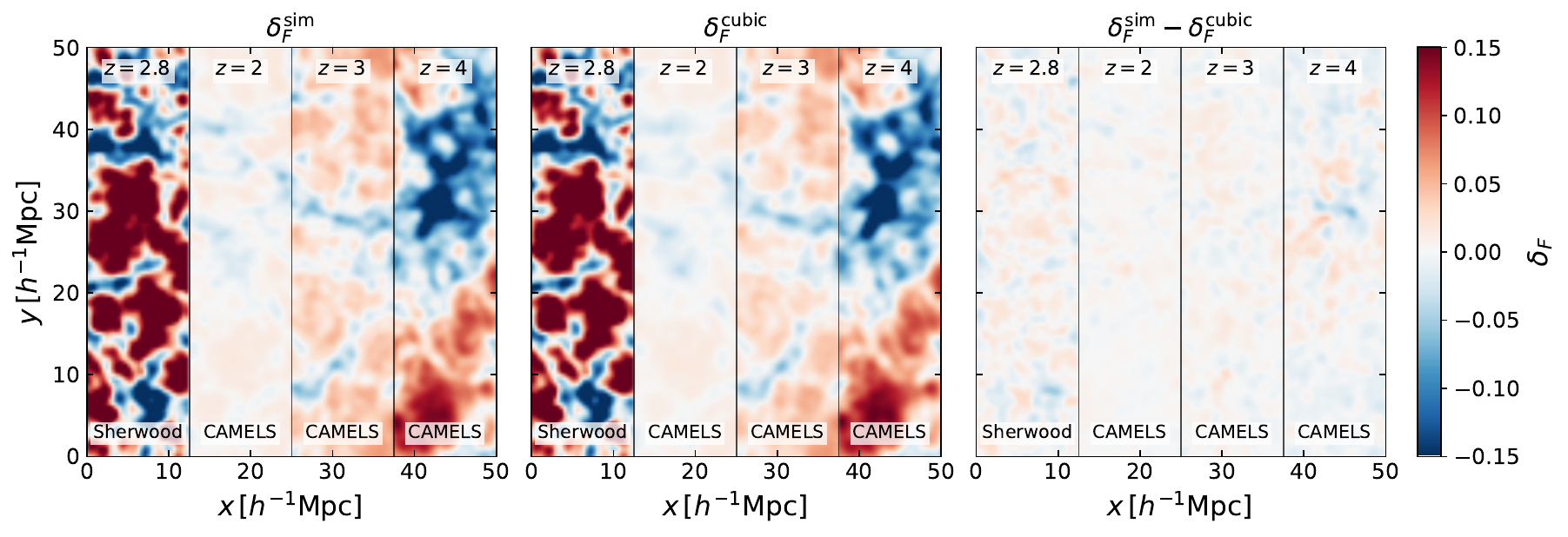}
    \caption{\textbf{Field-level comparison of simulated \lya flux field and our cubic model}. The left panel shows the simulation truth, $\delta^s_{F,\rm sim}$, the middle panel shows the corresponding cubic forward-model prediction, $\delta^s_{F,\rm cubic}$, and the right panel shows their residual. Each panel is divided into four vertical subpanels, corresponding to the Sherwood simulation at $z=2.8$, followed by a representative CAMELS-SB35 simulation at $z=2,3,4$. All fields are shown for the same LOS $=z$, smoothed with a 3D Gaussian filter of radius $R=1.5\,h^{-1}{\rm Mpc}$ for Sherwood and $R=0.5\,h^{-1}{\rm Mpc}$ for CAMELS, and averaged over a projection depth of $10\,h^{-1}{\rm Mpc}$.
    }
    \label{fig:cubic_fields}
\end{figure*}

Beyond the two-point statistics and phase-level correlation, we show 
in Fig. \ref{fig:cubic_fields} a visual comparison of the field-level performance of our forward model for the Sherwood simulation ($z=2.8$) and the near-Planck CAMELS-SB35 simulation at $z=2,3,4$. The left panel shows the true simulated \lya flux density contrast, $\delta^s_{F,\rm sim}$, the center panel shows the corresponding model prediction, $\delta^s_{F,\rm cubic}$, and the right panel shows the residual field $\epsilon \equiv \delta^s_{F,\rm sim}-\delta^s_{F,\rm cubic}$. 
Each panel is divided into four vertical strips, with Sherwood shown first, followed by CAMELS at increasing redshift, to demonstrate the redshift evolution alongside field-level reconstruction.

Our cubic model captures the major features of the \lya flux field for both Sherwood and CAMELS, including their spatial locations and morphologies - regions of positive and negative flux contrast appear at the same locations and shapes. This is visible across redshifts and spatial scales. Correspondingly, the residual field is mostly featureless, with tiny fluctuations on small scales not captured by our model. This visually showcases the same behavior quantified by $P_{\rm err}$ and $1-r_{\rm cc}^2$: the large-scale modes match well with simulations, while the mismatch increases with short-scale fluctuations. We test this approach on a simulation with a cosmology that is far away from Planck in App.~\ref{sec:RobustResults}.

\subsection{Transfer functions}\label{sec:TFandBiasResults}
Having validated the performance of our cubic-order field-level model, we now turn to the results of extracting the EFT bias parameters from our measured transfer functions.
The transfer functions $\beta_a(k, \mu)$ encode the contribution of each orthogonalized operator to the constructed flux field, and their scale dependence contains information about the underlying bias parameters and higher-order corrections. In particular, the large-scale behavior of these functions is important for extracting the scale-independent EFT bias coefficients, which can be obtained by matching the measured transfer functions to their perturbative predictions.

\begin{figure*}
    \centering
    \includegraphics[width=\textwidth]{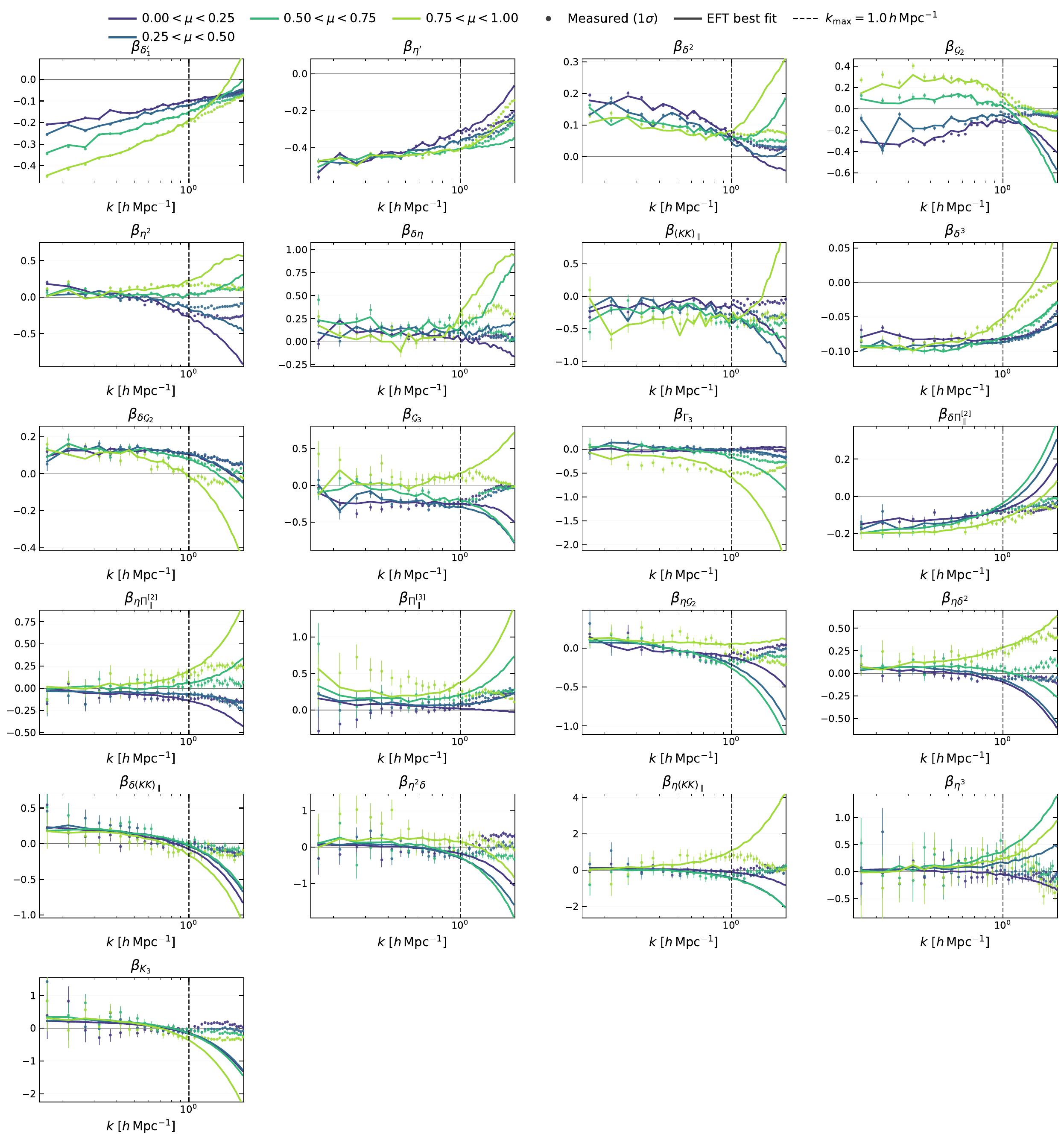}
    \caption{\textbf{Cubic-order \lya forest transfer functions and their EFT fits} in Sherwood ($z=2.8$). We show all the measured transfer functions $\beta_a(k,\mu)$ (points with $1\sigma$ error bars) for four line-of-sight angular bins, $0<\mu<0.25$, $0.25<\mu<0.5$, $0.5<\mu<0.75$, and $0.75<\mu<1$. The solid curves are the best-fit predictions of the cubic-order EFT model.
    The joint fit to all transfer functions yields a reduced chi-squared of $\chi^2/{\rm dof}=1.34$ over the fitting range $0.15\leq k\leq 1\,h\,{\rm Mpc}^{-1}$. The plots show the constant-approaching large scale behavior of the measured transfer functions, allowing us to extract the scale-independent EFT coefficients.
    }
    \label{fig:transfer_functions}
\end{figure*}

We first study their behavior using the Sherwood simulation. Compared to smaller CAMELS boxes, the larger Sherwood volume provides access to longer-wavelength Fourier modes, allowing us to clearly see their large-scale constant behavior. Fig. \ref{fig:transfer_functions} shows the full set of cubic-order transfer functions measured in Sherwood at $z=2.8$, for four line-of-sight angular bins, together with their statistical uncertainties. Their overall behavior is smooth and consistent with an approximately scale-independent limit as $k\rightarrow 0$ at fixed $\mu$. To illustrate this behavior, we overlay the best-fit of our model. We note that the fit to $\beta_{\Pi_\parallel^{[3]}}$ in Sherwood shows noticeable deviations at higher wavenumbers. We find that restricting the fit to a lower $k_{\max}$ improves the agreement, suggesting that the discrepancy is partly driven by the smaller scales included in the fit. The agreement may also be sensitive to the analytic covariance adopted in the fitting procedure, whose treatment could be further refined.
We also show the transfer functions for the near-Planck CAMELS simulation SB35-637 at $z=2$ in Fig. \ref{fig:camels_transfer_functions} of Appendix \ref{app:CAMELS_TF}, finding qualitatively consistent behavior, with better agreement between the measured $\beta_{\Pi_\parallel^{[3]}}$ transfer function and its EFT prediction compared to Sherwood.

\subsection{Simulation-based priors}\label{sec:SBP}
As detailed in Sec. \ref{sec:fitting}, we predict the shapes of the transfer functions in EFT and fit them, yielding a set of bias parameters.  
To compare across different cosmologies and redshifts, we rescale each bias coefficient $b_O$ by $\sigma_8^n(z)$, where $n$ is the perturbative order
of the corresponding operator and $\sigma_8(z)$ is the linear matter fluctuation amplitude at redshift $z$. This rescaling accounts for the leading dependence of operator amplitudes on normalization of the linear density field, allowing us to reduce the range of bias parameter values across the CAMELS suite.

\begin{figure*}
    \centering
    \includegraphics[width=\textwidth]{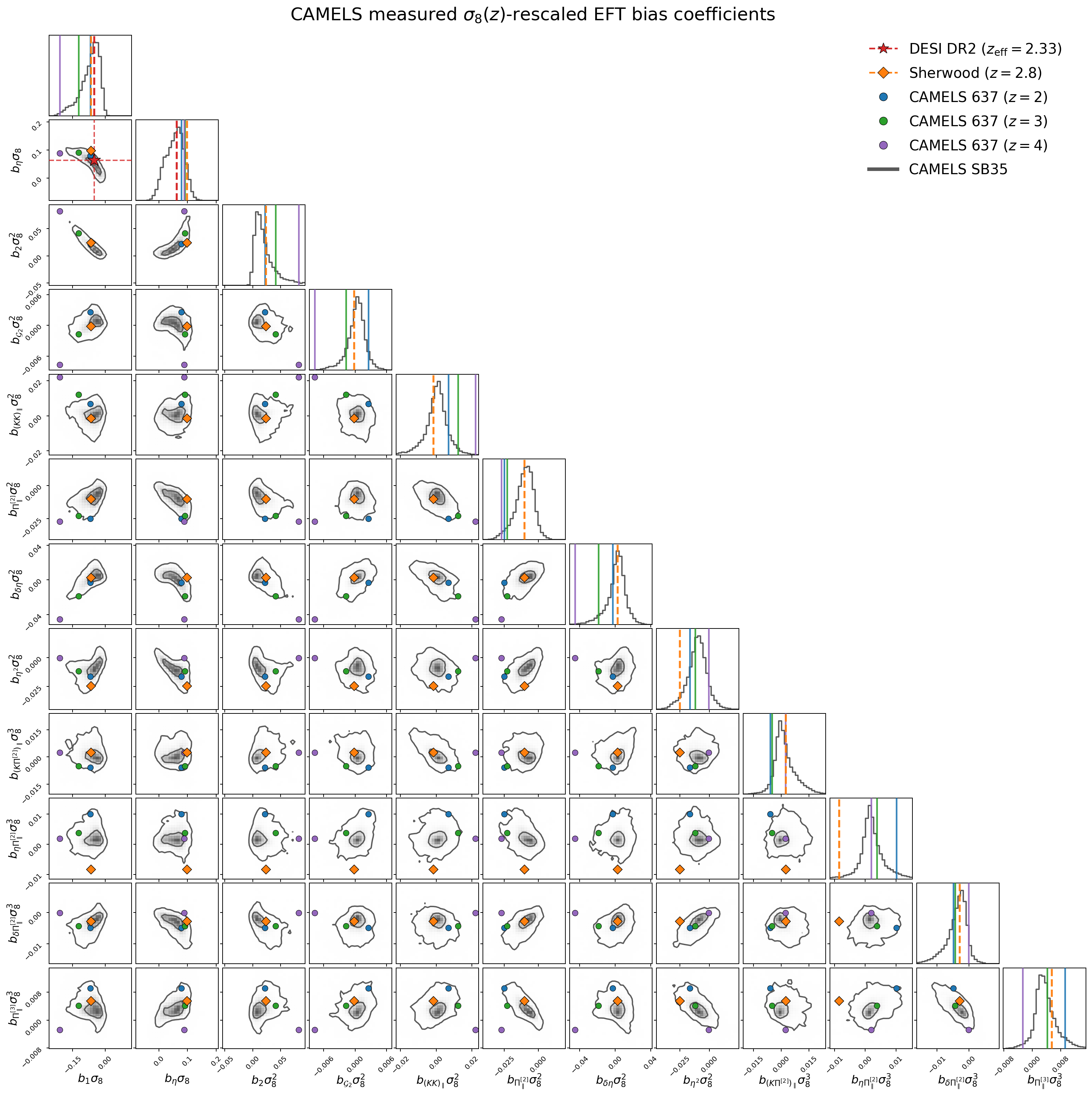}
    \caption{\textbf{Corner plot of the measured EFT $\sigma_8(z)$-rescaled bias coefficient across the CAMELS SB35 simulation suite.} Here, we show a subset of our full bias bases - we show the linear coefficients $b_1$ and $b_\eta$, the quadratic coefficients $b_2$, $b_{\mathcal{G}_2}$, $b_{(KK)_\parallel}$, $b_{\Pi_\parallel^{[2]}}$, $b_{\delta\eta}$, and $b_{\eta^2}$, and the cubic coefficients $b_{(K\Pi^{[2]})\parallel}$, $b{\eta\Pi_\parallel^{[2]}}$, $b_{\delta\Pi_\parallel^{[2]}}$, and $b_{\Pi_\parallel^{[3]}}$. $\sigma_8(z)$ is evaluated at the snapshot redshift, and an $n$th-order bias coefficient is rescaled by $\sigma_8^n(z)$.
    Contours indicate the $68\%$ and $95\%$ enclosed regions of the distribution across the suite and redshift snapshots. The blue, green, and purple points show the inferred bias parameters for the near-Planck simulation (CAMELS 637) at $z=2,3,4$ respectively. The orange diamond point shows the biases for the Sherwood snapshot, with $z=2.8$.
    The red star and dashed lines show the DESI DR2 linear \lya forest bias measurements at $z_{\rm eff} = 2.33$, rescaled by $\sigma_8(z_{\rm eff})$ for comparison \cite{DESI:2026lnd}.
    }
    \label{fig:bias_corner}
\end{figure*}

In Fig. \ref{fig:bias_corner} we show the joint distribution of a representative subset of these $\sigma_8(z)$-rescaled bias coefficients across all CAMELS simulations. We also show the best-fit values for the near-Planck simulation using blue, green, and purple points for redshifts $z=2,3,4$ respectively, alongside DESI DR2 linear \lya forest bias measurements (see Table V of \cite{DESI:2026lnd}) at $z_{\rm eff} = 2.33$ (red stars) and Sherwood's best-fit values (orange diamonds). Fig. \ref{fig:bias_corner} shows a tight distribution for the bias parameters that can be used to break degeneracies in the data, yielding potentially tighter cosmological constraints using physically informed priors~\cite[see, e.g.,~][]{Ivanov:2024hgq}.
Another particularly strong consistency check is that our linear bias values inferred for the near-Planck simulation at $z=2$, which is the available snapshot closest to the DESI $z_{\rm eff}$, are very close to the DESI DR2 measurement \cite{DESI:2026lnd}. This is also true for the Sherwood snapshot at $z=2.8$ with Planck cosmology. 
The complete set of best-fit EFT bias coefficients for Sherwood and the near-Planck CAMELS simulation at all three redshifts, together with the counterterms and stochastic parameters, is reported in Table \ref{tab:eft_bias_parameters} of Appendix \ref{app:eft_parameters}.

\subsection{Power spectrum reconstruction}\label{sec:PffReconstructionResults}
A complementary two-point validation of our methods is to compare the reconstructed power spectrum using the inferred bias parameters. The comparisons in Sec. \ref{sec:2ptResults} construct the power spectrum directly from the field-level forward model and hence test the 2-pt accuracy of the reconstructed field itself. Now, we use the bias parameters, as illustrated in Fig. \ref{fig:bias_corner}, as inputs to the one-loop power spectrum calculation described in Secs. \ref{sec:Pff_reconstruction} and \ref{sec:P1D} using the one-loop power spectrum code \texttt{CLASS-PT} \cite{Chudaykin:2020aoj, Ivanov:2023yla}.

\begin{figure*}[t]
    \centering
    \includegraphics[width=0.49\textwidth]{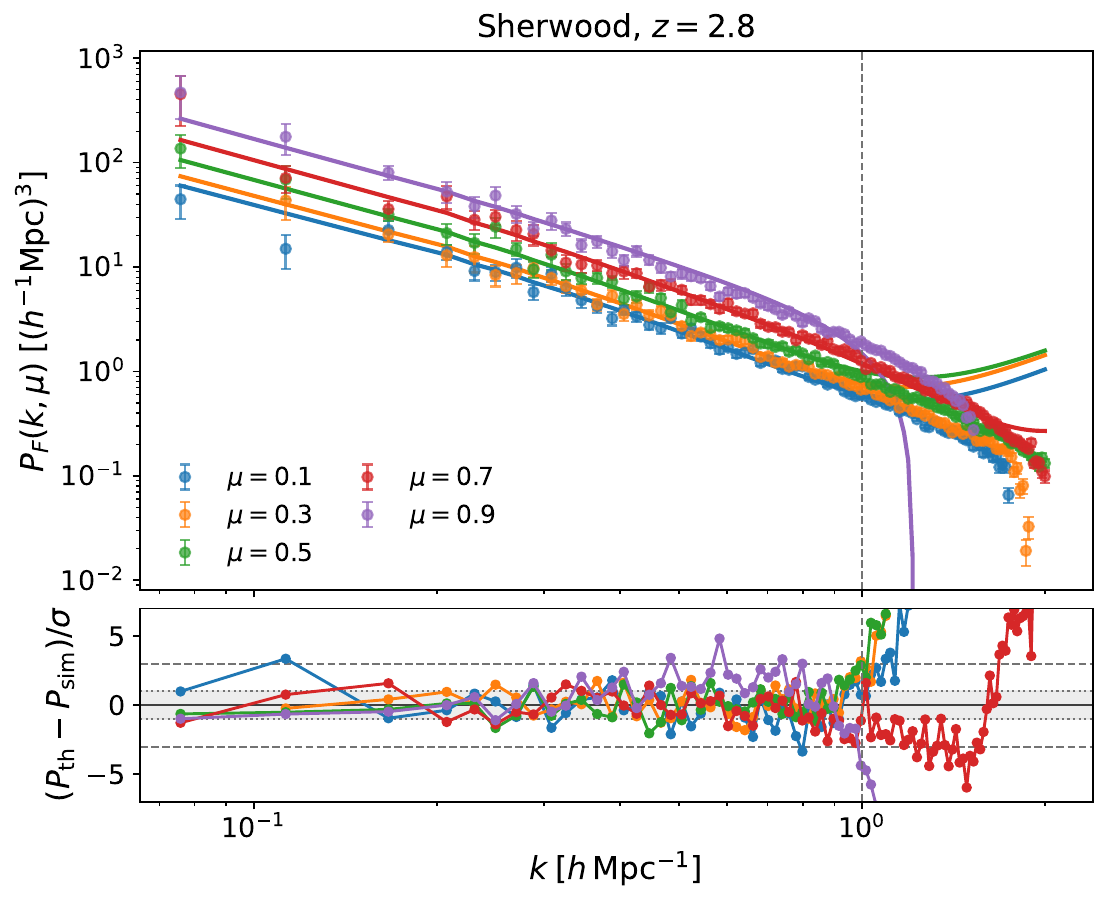}
    \hfill
    \includegraphics[width=0.49\textwidth]{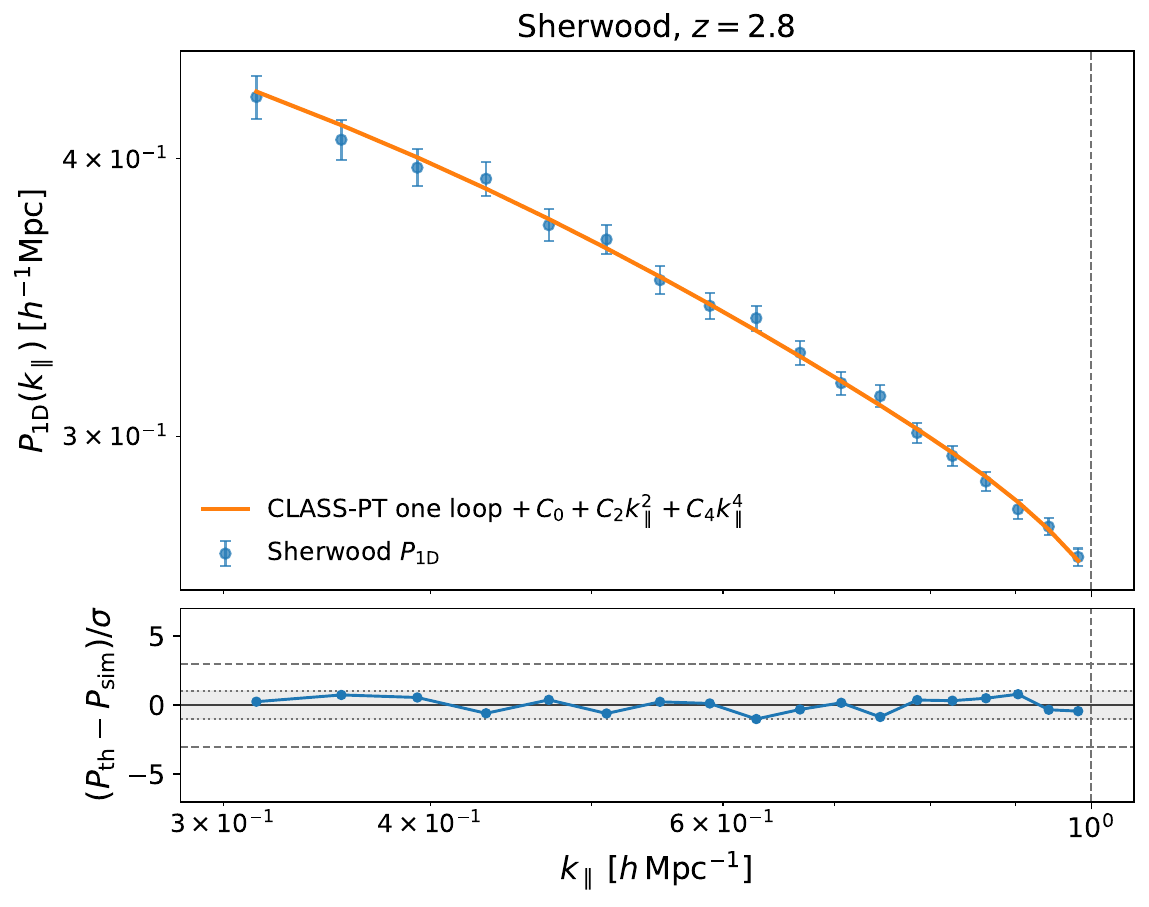}

    \caption{
    \textbf{Power spectrum reconstruction performance.} The field-level inferred bias parameters reproduce the three-dimensional (left) and one-dimensional (right) \lya flux power spectra of the Sherwood
    simulation at $z=2.8$ (circles with error bars) using the
    one-loop \texttt{CLASS-PT} predictions (solid lines).
    The left panel shows the three-dimensional flux power spectrum across five $\mu$ wedges.
    The right panel shows the one-dimensional  \texttt{CLASS-PT} predictions with contact terms.
    The lower panels show the residuals between the theoretical
    predictions and the simulation, normalized by the
    uncertainties in the measured power spectra,
    $(P_{\rm th}-P_{\rm sim})/\sigma$. The uncertainties in the measured three-dimensional power spectrum are estimated using the Gaussian mode-counting approximation.
    The one-dimensional uncertainties are estimated using a
    Gaussian covariance constructed from the measured
    three-dimensional power spectrum.
    The vertical dashed lines indicate the maximum wavenumber,
    $k_{\max}=1\,h\,{\rm Mpc}^{-1}$, used in the transfer
    function and one-dimensional contact-term fits.
    The horizontal dotted and dashed lines in the lower panels
    mark the $\pm1\sigma$ and $\pm3\sigma$ levels,
    respectively. The best-fit bias parameters are tabulated in Tab.~\ref{tab:eft_bias_parameters} for the $L=160\, h^{-1}\mathrm{Mpc}$ and in Tab.~\ref{tab:sherwood_L80_bias_parameters} for the $L=80\, h^{-1}\mathrm{Mpc}$ simulation.
    }
    \label{fig:Sherwood_classpt}
\end{figure*}

We first perform this comparison using the Sherwood simulation \cite{Bolton:2016bfs}. In Fig. \ref{fig:Sherwood_classpt}, 
we show in the left panel the three-dimensional power spectrum for five $\mu$ wedges, while the right panel shows the one-dimensional power spectrum. Note that for the P1D we additionally fit the 1D counterterms. The lower panels show the residuals between the perturbative predictions and the simulation, normalized by the uncertainty in the measured power spectra. The reconstructed three-dimensional power spectrum agrees with the Sherwood measurement to within $\approx 3\sigma$ up to $k \simeq 1\,h\,{\rm Mpc}^{-1}$ across the five $\mu$ wedges shown. The vertical dashed lines indicate the maximum scale used for the fits and beyond which our EFT model starts to break down. 
Similarly, the reconstructed one-dimensional power spectrum agrees with the Sherwood measurement to within $1-2\sigma$ for the full $k$-range, after including the fitted contact terms. 

\begin{figure*}[t]
\centering
\includegraphics[width=0.49\textwidth]{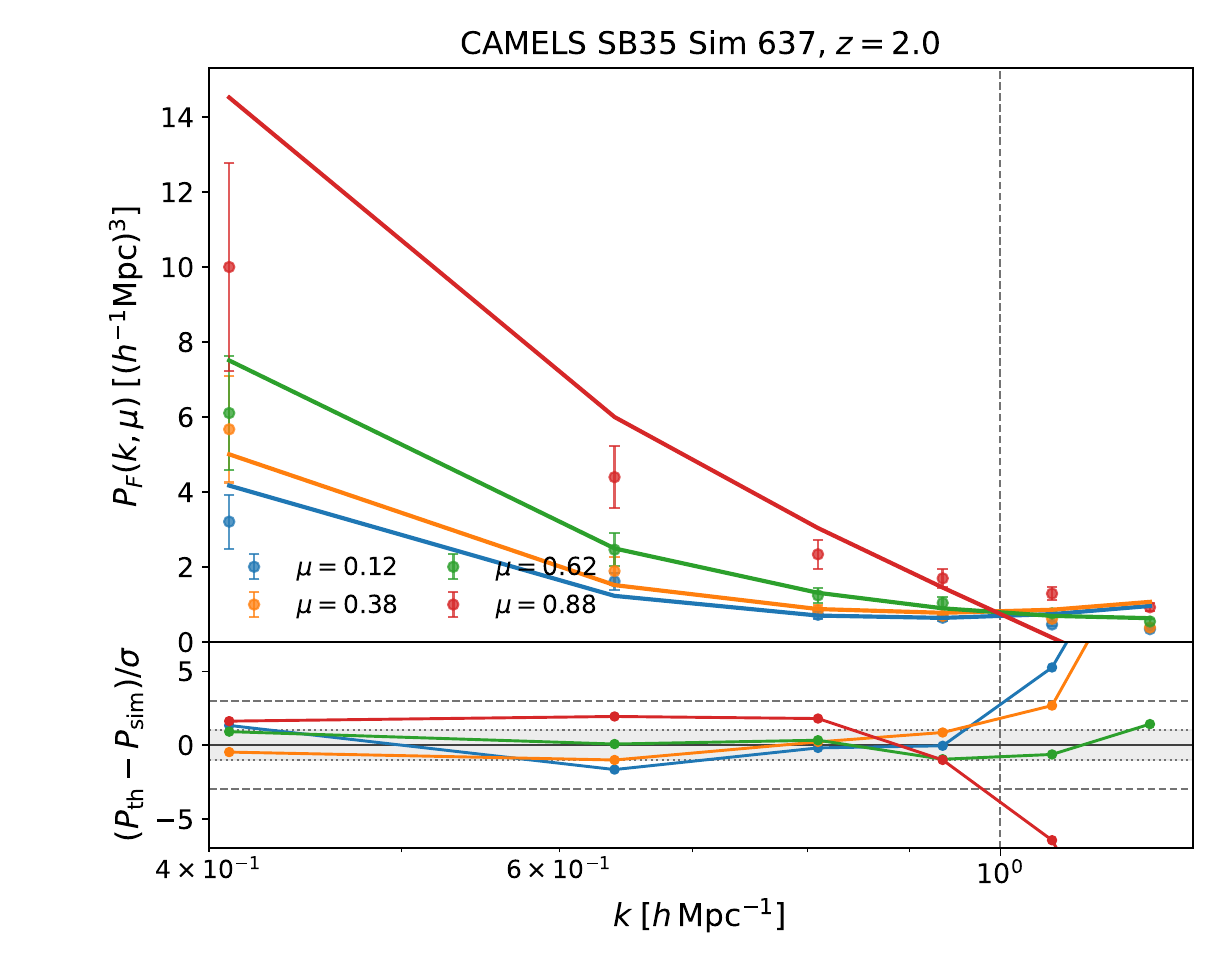}
\hfill
\includegraphics[width=0.49\textwidth]{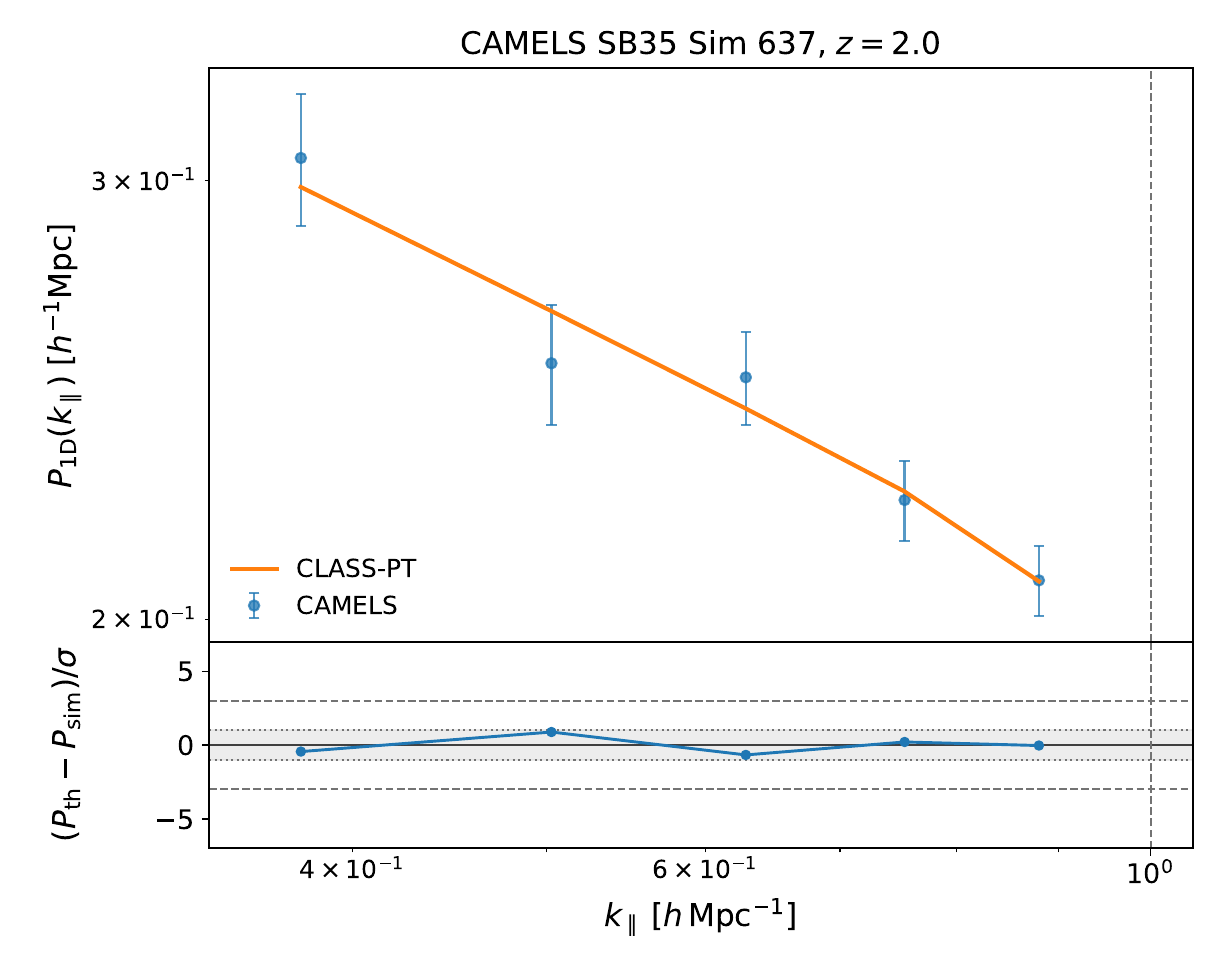}

\vspace{0.5em}

\includegraphics[width=0.49\textwidth]{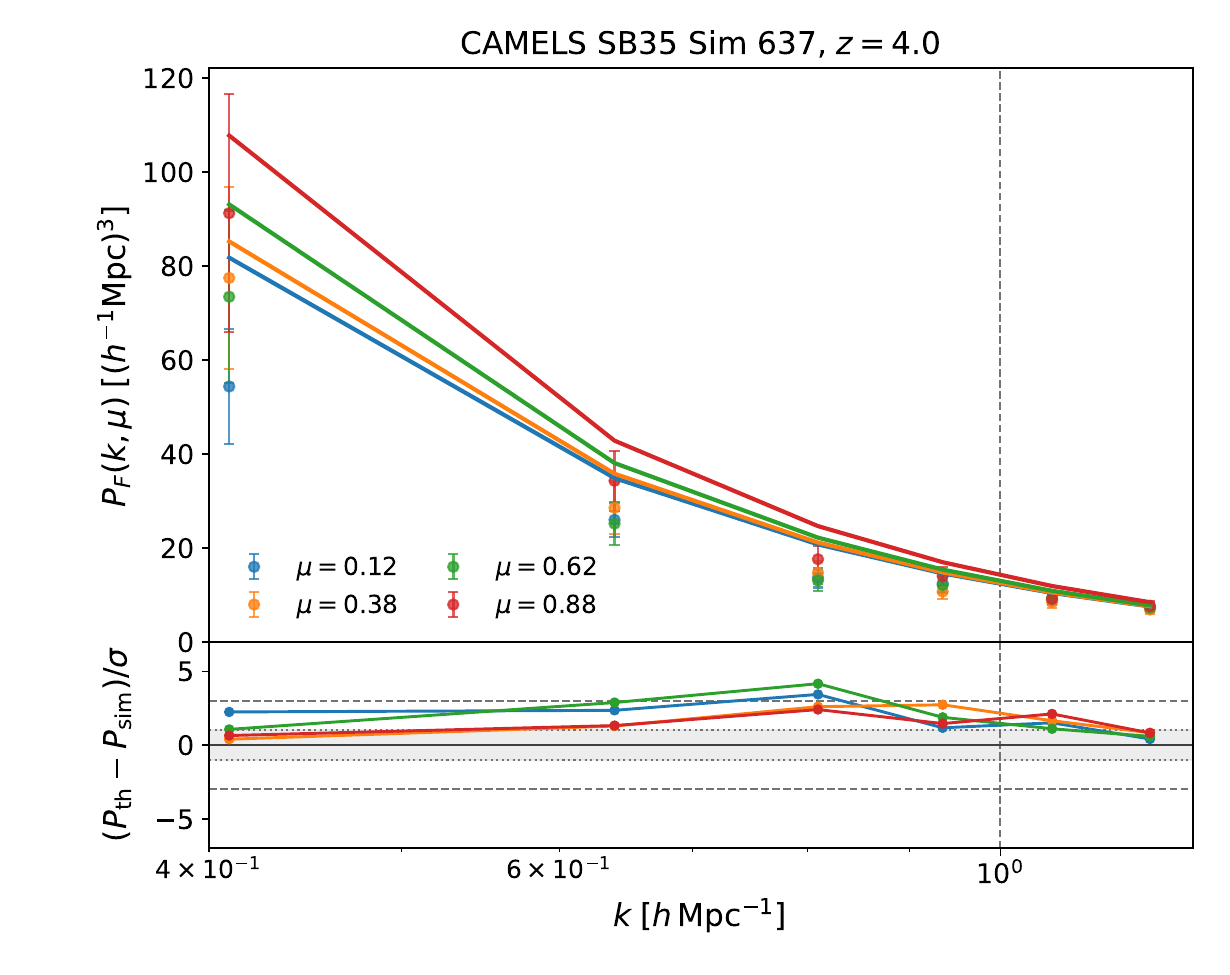}
\hfill
\includegraphics[width=0.49\textwidth]{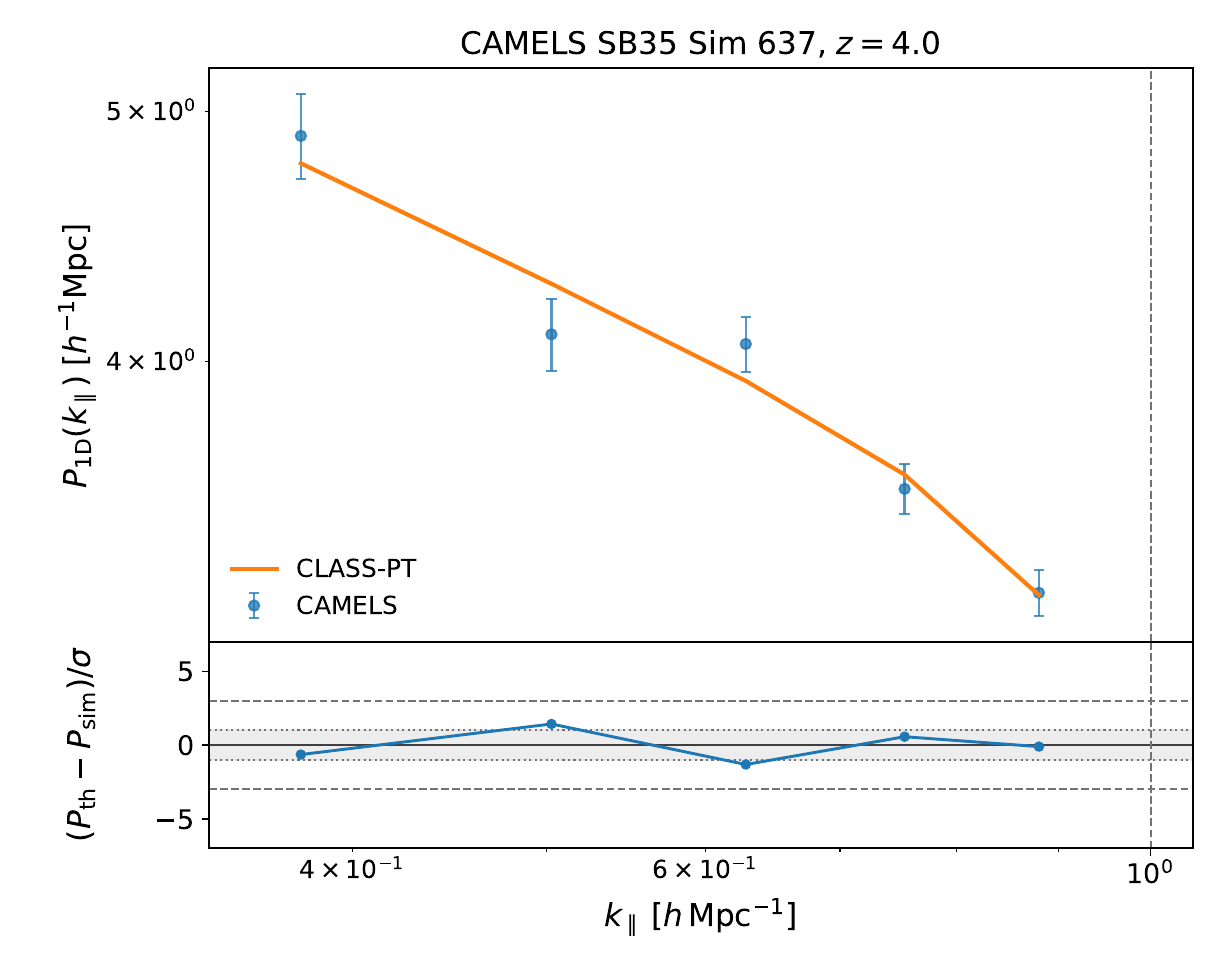}

\caption{Same as Fig. \ref{fig:Sherwood_classpt}, but for the near-Planck CAMELS simulation at $z=2$ and $z=4$, shown in the top and bottom panels, respectively.}
\label{fig:CAMELS_classpt_pff}
\end{figure*}

We next perform the same reconstructions for the near-Planck CAMELS simulation at $z=2$ and $z=4$, shown in Fig. \ref{fig:CAMELS_classpt_pff}.
At $z=2$, the reconstructed three-dimensional power spectrum agrees with the simulation within approximately $1$--$2\sigma$ up to $k\simeq 0.8\,h\,{\rm Mpc}^{-1}$ across most angular wedges, although larger deviations appear in the highest-$\mu$ wedge toward smaller scales, which we discuss below. At $z=4$, the agreement is generally better, with residuals remaining within approximately $2\sigma$ across all four $\mu$ wedges over the range shown, up to $k\simeq 1\,h\,{\rm Mpc}^{-1}$. The reconstructed one-dimensional power spectrum also agrees with the simulation within approximately $1\sigma$ at both redshifts after including the fitted contact terms.

However, residual differences, more pronounced than in Sherwood,
remain due to finite-grid effects which are significant in CAMELS. The transfer functions are measured using perturbative operators constructed on the same finite Fourier grid as the simulated flux field. Hence, when the power spectrum is constructed directly from the field-level model, as in the comparison above, the measured transfer functions and operator spectra incorporate the same finite-grid effects. In contrast, the \texttt{CLASS-PT} results evaluate the corresponding perturbative contributions in the continuum and hence do not capture the discreteness effects of the finite CAMELS grid. For instance, the nonlinear operators involve convolutions over Fourier modes whose ultraviolet support is truncated on the grid, so their measured auto- and cross-spectra differ from their continuum counterparts. This effect is particularly strong for CAMELS, where the $128^3$ grid used in our field-level analysis makes these discreteness effects non-negligible.

\subsection{BAO shift predictions}\label{sec:BAOshiftResults}
We now use the nonlinear EFT bias parameters to predict the nonlinear shifts of the \lya BAO feature in the radial and transverse directions, as detailed in Sec. \ref{sec:bao_shift}. Whilst our approach allows us to extract the relevant nonlinear physics from relatively small hydrodynamical simulation boxes, which do not themselves resolve the BAO scale, we caution the reader that these results should be taken indicatively due to the small volumes and resolutions. We use the inferred bias parameters to construct the large-scale power spectrum that does resolve the BAO feature, validated in the previous section.

\begin{figure}
    \centering
    \includegraphics[width=\columnwidth]{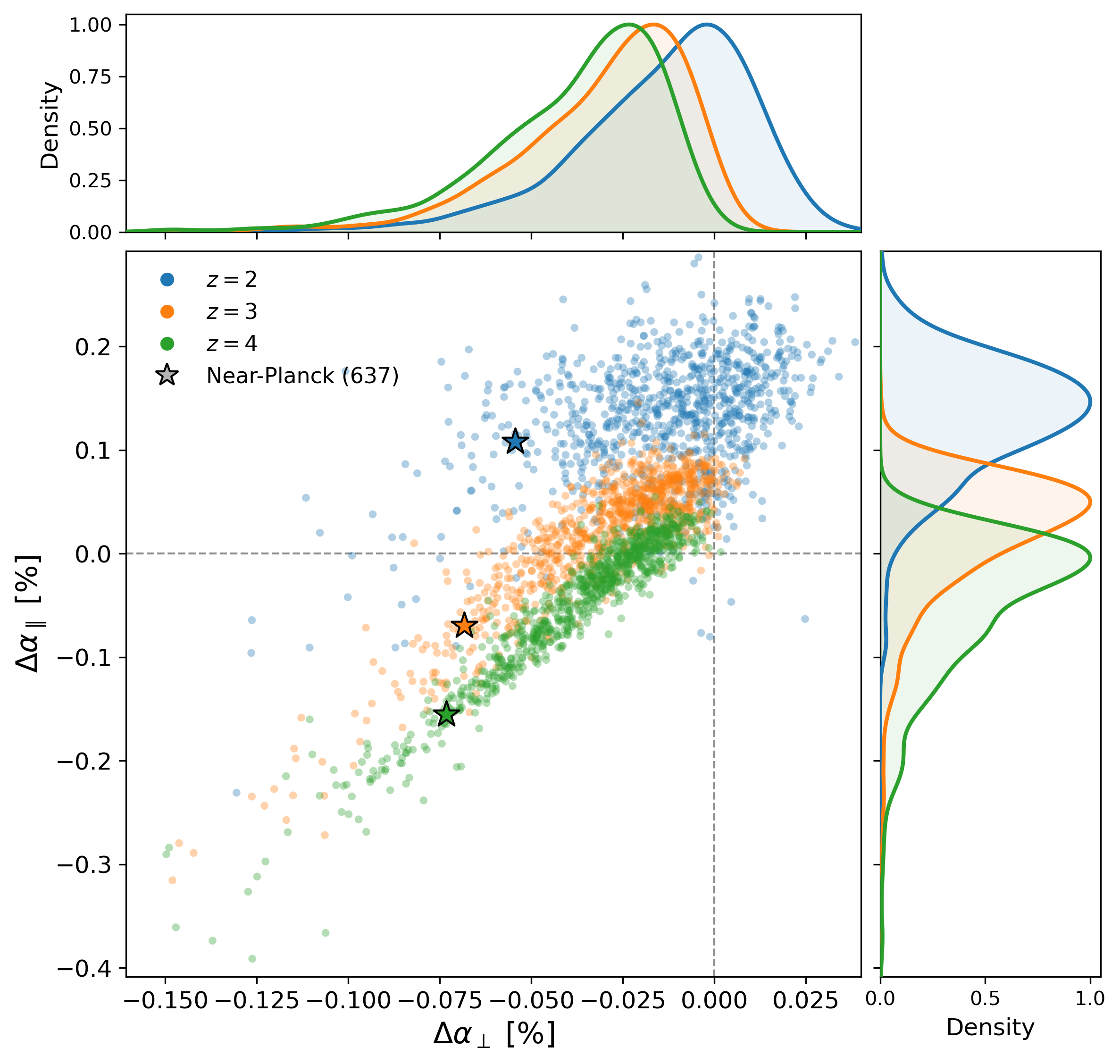}
    \caption{\textbf{Joint distribution of the nonlinear BAO shifts}, $\Delta\alpha_\parallel$ and $\Delta\alpha_\perp$ across CAMELS-SB35 at $z=2,3,4$. Each point is the value for one simulation in the suite. Stars show the values for the near-Planck simulation, and the marginalized distributions are shown along the top and right panels. 
    }
    \label{fig:bao_shift}
\end{figure}

Fig. \ref{fig:bao_shift} shows our measured joint distribution of the radial and transverse shifts, $\Delta\alpha_\parallel$ and $\Delta\alpha_\perp$, at the CAMELS redshifts $z=2,3,4$. Each point is the prediction for one simulation, while the marginalized distributions are shown along the top and right panels. The distribution is qualitatively similar to that shown in Fig. 7 of Ref. \cite{deBelsunce:2024rvv} using the ACCEL$^2$ simulations, although the predicted shifts differ in magnitude and, in some cases, sign.

The predicted shifts are small, at the sub-percent level. Fig. \ref{fig:bao_shift} also illustrates a positive correlation between the shift parameters: simulations predicting more negative transverse shifts tend to predict more negative radial shifts. The Pearson correlation coefficient increases from $r=0.42$ at $z=2$ to $r=0.88$ and $r=0.85$ at $z=3$ and $z=4$, respectively. This is a consistent consequence of the correlated nonlinear biases entering both components of the BAO shift.

We also find a clear redshift evolution of the BAO shift distribution. At $z=2$, the median transverse and radial shifts are $\Delta\alpha_\perp=-0.0095\%$ and $\Delta\alpha_\parallel=+0.137\%$, respectively. With increasing redshift, both shift towards more negative values, reaching $(-0.025\%,+0.033\%)$ at $z=3$ and $(-0.033\%,-0.030\%)$ at $z=4$. The marginalized distributions make this evolution clear: the transverse shift becomes progressively more negative, while the radial shift changes from predominantly positive at $z=2$ to predominantly negative at $z=4$.

The stars in Fig. \ref{fig:bao_shift} show the predictions for the near-Planck CAMELS simulation SB35-637. For this simulation, we obtain $(\Delta\alpha_\perp,\Delta\alpha_\parallel)=(-0.054\%,+0.108\%)$, $(-0.068\%,-0.070\%)$, and $(-0.073\%,-0.155\%)$ at $z=2,3,4$, respectively. These predictions follow the same overall redshift evolution as the ensemble, although they are generally more negative than the corresponding ensemble medians.

We compare these near-Planck predictions with the nonlinear \lya BAO shifts obtained from the Planck-like ACCEL$^2$ simulations in Ref. \cite{deBelsunce:2024rvv} \cite{Chabanier:2024knr}. At $z=2$, the CAMELS near-Planck simulation predicts a positive radial shift of $\Delta\alpha_\parallel=+0.108\%$, in contrast to the negative shifts of $-0.20\pm0.09\%$ and $-0.08\pm0.05\%$ found for the baseline and larger ACCEL$^2$ simulation boxes, respectively. The transverse shifts, however, agree in sign, with the CAMELS prediction of $\Delta\alpha_\perp=-0.054\%$ being close to the larger-box ACCEL$^2$ result of $-0.05\pm0.05\%$. At higher redshift, the near-Planck CAMELS radial shift becomes negative, reaching $-0.070\%$ at $z=3$, consistent in sign with the ACCEL$^2$ prediction of $-0.16\pm0.05\%$ at $z=2.6$. Since both simulations adopt Planck-like cosmologies, the remaining differences may reflect variations in astrophysical modeling, simulation resolution and volume, or the sampled nonlinear bias coefficients. A direct comparison of the underlying bias combinations would be needed to establish the origin of this difference. In this work, the CAMELS suite allows us to quantify the variation in the predicted shifts and their correlations, providing a basis for simulation-based priors that can be used in future \lya BAO analyses.

\section{Conclusions}\label{sec:conclusions}
In this paper, we develop and validate a cubic-order EFT description of the \lya forest flux decrement at the field level. We extend the previous field-level analysis \cite{deBelsunce:2025bqc, deBelsunce:2026tks} by including the cubic operators and test this framework against hydrodynamical simulations across a broad range of cosmological and astrophysical conditions using the CAMELS-SB35 suite at $z=2.0,3.0,4.0$, alongside the Sherwood simulation for two volumes for a similar redshift coverage. 

We show the expected order-by-order improvement at the field level by extending the perturbative analysis from linear- to quadratic- to cubic-order. Going from linear to quadratic and finally cubic models progressively causes the mismatch in the three-dimensional flux power spectrum as well as the residual field-level error to decrease.

Our model reproduces the true three-dimensional power spectrum at the $\sim 5\%$ level down to $k\simeq (1-1.5)~h{\rm Mpc}^{-1}$ across the CAMELS redshifts. Furthermore, our agreement is not limited to the two-point amplitudes: we show that the EFT model
correctly predicts the \lya forest flux decrement at the pixel level. 
The decorrelation between the truth and our fields remains at the few-percent level over comparable scale ranges, displaying that we also retain phase information of the nonlinear hydrodynamical flux field. 
Beyond the two-point comparisons, we directly compare the redshift-space flux maps and find that the perturbative model reconstructs the locations and morphologies of the large-scale structures present in the simulated field, with the residual dominated by fluctuations on small scales not captured by our perturbative treatment.

The field-level modeling allows us to extract EFT bias coefficients with high precision, owing to cosmic-variance cancellation \cite{Schmittfull:2018yuk, deBelsunce:2026tks}.
By measuring these, our methods allow us to perform independent consistency checks by computing the one-loop \lya power spectrum in \texttt{CLASS-PT} \cite{Chudaykin:2020aoj} and comparing it against the truth from simulations. 
Importantly, we detect the entire set of cubic EFT parameters in all simulations. 
This strengthens the notion that all EFT operators and loop contributions are needed in order to accurately describe the data~\cite{Ivanov:2023yla,deBelsunce:2024rvv,deBelsunce:2025bqc,deBelsunce:2026tks}. 

We measure EFT parameters of the forest across a wide range of cosmological and astrophysical parameters of the CAMELS SB35 suite to obtain simulation-based priors. We provide these to obtain tighter constraints on EFT nuisance parameters and, in turn, on cosmological parameters \cite{Ivanov:2024hgq}. The correlations we find open up new avenues to use physics-informed priors to constrain \lya physics, such as joint nuisance-parameter priors in DESI full shape $P_{\rm 3D}$ or $P_{\rm 1D}$ analyses. Similar simulation-based approaches have shown how correlations among EFT parameters can be used as informative joint priors in galaxy clustering analyses \cite{Ivanov:2024xgb, Chudaykin:2026nls, Chen:2025jnr, Ivanov:2024hgq}. Our results extend this to the \lya forest, with cosmological constraints left to future work.

We further use the nonlinear bias parameters extracted from the CAMELS suite to predict the nonlinear shifts of the \lya forest BAO feature in the radial and transverse directions. We find sub-percent shifts, with positive correlations between the radial and transverse components, characterized by Pearson correlation coefficients of $r=0.42$, $0.88$, and $0.85$ at $z=2,3,4$, respectively. The predicted shifts also exhibit clear redshift evolution: the median transverse and radial shifts evolve from $(\Delta\alpha_\perp,\Delta\alpha_\parallel)=(-0.0095\%,+0.137\%)$ at $z=2$ to $(-0.025\%,+0.033\%)$ at $z=3$ and $(-0.033\%,-0.030\%)$ at $z=4$. For the near-Planck CAMELS simulation, we obtain $(\Delta\alpha_\perp,\Delta\alpha_\parallel)=(-0.054\%,+0.108\%)$, $(-0.068\%,-0.070\%)$, and $(-0.073\%,-0.155\%)$ at $z=2,3,4$, respectively, following the same overall redshift evolution. These projections are broadly consistent with previous results from ACCEL$^2$ simulations \cite{deBelsunce:2024rvv}. 

Another important aspect of our methods is robustness across the simulation parameter space. A major motivation for using the CAMELS-SB35 suite is the large variations in both cosmology and astrophysical prescriptions. 
Nonetheless, we find comparable field-level performance for simulations occupying widely separated regions of the parameter space. Different simulation parameters simply change the EFT parameters, as expected in theory. An interesting result of our analysis is that the range of scales
where the forest is quasi-linear is not particularly sensitive to the underlying cosmology or 
astrophysics model. 

Several directions follow naturally from this work. First, extending this analysis to additional hydrodynamical simulations, with different resolutions, volumes, and IGM models, will allow us to quantify the dependence of EFT parameters on the subgrid models.

The broad coverage of the CAMELS suite used here also provides an immediate opportunity to construct conditional models for EFT parameters as functions of cosmological and astrophysical parameters. 
These conditional models can be used to understand the physics behind the EFT parameters 
and also in \lya analyses of DESI data using simulation-based priors.

Our work also provides significant opportunity for the generation of accurate covariance matrices. Our cubic model contains the tree-level trispectrum, which fully captures the connected covariance at the leading order. The model can be used to generate large sets of mock Lyman-$\alpha$
forest maps with correct 
four-point function correlations. Indeed, this will only require a large range of initial dark matter fields, which then can be forward-evolved into the forest maps using the transfer functions calibrated on high-fidelity small-scale hydrodynamical simulations along the lines of~\cite{deBelsunce:2026tks}.
As such, the field-level technique can significantly improve accuracy and robustness of covariance matrix estimation for cosmic surveys. The same technique can be used for galaxies, to which our model is readily applicable once we remove all line-of-sight dependent operators from the basis.  

These and other research directions are deferred to
future work. 

\begin{acknowledgments}
We thank Jahmour Givans for sharing the CAMELS simulations. 
We thank the Ly$\alpha$Fers Jamie Sullivan, 
Stephen Chen, and Kazuyuki Akitsu for fruitful discussions. 
This work is partly supported by the National Science Foundation under Cooperative Agreement PHY-2019786 (The NSF AI Institute for Artificial Intelligence and Fundamental Interactions, \url{http://iaifi.org/}). 
This research used resources of the National Energy Research Scientific Computing Center (NERSC), a U.S. Department of Energy Office of Science User Facility operated under Contract No. DE-AC02-05CH11231.
MMI is supported by the Office of High Energy Physics of U.S. Department of Energy under grant Contract Number DE-SC0012567.
This research has made use of NASA’s Astrophysics Data System and the arXiv preprint server.

\textit{AI usage statement:} This work was entirely conceptualized and written by the authors. Claude (Opus 5.5 and Fable 5.1; Anthropic) was used, in part, to perform numerical computations and for code optimization in \texttt{python} as well as to edit the manuscript for clarity. The authors have checked and validated all the results, taking full responsibility for all results presented herein.

\end{acknowledgments}

% \appendix

\newpage 

%%%%%%%%%% Merge with supplemental materials %%%%%%%%%%
\pagebreak
\widetext

\appendix

\section{Lagrangian cubic model for the Lyman-$\alpha$ forest}\label{sec:Lmodel}

Let us start with conventions. $\delta_1$ is the linear density and
\be
\eta \equiv \Pi_{zz} = \partial_z^2\nabla^{-2}\delta_1
\ee
is the line-of-sight tidal component, so that the linear velocity gradient is
$\partial_zu_z^{(1)} = -f\eta$ (this is the $\eta_1$ of the rest of this note,
$\eta_1 = -f\eta$). Line-of-sight composites of tensors are written 
as $(KK)_\parallel$, $(K\Pi^{[2]})_\parallel$,
$\Pi^{[3]}_\parallel$.

\textbf{Purely Lagrangian model.}
We start by formulating the \lya field model directly in Lagrangian space, in
analogy with other tracers such as galaxies. 
The previous EFT formulations
of the forest fluctuation models 
were given in the Eulerian space.
The goal of this section is to develop an alternative Lagrangian formulation, which is better suited to the field-level analysis.

The Lagrangian weight $w$ is a function of
the distortion tensors $L^{(n)}_{ij} = \partial_i\psi_j^{(n)}$ built from the $n$-th order
LPT displacements $\psi_i^{(n)}$. The symmetries of the forest allow for operators built by
contracting $L^{(n)}_{ij}$ with the line-of-sight vector $\hat z_i$. The observed (Eulerian) \lya field is then given by
\be
\delta_F^s(\vk) = \int d^3q\,\big[1 + w(\vq)\big]\,e^{-i\vk\cdot(\vq + \bpsi^s(\vq))} \cs
\bpsi^s = \sum_nR^{[n]}\bpsi^{(n)} \cs w = \sum_{a=1}^{22}c_a\,O_a\,,
\label{v2:LM}
\ee
with $R^{[n]}_{ij} = \delta_{ij} + nf\,\hat z_i\hat z_j$ and $O_a$ the following Lagrangian
operators through cubic order ($L_n\equiv L^{(n)}$, $\zz{A} = \hat z_iA_{ij}\hat z_j$):
\be
\begin{array}{lll}
n=1: & O_1 = \operatorname{tr}L_1 & O_2 = \zz{L_1}\\[3pt]
n=2: & O_3 = (\operatorname{tr}L_1)^2 \cs O_4 = \operatorname{tr}L_1^2 & O_5 = \zz{L_1}\operatorname{tr}L_1 \cs O_6 = \big[\zz{L_1}\big]^2\\
 & O_7 = \zz{L_1^2} & O_8 = \zz{L_2}\\[3pt]
n=3: & O_9 = (\operatorname{tr}L_1)^3 \cs O_{10} = \operatorname{tr}L_1\operatorname{tr}L_1^2 & O_{11} = \operatorname{tr}L_1^3\\
 & O_{12} = \zz{L_1}(\operatorname{tr}L_1)^2 \cs O_{13} = \zz{L_1}\operatorname{tr}L_1^2 & O_{14} = \big[\zz{L_1}\big]^2\operatorname{tr}L_1 \cs O_{15} = \big[\zz{L_1}\big]^3\\
 & O_{16} = \zz{L_1^2}\operatorname{tr}L_1 \cs O_{17} = \zz{L_1^2}\zz{L_1} & O_{18} = \operatorname{tr}(L_1L_2)\\
 & O_{19} = \zz{L_2}\operatorname{tr}L_1 \cs O_{20} = \zz{L_2}\zz{L_1} & O_{21} = \tfrac12\zz{\{L_1,L_2\}} \cs O_{22} = \zz{L_3}
\end{array}
\label{v2:D}
\ee
where $\{L_1,L_2\} = L_1L_2 + L_2L_1$.

\textbf{Proto-Eulerian model.}
The above model is perturbatively equivalent to the Eulerian \lya model of
Ref.~\cite{Ivanov:2023yla}. To see this we use a different basis, which is more easily
mapped onto the Eulerian one of~\cite{Ivanov:2023yla}; we call it ``proto-Eulerian.'' It is
formulated in terms of the operators $\Pi^{[n]}_{ij}$~\cite{Mirbabayi:2014zca,Desjacques:2018pfv},
\be
\begin{aligned}
&\Pi^{[1]}_{ij} = \frac{\partial_i\partial_j}{\nabla^2}\,\delta \cs
\Pi^{[n]}_{ij} = \frac{1}{(n-1)!}\Big[\frac{D}{D\ln D}\,\Pi^{[n-1]}_{ij} - (n-1)\,\Pi^{[n-1]}_{ij}\Big]\quad(n\ge2)\,,\\
&\frac{D}{D\ln D} = \frac{\partial}{\partial\ln D}\Big|_{\vx} + \dot\psi_k\,\partial_k\,,
\end{aligned}
\label{v2:Pirec}
\ee
with $\delta$ denoting the nonlinear matter density. In the monomial basis, the 22 operators are Eulerian
operators evaluated on the linear field,
\be
\begin{aligned}
&\text{linear:} && \delta_1,\ \eta\,,\\
&\text{quadratic:} && \delta_1^2,\ \GG_2,\ \delta_1\eta,\ \eta^2,\ (KK)_\parallel,\ \Pi^{[2]}_\parallel\,,\\
&\text{cubic:} && \delta_1^3,\ \delta_1\GG_2,\ \GG_3,\ \Gamma_3,\ \eta\delta_1^2,\ \eta^2\delta_1,\ \eta^3,\\
& && \delta_1(KK)_\parallel,\ \eta(KK)_\parallel,\ \delta_1\Pi^{[2]}_\parallel,\ \eta\Pi^{[2]}_\parallel,\ \eta\GG_2,\ (K\Pi_2)_\parallel,\ \Pi^{[3]}_\parallel\,,
\end{aligned}
\label{v2:tmono}
\ee
where the products are local products, and the remaining operators are
\be
\begin{aligned}
&\delta_1 = \Pi_{kk} \cs
\eta = \hat z_i\hat z_j\Pi_{ij} \cs
\GG_2 = \Pi_{ij}\Pi_{ji} - (\Pi_{kk})^2 \cs
(KK)_\parallel = \hat z_iK_{ik}K_{kj}\hat z_j\,,\\
&\Pi_{\parallel}^{[2]} = \hat z_i\hat z_j\Pi^{[2]}_{ij} \cs
\GG_3 = \tfrac12\big[(\Pi_{kk})^3 - 3\,\Pi_{kk}\Pi_{ij}\Pi_{ji} + 2\,\Pi_{ij}\Pi_{jk}\Pi_{ki}\big] \cs
\Gamma_3 = -\tfrac47\big[\Pi_{kk}\,\GG_2 - \Pi_{ij}\Pi^{\GG_2}_{ji}\big]\,,\\
&(K\Pi^{[2]})_\parallel = \tfrac12\,\hat z_i\big(K_{ik}\Pi^{[2]}_{kj} + \Pi^{[2]}_{ik}K_{kj}\big)\hat z_j \cs
\Pi^{[3]}_\parallel = \hat z_i\hat z_j\Pi^{[3]}_{ij} = \hat z_i\Pi_{ik}\Pi_{kl}\Pi_{lj}\hat z_j - \tfrac{13}{14}\,\KK_3
- 7\,\hat z_i\hat z_j\,\partial_i\psi^{(3)}_j\,,
\end{aligned}
\label{v2:tdefs}
\ee
built from the linear tidal tensor and its descendants
\be
\begin{aligned}
&\Pi_{ij} = \partial_i\partial_j\nabla^{-2}\delta_1 \cs
K_{ij} = \Pi_{ij} - \tfrac13\,\delta_{ij}\,\Pi_{kk} \cs
\Pi^{\GG_2}_{ij} = \partial_i\partial_j\nabla^{-2}\GG_2 \cs
\Pi^{[2]}_{ij} = \Pi_{ik}\Pi_{kj} - \tfrac57\,\Pi^{\GG_2}_{ij}\,,\\
&\KK_3 = \hat z_i\Pi_{ik}\Pi^{\GG_2}_{kj}\hat z_j \cs
\psi^{(1)}_i = -\partial_i\nabla^{-2}\delta_1 \cs
\psi^{(2)}_i = \tfrac{3}{14}\,\partial_i\nabla^{-2}\GG_2\,,\\
&\psi^{(3)}_i = \partial_i\nabla^{-2}W_3 - \nabla^{-2}\epsilon_{ijk}\partial_jC_k \cs
W_3 = \tfrac19\,\GG_3 - \tfrac{5}{24}\,\Gamma_3 \cs
C_i = -\tfrac{1}{14}\,\epsilon_{ijk}\,\Pi_{jl}\Pi^{\GG_2}_{kl}\,.
\end{aligned}
\label{v2:tblocks}
\ee
Since $\Pi_{ij}$ is linear in $\delta_1$, every operator in \eqref{v2:tmono} is a
fixed-order functional of the linear density. The nonlinear Eulerian operators, built
from the full $\delta$ and the nonlinear $\etaE = \partial_zu_z$, reduce to the above
operators perturbatively ($\etaE \to -f\eta$ at linear order). The 22 Lagrangian
distortion scalars are exact linear combinations of the above operators:
\be
\begin{aligned}
&\operatorname{tr}L_1 = -\delta_1 \cs \zz{L_1} = -\eta \cs (\operatorname{tr}L_1)^2 = \delta_1^2 \cs
\operatorname{tr}L_1^2 = \delta_1^2 + \GG_2\,,\\
&\zz{L_1}\operatorname{tr}L_1 = \delta_1\eta \cs \big[\zz{L_1}\big]^2 = \eta^2\,,\\
&\zz{L_1^2} = (KK)_\parallel + \tfrac23\,\delta_1\eta - \tfrac19\,\delta_1^2\,,\\
&\zz{L_2} = -\tfrac1{30}\,\delta_1^2 + \tfrac15\,\delta_1\eta + \tfrac{3}{10}\,(KK)_\parallel - \tfrac{3}{10}\,\Pi^{[2]}_\parallel\,,
\end{aligned}
\label{v2:Mquad}
\ee
\be
\begin{aligned}
(\operatorname{tr}L_1)^3 &= -\delta_1^3 \cs
\operatorname{tr}L_1\operatorname{tr}L_1^2 = -\delta_1^3 - \delta_1\GG_2 \cs
\operatorname{tr}L_1^3 = -\delta_1^3 - \tfrac32\,\delta_1\GG_2 - \GG_3\,,\\
\zz{L_1}(\operatorname{tr}L_1)^2 &= -\eta\delta_1^2 \cs
\zz{L_1}\operatorname{tr}L_1^2 = -\eta\delta_1^2 - \eta\GG_2\,,\\
\big[\zz{L_1}\big]^2\operatorname{tr}L_1 &= -\eta^2\delta_1 \cs \big[\zz{L_1}\big]^3 = -\eta^3\,,\\
\zz{L_1^2}\operatorname{tr}L_1 &= \tfrac19\,\delta_1^3 - \tfrac23\,\eta\delta_1^2 - \delta_1(KK)_\parallel\,,\\
\zz{L_1^2}\zz{L_1} &= \tfrac19\,\eta\delta_1^2 - \tfrac23\,\eta^2\delta_1 - \eta(KK)_\parallel \cs
\operatorname{tr}(L_1L_2) = -\tfrac{3}{14}\,\delta_1\GG_2 - \tfrac38\,\Gamma_3\,,\\
\zz{L_2}\operatorname{tr}L_1 &= \tfrac1{30}\,\delta_1^3 - \tfrac15\,\eta\delta_1^2 - \tfrac{3}{10}\,\delta_1(KK)_\parallel + \tfrac{3}{10}\,\delta_1\Pi^{[2]}_\parallel\,,\\
\zz{L_2}\zz{L_1} &= \tfrac1{30}\,\eta\delta_1^2 - \tfrac15\,\eta^2\delta_1 - \tfrac{3}{10}\,\eta(KK)_\parallel + \tfrac{3}{10}\,\eta\Pi^{[2]}_\parallel\,,\\
\tfrac12\zz{\{L_1,L_2\}} &= -\tfrac{3}{14}\,\KK_3
= \tfrac1{30}\,\delta_1^3 - \tfrac1{10}\,\GG_3 - \tfrac15\,\eta\delta_1^2 - \tfrac{3}{10}\,\delta_1(KK)_\parallel
+ \tfrac1{10}\,\delta_1\Pi^{[2]}_\parallel - \tfrac{3}{20}\,\eta\GG_2 + \tfrac{3}{10}\,(K\Pi^{[2]})_\parallel\,,\\
\zz{L_3} &= \tfrac1{210}\,\delta_1^3 - \tfrac1{70}\,\GG_3 - \tfrac1{35}\,\eta\delta_1^2 - \tfrac{3}{70}\,\delta_1(KK)_\parallel
+ \tfrac{13}{210}\,\delta_1\Pi^{[2]}_\parallel - \tfrac{3}{140}\,\eta\GG_2 + \tfrac{13}{70}\,(K\Pi^{[2]})_\parallel - \tfrac17\,\Pi^{[3]}_\parallel\,.
\end{aligned}
\label{v2:Mcub}
\ee

Introduce now the new Lagrangian operators
\be
\begin{aligned}
&\delta_L = -\operatorname{tr}L_1 \cs
\eta_L = f\,\zz{L_1 + 2L_2 + 3L_3} = -f\eta + \tfrac{3f}{7}\,\GG_2^\parallel + \tfrac f3\,\mu^2\GG_3
- \tfrac{5f}{8}\,\mu^2\Gamma_3 + 3f\,\TT_3\,,\\
&(\delta^2)_L = (\operatorname{tr}L_1)^2 \cs (\GG_2)_L = \operatorname{tr}L_1^2 - (\operatorname{tr}L_1)^2 \cs
(\delta\eta)_L = \delta_L\eta_L \cs (\eta^2)_L = \eta_L^2\,,\\
&\big((KK)_\parallel\big)_L = \zz{L_1^2} - \tfrac23\operatorname{tr}L_1\zz{L_1} + \tfrac19(\operatorname{tr}L_1)^2 \cs
(\Pi_2)_L = \zz{L_1^2} - \tfrac{10}{3}\zz{L_2}\,,
\end{aligned}
\label{v2:OL}
\ee
($\eta_L = \hat z_i\hat z_j\partial_{q_i}(f\dot\psi_j)$ is the Lagrangian gradient of the
line-of-sight velocity; $\GG_2^\parallel$, $\TT_3$ are defined in \eqref{v2:shapes} below)
and the 14 cubic operators $O_L = (-f)^{k}O$, with $k$ the number of $\eta$ legs of $O$
(the linear part $-f\eta$ of $\eta_L$ on every $\eta$ leg), each a combination of
$O_9,\dots,O_{22}$ obtained by inverting \eqref{v2:Mcub}. 
We arrive at a second Lagrangian model,
equivalent to the first one, which is mapped more easily onto the Eulerian operators. The
Lagrangian weight now is
\be
w = \sum_Ob^{\rm L}_O\,O_L(\vq)\,,
\ee
where $b^{\rm L}_O$ are the Lagrangian bias parameters of the forest.

\textbf{The Lagrangian field-level model with shifted operators.}
We now split $\bpsi^s = R^{[1]}\bpsi^{(1)} + \bar\bpsi$ and expand in $\bar\bpsi$ to cubic
order. Upon integration by parts we arrive at
\be
\delta_F^s(\vk) = \int d^3q\;\mathcal B(\vq)\,e^{-i\vk\cdot(\vq + R^{[1]}\bpsi^{(1)}(\vq))}\,,
\ee
where
\be
\begin{aligned}
\mathcal B &= 1 + w - \big(1 + w^{(1)}\big)R^{[2]}_{ij}L^{(2)}_{ji} - R^{[3]}_{ij}L^{(3)}_{ji}
+ R^{[2]}_{ij}L^{(2)}_{jk}R^{[1]}_{kl}L^{(1)}_{li} - R^{[2]}_{ij}\psi^{(2)}_j\partial_i\big(\delta_1 + f\eta + w^{(1)}\big)\\
&= 1 + w - \tfrac{3}{14}\,\GG_2 - \tfrac{3f}{7}\,\GG_2^\parallel
- \tfrac{3}{14}\big(1+b^{\rm L}_\delta\big)\delta_1\GG_2 - \tfrac{3f}{7}\,b^{\rm L}_\delta\,\delta_1\GG_2^\parallel
+ \tfrac{3f}{14}\,b^{\rm L}_\eta\,\eta\GG_2 - \tfrac{3f^2}{7}\big(1 - b^{\rm L}_\eta\big)\eta\GG_2^\parallel\\
&\quad - \tfrac19\,\GG_3 - \tfrac16\,\Gamma_3 - f\mu^2\Big(\tfrac13\,\GG_3 - \tfrac58\,\Gamma_3\Big)
- \tfrac{9f}{14}\,\KK_3 - 3f\,\TT_3
- \big(1+b^{\rm L}_\delta\big)S_3^\delta - \big(1 - b^{\rm L}_\eta\big)S_3^\eta\,,
\end{aligned}
\label{v2:B}
\ee
with $w^{(1)} = b^{\rm L}_\delta\,\delta_1 - f\,b^{\rm L}_\eta\,\eta$ the linear part of the
proto-Eulerian weight. The new operators above are
\be
\begin{aligned}
&\GG_2^\parallel = \hat z_i\hat z_j\Pi^{\GG_2}_{ij} \cs
\KK_3 = \hat z_i\Pi_{ik}\Pi^{\GG_2}_{kj}\hat z_j \cs
\TT_3 = \hat z_i\hat z_j\partial_i\psi^{(3)T}_j = -\hat z_i\hat z_j\,\partial_i\nabla^{-2}\epsilon_{jab}\partial_aC_b\,,\\
&S_3 = \psi^{(2)}_i\partial_i\delta_1 \cs
S_3^\delta = R^{[2]}_{ij}\psi^{(2)}_j\partial_i\delta_1 \cs
S_3^\eta = f\,R^{[2]}_{ij}\psi^{(2)}_j\partial_i\eta \cs
\mu^2O \equiv \hat z_i\hat z_j\partial_i\partial_j\nabla^{-2}O\,,
\end{aligned}
\label{v2:shapes}
\ee
together with the local products $\delta_1\GG_2^\parallel$, $\eta\GG_2^\parallel$,
$\bpsi_2\cdot\nabla\eta$ and $\psi_2^z\partial_z\eta$ ($\bpsi_2 \equiv \bpsi^{(2)}$).

Let us make a few comments. $S_3$ is the linear density shifted by the 2LPT displacement;
the coefficient in front of it is protected by the equivalence principle. $S_3^\delta$ is
its redshift-space version, and $S_3^\eta$ is the analog of $S_3^\delta$ for the
line-of-sight velocity gradient. $\TT_3$ is the transverse part of the 3LPT displacement
projected onto the line of sight. Unlike the real-space situation, $\TT_3$ does not drop
out, because $k_z\psi^{(3)T}_z$ does not vanish. This operator obeys the kernel identity
\be
\TT_3 = \tfrac1{14}\,\KK_3 + \tfrac13\,\bpsi_2\cdot\nabla\eta - \tfrac13\,\mu^2S_3 -
\tfrac18\,\mu^2\Gamma_3 - \tfrac1{14}\,\mu^2\delta_1\GG_2\,,
\ee
which allows us to rewrite the Lagrangian model as

\be
\begin{aligned}
\delta_F^s = \int d^3q\,\Big[\,
&1 + b^{\rm L}_\delta\,\delta_1 - f\,b^{\rm L}_\eta\,\eta
+ b^{\rm L}_{\delta^2}\,\delta_1^2 + \big(b^{\rm L}_{\GG_2} - \tfrac{3}{14}\big)\GG_2
- \tfrac{3f}{7}\big(1 - b^{\rm L}_\eta\big)\GG_2^\parallel
- f\,b^{\rm L}_{\delta\eta}\,\delta_1\eta + f^2\,b^{\rm L}_{\eta^2}\,\eta^2\\
&+ b^{\rm L}_{(KK)_\parallel}\,(KK)_\parallel + b^{\rm L}_{\Pi^{[2]}_\parallel}\,\Pi^{[2]}_\parallel
+ b^{\rm L}_{\delta^3}\,\delta_1^3
+ \Big[b^{\rm L}_{\delta\GG_2} - \tfrac{3}{14}\big(1 + b^{\rm L}_\delta\big) + \tfrac{3f}{14}\big(1 - b^{\rm L}_\eta\big)\mu^2\Big]\delta_1\GG_2\\
&+ \Big[b^{\rm L}_{\GG_3} - \tfrac19 - \tfrac f3\big(1 - b^{\rm L}_\eta\big)\mu^2\Big]\GG_3
+ \Big[b^{\rm L}_{\Gamma_3} - \tfrac16 + f\big(1 - b^{\rm L}_\eta\big)\mu^2\Big]\Gamma_3\\
&- \tfrac{3f}{7}\big(b^{\rm L}_\delta - b^{\rm L}_{\delta\eta}\big)\delta_1\GG_2^\parallel
- \tfrac{3f^2}{7}\big(1 - b^{\rm L}_\eta + 2\,b^{\rm L}_{\eta^2}\big)\eta\GG_2^\parallel
- f\Big[b^{\rm L}_{\eta\GG_2} - \tfrac{3}{14}\,b^{\rm L}_\eta\Big]\eta\GG_2\\
&- f\,b^{\rm L}_{\delta^2\eta}\,\eta\delta_1^2 + f^2\,b^{\rm L}_{\delta\eta^2}\,\eta^2\delta_1 - f^3\,b^{\rm L}_{\eta^3}\,\eta^3
+ b^{\rm L}_{\delta(KK)_\parallel}\,\delta_1(KK)_\parallel - f\,b^{\rm L}_{\eta(KK)_\parallel}\,\eta(KK)_\parallel\\
&+ b^{\rm L}_{\delta\Pi^{[2]}_\parallel}\,\delta_1\Pi^{[2]}_\parallel - f\,b^{\rm L}_{\eta\Pi^{[2]}_\parallel}\,\eta\Pi^{[2]}_\parallel
+ b^{\rm L}_{(K\Pi^{[2]})_\parallel}\,(K\Pi^{[2]})_\parallel + b^{\rm L}_{\Pi^{[3]}_\parallel}\,\Pi^{[3]}_\parallel
- \Big(\tfrac{6f}{7} - \tfrac{3f}{14}\,b^{\rm L}_\eta\Big)\KK_3\\
&- \big(1 + b^{\rm L}_\delta\big)S_3^\delta
- \big(1 - b^{\rm L}_\eta\big)\Big[2f\,\bpsi_2\cdot\nabla\eta + 2f^2\,\psi_2^z\partial_z\eta - f\mu^2S_3\Big]
\;\Big]\;e^{-i\vk\cdot(\vq + R^{[1]}\bpsi^{(1)})}\,.
\end{aligned}
\label{v2:57}
\ee
We define a shifted operator following~\cite{Schmittfull:2018yuk}
\be
\td O(\vk) = \int d^3q\,O(\vq)\,e^{-i\vk\cdot(\vq + R^{[1]}\bpsi^{(1)}(\vq))}\,.
\ee
Removing $\td\eta$ with the exact identity
\be
f\,\td\eta = \td 1 - \td\delta_1 - \tfrac12\,\td\GG_2 - f\,\td{(KK)_\parallel}
+ \tfrac f3\,\td{\delta_1\eta} + \tfrac f9\,\td{\delta_1^2} - \tfrac{1+f}{3}\,\td{\GG}_3\,,
\ee
expressing 
$(K\Pi^{[2]})_\parallel$
through
$\KK_3$
and other operators via~\eqref{v2:tdefs} and introducing
\be
\td\delta'_1= \td\delta_1 - \td S_3^\delta \cs
\td\eta^{\prime} = \td 1 - \tfrac{3f}{7}\,\td\GG_2^\parallel - \td S_3^\delta
- 2f\,\td{\bpsi_2\cdot\nabla\eta} - 2f^2\,\td{\psi_2^z\partial_z\eta} + f\,\td{\mu^2S_3}\,,
\label{v2:linops}
\ee
we arrive at
\be
\begin{aligned}
\delta_F^s ={}& \beta_1\,\td\delta_1+ \beta_\eta\,\td\eta^{\prime}\\
&+ \beta_{\delta^2}\,\td{\delta_1^2} + \beta_{\GG_2}\,\td\GG_2
+ \beta_{\delta\eta}\,\td{\delta_1\eta} + \beta_{\eta^2}\,\td{\eta^2}
+ \beta_{(KK)_\parallel}\,\td{(KK)_\parallel}\\
&+ \beta_{\delta^3}\,\td{\delta_1^3} + \beta_{\delta\GG_2}\,\td{\delta_1\GG_2}
+ \beta_{\GG_3}\,\td\GG_3 + \beta_{\Gamma_3}\,\td\Gamma_3
+ \beta_{\eta\delta^2}\,\td{\eta\delta_1^2} + \beta_{\eta^2\delta}\,\td{\eta^2\delta_1}
+ \beta_{\eta^3}\,\td{\eta^3}\\
&+ \beta_{\delta(KK)_\parallel}\,\td{\delta_1(KK)_\parallel}
+ \beta_{\eta(KK)_\parallel}\,\td{\eta(KK)_\parallel}
+ \beta_{\delta\Pi^{[2]}_\parallel}\,\td{\delta_1\Pi^{[2]}_\parallel}
+ \beta_{\eta\Pi^{[2]}_\parallel}\,\td{\eta\Pi^{[2]}_\parallel}\\
&+ \beta_{\eta\GG_2}\,\td{\eta\GG_2}
+ \beta_{\KK_3}\,\td\KK_3
+ \beta_{\Pi^{[3]}_\parallel}\,\td\Pi^{[3]}_\parallel+ \beta_{\Pi^{[2]}_\parallel}\,\td\Pi^{[2]}_\parallel\,,
\end{aligned}
\label{v2:FL}
\ee
with two linear, six quadratic and fourteen cubic operators. This is our final Lagrangian
shifted-field model. The coefficients $\beta_O$ are a rotation of $b^{\rm L}_O$. While
straightforward to obtain, we do not need this map in practice, since below we
express the $\beta_O$ directly through the Eulerian bias parameters.

% We define a shifted operator following~\cite{Schmittfull:2018yuk}
% \be
% \td O(\vk) = \int d^3q\,O(\vq)\,e^{-i\vk\cdot(\vq + R^{[1]}\bpsi^{(1)}(\vq))}\,.
% \ee
% Removing $\td\eta$ with the exact identity
% \be
% f\,\td\eta = \td 1 - \td\delta_1 - \tfrac12\,\td\GG_2 - f\,\td{(KK)_\parallel}
% + \tfrac f3\,\td{\delta_1\eta} + \tfrac f9\,\td{\delta_1^2} - \tfrac{1+f}{3}\,\td{\GG}_3\,,
% \ee
% expressing $\KK_3$ through 
% $\Pi_\parallel^{[3]}$
% and other operators via~\eqref{v2:tdefs} and introducing
% \be
% \td\delta_1^{\rm new} = \td\delta_1 - \td S_3^\delta \cs
% \td\eta^{\prime}_{\rm new} = \td 1 - \tfrac{3f}{7}\,\td\GG_2^\parallel - \td S_3^\delta
% - 2f\,\td{\bpsi_2\cdot\nabla\eta} - 2f^2\,\td{\psi_2^z\partial_z\eta} + f\,\td{\mu^2S_3}\,,
% \label{v2:linops}
% \ee
% we arrive at
% \be
% \delta_F^s = \beta_1\,\td\delta_1^{\rm new} + \beta_\eta\,\td\eta^{\prime}_{\rm new}
% + \sum_{O\,\in\,\{\delta_1^2,\,\GG_2,\,\delta_1\eta,\,\eta^2,\,(KK)_\parallel,\,\Pi_2\}}\beta_O\,\td O
% + \sum_{O\,\in\,14\ {\rm cubic}}\beta_O\,\td O\,,
% \label{v2:FL}
% \ee
% where the cubic operators are those of Eq.~\eqref{v2:tmono}. This is our final Lagrangian
% shifted-field model. The coefficients $\beta_O$ are a rotation of $b^{\rm L}_O$. While straightforward to obtain, we do not need this map in practice, since below we
% express the $\beta_O$ directly through the Eulerian bias parameters.

\textbf{Renormalized operators.}
One important observation is that the cubic operators above have formally infinite
correlations with the linear field $\delta_1$. At one loop
\be
\big\langle O(\vk)\,\delta_1(\vk')\big\rangle' \Big|_{k\to0}=\Sigma_O(\mu)\,\sigma^2P_{11}(k) \cs
\sigma^2 = \int_{\vp}P_{11}(p) \cs \Sigma_O(\mu) = 3\,\big\langle K_O^{\rm sym}(\hat\vk,\hat\vp,-\hat\vp)\big\rangle_{\hat\vp}\,,
\ee
the UV-sensitive part of the $P_{13}$-type integral. Note that $\Sigma_O$ above is a polynomial in $\mu^2$.
We therefore use the renormalized operators
\be
[O]_{R} \equiv O - \Sigma_O(\mu)\,\sigma^2\,\delta_1\,,
\label{v2:renorm}
\ee
subtracted in Lagrangian space before the shift ($\sigma^2$ is the variance computed on the grid, and $\mu^2$ acts
on $\delta_1(\vk)$). The cubic products are built from raw fields. 
For the 14 cubic operators,
\be
\begin{aligned}
&\Sigma_{\delta_1^3} = 3 \cs \Sigma_{\delta_1\GG_2} = -\tfrac43 \cs \Sigma_{\GG_3} = \Sigma_{\Gamma_3} = 0\,,\\
&\Sigma_{\eta\delta_1^2} = \tfrac23 + \mu^2 \cs \Sigma_{\eta^2\delta_1} = \tfrac15 + \tfrac23\,\mu^2 \cs \Sigma_{\eta^3} = \tfrac35\,\mu^2\,,\\
&\Sigma_{\delta_1(KK)_\parallel} = \tfrac29 \cs \Sigma_{\eta(KK)_\parallel} = -\tfrac{8}{135} + \tfrac25\,\mu^2 \cs
\Sigma_{\eta\GG_2} = -\tfrac{8}{15} + \tfrac{4}{15}\,\mu^2\,,\\
&\Sigma_{\delta_1\Pi^{[2]}_\parallel} = \tfrac57 + \tfrac{10}{21}\,\mu^2 \cs \Sigma_{\eta\Pi^{[2]}_\parallel} = \tfrac{12}{49} + \tfrac{419}{735}\,\mu^2\,,\\
&\Sigma_{\KK_3} = -\tfrac{8}{15} + \tfrac{4}{15}\,\mu^2 \cs \Sigma_{\Pi^{[3]}_\parallel} = \tfrac{59}{105} + \tfrac{58}{105}\,\mu^2\,.
\end{aligned}
\label{v2:Sigma}
\ee

The shifted quadratic operators (mean-subtracted in Lagrangian space) and the
Zel'dovich parts of the linear operators have no overlap with $\delta_1$ to any order in
the displacement; only the $\bpsi_2$ shapes inside \eqref{v2:linops} need subtraction, which gives
\be
\begin{aligned}
[\td\delta_1^{{\prime}}]_R &= \td\delta_1^{\prime} + \Big[\tfrac27 + \tfrac{8f}{35} - \tfrac{4f}{35}\,\mu^2\Big]\sigma^2\,\delta_1\,,\\
[\td\eta^{{\prime}}]_R &= \td\eta^{\prime} + \Big[\tfrac27 + \tfrac{16f}{35} + \tfrac{36f^2}{245}
- \Big(\tfrac{18f}{35} + \tfrac{24f^2}{245}\Big)\mu^2\Big]\sigma^2\,\delta_1\,.
\end{aligned}
\label{v2:renormlin}
\ee
Since $\sum_O\beta_OO = \sum_O\beta_O[O]_{R} + \big[\sum_O\beta_O\Sigma_O(\mu)\big]\sigma^2\delta_1$,
renormalization only moves a $\mu$-dependent, $\sigma^2$-proportional piece into the
linear transfer functions and leaves the coefficients of the composite operators, and hence the
mapping between the Lagrangian and Eulerian bias parameters (below) unchanged. 

\textbf{Mapping to the CLASS-PT-based Eulerian model.}
Let us connect this model with the Eulerian model
of~\cite{Ivanov:2023yla,Desjacques:2018pfv}. We start with the following expression for the \lya flux overdensity in the rest frame (all fields below are nonlinear Eulerian
fields, and we use $\etaE = \partial_zu_z$):
\be
\begin{aligned}
\delta_F^{\rm rf} ={}& b_1\,\delta + b^{\rm rf}_\eta\,\etaE + \tfrac{b_2}{2}\,\delta^2 + b_{\GG_2}\,\GG_2
+ b^{\rm rf}_{\delta\eta}\,\delta\etaE + b^{\rm rf}_{\eta^2}\,(\etaE)^2\\
&+ b_{(KK)_\parallel}\,(KK)_\parallel + b_{\Pi^{[2]}_\parallel}\,\Pi^{[2]}_\parallel
+ b_{\Gamma_3}\,\Gamma_3^{\rm CPT} + b_{\delta\Pi^{[2]}_\parallel}\,\delta\,\Pi^{[2]}_\parallel\\
&+ b^{\rm rf}_{\eta\Pi^{[2]}_\parallel}\,\etaE\,\Pi^{[2]}_\parallel
+ b_{(K\Pi^{[2]})_\parallel}\,(K\Pi^{[2]})_\parallel + b_{\Pi^{[3]}_\parallel}\,\Pi^{[3]}_\parallel\\
&+ b_{\delta^3}\,\delta^3 + b_{\delta\GG_2}\,\delta\GG_2 + b_{\GG_3}\,\GG_3
+ b_{\delta^2\eta}\,\delta^2\etaE + b_{\delta\eta^2}\,\delta(\etaE)^2\\
&+ b_{\eta^3}\,(\etaE)^3 + b_{\delta(KK)_\parallel}\,\delta\,(KK)_\parallel
+ b_{\eta(KK)_\parallel}\,\etaE(KK)_\parallel + b_{\eta\GG_2}\,\etaE\GG_2\,,
\end{aligned}
\label{v2:Ffull}
\ee
where $\Gamma_3^{\rm CPT} = -\Gamma_3$ accounts for the convention mismatch
between~\cite{Ivanov:2023yla} and our Lagrangian model (stemming from the definition used
in~\cite{Schmittfull:2020trd}). Applying the redshift-space distortions and expanding the model perturbatively we get
\be
\begin{aligned}
\delta_F^s ={}& b_1\,\delta + b_\eta\,\etaE + \tfrac{b_2}{2}\,\delta^2 + b_{\GG_2}\,\GG_2
+ \big(b_{\delta\eta} + b_1\big)\delta\etaE + \big(b_{\eta^2} + b_\eta\big)(\etaE)^2\\
&+ b_{(KK)_\parallel}\,(KK)_\parallel + b_{\Pi^{[2]}_\parallel}\,\Pi^{[2]}_\parallel
+ b_{\Gamma_3}\,\Gamma_3^{\rm CPT} + b_{\delta\Pi^{[2]}_\parallel}\,\delta\,\Pi^{[2]}_\parallel\\
&+ \big(b_{\eta\Pi^{[2]}_\parallel} + b_{\Pi^{[2]}_\parallel}\big)\etaE\,\Pi^{[2]}_\parallel
+ b_{(K\Pi^{[2]})_\parallel}\,(K\Pi^{[2]})_\parallel + b_{\Pi^{[3]}_\parallel}\,\Pi^{[3]}_\parallel\\
&+ b_{\delta^3}\,\delta^3 + b_{\delta\GG_2}\,\delta\GG_2 + b_{\GG_3}\,\GG_3
+ b_{\delta^2\eta}\,\delta^2\etaE + b_{\delta\eta^2}\,\delta(\etaE)^2\\
&+ b_{\eta^3}\,(\etaE)^3 + b_{\delta(KK)_\parallel}\,\delta\,(KK)_\parallel
+ b_{\eta(KK)_\parallel}\,\etaE(KK)_\parallel + b_{\eta\GG_2}\,\etaE\GG_2\\
&- \partial_z\Big[u_z\Big(b_1\,\delta + b_\eta\,\etaE + \tfrac{b_2}{2}\,\delta^2 + b_{\GG_2}\,\GG_2
+ \big(b_{\delta\eta} + b_1\big)\delta\etaE\\
&\qquad\quad + \big(b_{\eta^2} + b_\eta\big)(\etaE)^2 + b_{(KK)_\parallel}\,(KK)_\parallel
+ b_{\Pi^{[2]}_\parallel}\,\Pi^{[2]}_\parallel\Big)\Big]\\
&+ \tfrac12\,\partial_z^2\Big[u_z^2\big(b_1\,\delta + b_\eta\,\etaE\big)\Big] + \mathcal O(\delta_1^4)\,.
\end{aligned}
\label{v2:Fs}
\ee
Following the conventions of~\cite{Ivanov:2023yla} we have defined
\be
b_\eta = b^{\rm rf}_\eta - 1 \cs b_{\delta\eta} = b^{\rm rf}_{\delta\eta} - b_1 \cs
b_{\eta^2} = b^{\rm rf}_{\eta^2} - b_\eta \cs
b_{\eta\Pi^{[2]}_\parallel} = b^{\rm rf}_{\eta\Pi^{[2]}_\parallel} - b_{\Pi^{[2]}_\parallel}\,.
\label{v2:obs}
\ee
so that all coefficients in the model~\eqref{v2:Fs} are the observed coefficients.
Now we match the field-level model and the CLASS-PT Eulerian model perturbatively.
Expanding both through $\delta_1^3$ and demanding that the kernels match at each order in $\delta_1$ we
obtain the map
\be
\begin{aligned}
&\beta_1 = b_1 + b_\eta \cs \beta_\eta = -b_\eta \cs \beta_{\delta^2} = \tfrac{b_2}{2}\cs\beta_{\GG_2} = b_{\GG_2} + \tfrac27\,b_1 + \tfrac12\,b_\eta \cs
\beta_{\delta\eta} = -f\,\big(b_1 + b_{\delta\eta}\big)\,,\\
&\beta_{\eta^2} = f^2\big(b_\eta + b_{\eta^2}\big) \cs
\beta_{(KK)_\parallel} = b_{(KK)_\parallel} \cs \beta_{\Pi^{[2]}_\parallel} = b_{\Pi^{[2]}_\parallel}\cs \beta_{\Gamma_3} = -b_{\Gamma_3} - \tfrac16\,b_1 - \tfrac52\,b_{\GG_2} - f\mu^2\,b_\eta\,,\\
&\beta_{\delta\Pi^{[2]}_\parallel} = b_{\delta\Pi^{[2]}_\parallel} - b_{\Pi^{[2]}_\parallel} - \tfrac{3f}{5}\,b_{\delta\eta} + \tfrac{3f}5\,b_\eta - \tfrac23\,b_{(KK)_\parallel} - \tfrac13\,b_{(K\Pi^{[2]})_\parallel}\,,\\
&\beta_{\eta\Pi^{[2]}_\parallel} = -f\,b_{\eta\Pi^{[2]}_\parallel} - f\,b_{\Pi^{[2]}_\parallel} + \tfrac{3f^2}{5}\,b_\eta + \tfrac{6f^2}{5}\,b_{\eta^2}\,,\\
&\beta_{\KK_3} = -\tfrac57\,b_{(K\Pi^{[2]})_\parallel} - \tfrac{10}{7}\,b_{(KK)_\parallel} + \tfrac{6f}{7}\,b_\eta\cs \beta_{\Pi^{[3]}_\parallel} = b_{\Pi^{[3]}_\parallel} + 2\,b_{\Pi^{[2]}_\parallel}\,,\\[3pt]
&\beta_{\delta^3} = b_{\delta^3} + \tfrac{f}{9}\,b_1 + \tfrac{b_2}{2} + \tfrac{2f}{45}\,b_{\delta\eta} + \tfrac{f}{15}\,b_\eta - \tfrac19\,b_{(K\Pi^{[2]})_\parallel}\cs \beta_{\delta\GG_2} = b_{\delta\GG_2} - \tfrac3{14}\,b_1 + \tfrac27\,b_2 + \tfrac{4}{63}\,b_{(KK)_\parallel} - \tfrac{3f\mu^2}{14}\,b_\eta\,,\\
&\beta_{\GG_3} = b_{\GG_3} + \tfrac29\,b_1 + 2\,b_{\GG_2} + \Big(\tfrac13 + \tfrac{f\mu^2}{3}\Big)b_\eta + \tfrac23\,b_{(KK)_\parallel} + \tfrac13\,b_{(K\Pi^{[2]})_\parallel}\,,\\
&\beta_{\delta^2\eta} = -f\,b_{\delta^2\eta} - \tfrac{2f}{3}\,b_1 - \tfrac{4f}{15}\,b_{\delta\eta}
- \Big(\tfrac{2f}{5} + \tfrac{7f^2}{45}\Big)b_\eta- \tfrac{4f^2}{45}\,b_{\eta^2} + \tfrac23\,b_{(KK)_\parallel} + \tfrac23\,b_{(K\Pi^{[2]})_\parallel}\,,\\
&\beta_{\delta\eta^2} = f^2\,b_{\delta\eta^2} - \tfrac{f^2}{15}\,b_\eta - \tfrac{7f^2}{15}\,b_{\eta^2} \cs
\beta_{\eta^3} = -f^3\,b_{\eta^3}\,,\\
&\beta_{\delta(KK)_\parallel} = b_{\delta(KK)_\parallel} - f\,b_1 + b_{(KK)_\parallel} - \tfrac{2f}{5}\,b_{\delta\eta} - \tfrac{3f}{5}\,b_\eta + b_{(K\Pi^{[2]})_\parallel}\,,\\
&\beta_{\eta(KK)_\parallel} = -f\,b_{\eta(KK)_\parallel} + \tfrac{7f^2}{5}\,b_\eta + \tfrac{4f^2}{5}\,b_{\eta^2}\cs
\beta_{\eta\GG_2} = -f\,b_{\eta\GG_2} - \tfrac{2f}{7}\,\big(b_1 + b_{\delta\eta}\big) + \tfrac{17}{21}\,b_{(KK)_\parallel} + \tfrac12\,b_{(K\Pi^{[2]})_\parallel}\,,
\end{aligned}
\label{v2:beta}
\ee
% where $\beta_O$ is the coefficient of the operator $O$ of the field-level
% model~\eqref{v2:FL}, built from the linear fields $\delta_1$, $\eta$ (e.g.\ $\beta_{\delta\eta}$
% multiplies $\td{\delta_1\eta}$ and $\beta_{\delta^2\eta}$ multiplies $\eta\delta_1^2$), and
% $b_O$ are the observed CLASS-PT coefficients of~\eqref{v2:Fs}. The operator
% $(K\Pi^{[2]})_\parallel$ of~\eqref{v2:Fs} is not part of the basis~\eqref{v2:FL}: it is
% expressed through $\KK_3$ and the other cubic operators with the first identity
% of~\eqref{id:all}, so that $b_{(K\Pi^{[2]})_\parallel}$ enters the coefficients of $\KK_3$,
% $\delta_1^3$, $\GG_3$, $\eta\GG_2$, $\eta\delta_1^2$, $\delta_1(KK)_\parallel$ and
% $\delta_1\Pi^{[2]}_\parallel$.
where $\beta_O$ is the coefficient of the operator $O$ of the field-level
model~\eqref{v2:FL}, built from the linear fields $\delta_1$, $\eta$ (e.g.\ $\beta_{\delta\eta}$
multiplies $\td{\delta_1\eta}$ and $\beta_{\delta^2\eta}$ multiplies $\eta\delta_1^2$), and
$b_O$ are the observed CLASS-PT coefficients of~\eqref{v2:Fs}.

Two operators of~\eqref{v2:Fs} require a comment. The operator $(K\Pi^{[2]})_\parallel$
is not part of the basis~\eqref{v2:FL}. We trade it for $\KK_3$ by virtue of the identity
\be
(K\Pi^{[2]})_\parallel = -\tfrac57\,\KK_3 - \tfrac19\,\delta_1^3 + \tfrac13\,\GG_3
+ \tfrac12\,\eta\GG_2 + \tfrac23\,\eta\delta_1^2 + \delta_1(KK)_\parallel
- \tfrac13\,\delta_1\Pi^{[2]}_\parallel\,,
\label{id:KPi2}
\ee
so that $b_{(K\Pi^{[2]})_\parallel}$ enters only through the coefficients of the seven operators on the
right-hand side.

The operator $\Pi^{[2]}_\parallel$ is part of the model~\eqref{v2:FL}, but its contribution is fully redundant once the bias parameters are promoted to transfer functions. At quadratic order it obeys
\be
\Pi^{[2]}_\parallel = (KK)_\parallel + \tfrac23\,\delta_1\eta - \tfrac19\,\delta_1^2
- \tfrac57\,\GG_2^\parallel \,,
\label{id:Pi2}
\ee
and $\td{\GG_2^\parallel}=\mu^2\,\td{\GG}_2$ at this order. With $(k,\mu)$-dependent
transfer functions $\Pi^{[2]}_\parallel$ is absorbed by
\be
\beta_{(KK)_\parallel}\to\beta_{(KK)_\parallel}+\beta_{\Pi^{[2]}_\parallel} \cs
\beta_{\delta\eta}\to\beta_{\delta\eta}+\tfrac23\,\beta_{\Pi^{[2]}_\parallel} \cs
\beta_{\delta^2}\to\beta_{\delta^2}-\tfrac19\,\beta_{\Pi^{[2]}_\parallel} \cs
\beta_{\GG_2}\to\beta_{\GG_2}-\tfrac57\,\mu^2\,\beta_{\Pi^{[2]}_\parallel}\,.
\label{v2:Pi2abs}
\ee
At quadratic order $b_{\Pi^{[2]}_\parallel}$ is thus distinguished from $b_{(KK)_\parallel}$,
$b_{\delta\eta}$, $b_2$ and $b_{\GG_2}$ only by the $\mu^2$ dependence of the $\GG_2$ transfer
function.

At cubic order the contribution of $b_{\Pi^{[2]}_\parallel}$ is accounted for by the
map~\eqref{v2:beta}. The Eulerian operator
of~\eqref{v2:Fs} and the shifted operator of~\eqref{v2:FL} differ by cubic terms,
\be
\Pi^{[2]}_\parallel\Big|_{\rm Eulerian, (3)} = \td\Pi^{[2]}_\parallel
- \delta_1\Pi^{[2]}_\parallel - f\,\eta\,\Pi^{[2]}_\parallel + 2\,\Pi^{[3]}_\parallel\,,
\ee
which come from the mass weighting of the shifted field, the Jacobian of the redshift-space
map, and the convective part of $\Pi^{[2]}$, respectively. These are the
$b_{\Pi^{[2]}_\parallel}$ terms in $\beta_{\delta\Pi^{[2]}_\parallel}$,
$\beta_{\eta\Pi^{[2]}_\parallel}$ and $\beta_{\Pi^{[3]}_\parallel}$ of~\eqref{v2:beta}.
The advection of $\Pi^{[2]}_\parallel$ by the displacement is common to both descriptions
and is carried by the shifted operator $\td\Pi^{[2]}_\parallel$, so that nothing in
the cubic content of $\Pi^{[2]}_\parallel$ is left unaccounted for if 
this operator is included in the model. Its cubic content can be decomposed 
\be
\begin{aligned}
\big(\td\Pi^{[2]}_\parallel\big)^{(3)} ={}&
\Big(\td{(KK)_\parallel}+\tfrac23\,\td{\delta_1\eta}-\tfrac19\,\td{\delta_1^2}
-\tfrac57\,\mu^2\,\td{\GG}_2\Big)^{(3)}
+\delta_1\Pi^{[2]}_\parallel - f\,\eta\,\Pi^{[2]}_\parallel
-\tfrac{10}{7}\,\Pi^{[3]}_\parallel
-\Big(\tfrac{30}{49}+\tfrac{10f}{7}\,\mu^2\Big)\KK_3\\
&+\Big(\tfrac{10}{21}-\tfrac{10}{9}\,\mu^2\Big)\GG_3+\tfrac56\,\mu^2\,\Gamma_3
+\tfrac57\big(1+f\mu^2\big)\,\eta\GG_2
+\Big(\tfrac27-\tfrac f9\Big)\eta\delta_1^2+\tfrac{2f}{3}\,\eta^2\delta_1\\
&+\tfrac37\,\delta_1(KK)_\parallel+f\,\eta(KK)_\parallel-\tfrac1{21}\,\delta_1^3
+\tfrac{10f}{3}\,\mathcal V_\parallel\,,
\end{aligned}
\label{v2:Pi2cubic}
\ee
which is fully absorbed
by other operators except
\be
\mathcal V_\parallel\equiv\psi_2^\parallel\,\partial_\parallel\eta
-\mu^2\,\psi_2^\parallel\,\partial_\parallel\delta_1 \cs
\bpsi_2=\tfrac{3}{14}\,\nabla\nabla^{-2}\GG_2\,,
\ee
which do appear in $\eta'$
and $\delta'_1$, but with fixed coefficients. In practice, 
the projection of $\td\Pi^{[2]}_\parallel$
on other operators
in the basis is modeled with our fitting procedure. The contribution from $\mathcal{V}_\parallel$ is negligibly small at the numerical level. It does not improve constraints on 
$b_{\Pi^{[2]}_\parallel}$
or the cross-correlation coefficient in practice.
We exclude this operator for  
the purpose of numerical stability of the fit.
Because of exact degeneracy at the 2nd order and a strong degeneracy at the 3rd order, we do not use a separate transfer function for
$\td\Pi^{[2]}_\parallel$, but keep it as a model operator in the fit, which allows us to accurately measure the corresponding coefficient.

\section{Details of the fitting procedure}
\label{app:fit}

The field-level model \eqref{v2:FL} is fitted to simulations through its transfer
functions. The forward model produces, for one box and one line of sight, the $n_f$
operator fields $O_a(\vk)$ of \eqref{v2:FL}, built from the linear density of the box:
the two linear operators $\td\delta'_1$, $\td\eta^{\prime}$, the shifted quadratic
operators and the cubic ones, all renormalized as in \eqref{v2:renorm}--\eqref{v2:renormlin}.
In the pipeline the $(K\Pi_2)_\parallel$ operator is replaced by $\KK_3$ and
$\Pi^{[2]}_\parallel$ is kept in the Gram matrix but not among the fitted transfer functions because of the aforementioned degeneracy.
This implies $n_f=21$.
All
quantities are binned in wedges $(k,\mu)$, and in each bin we measure
\be
M_{ab} = \langle O_aO_b^*\rangle \cs X_a = \langle\delta_FO_a^*\rangle \cs
P_{FF} = \langle\delta_F\delta_F^*\rangle\,,
\ee
where $\langle\cdot\rangle$ is the average over the $N(k,\mu)$ independent Fourier modes of
the bin.

\textbf{Transfer functions and their errors.}
The operators are strongly correlated, so the transfer functions are defined on the
sequentially orthogonalized (Gram--Schmidt) fields. With the Cholesky factorization
\be
M = L_t\,D\,L_t^T \cs L_t\ \text{unit lower triangular} \cs D = {\rm diag}(D_1,\dots,D_{n_f})\,,
\ee
the orthogonalized fields $O^\perp = L_t^{-1}O$ satisfy
$\langle O^\perp_iO^{\perp*}_j\rangle = D_i\,\delta_{ij}$, and the measured transfer functions are
\be
\beta^\perp_i(k,\mu) = \frac{\langle\delta_F\,O^{\perp*}_i\rangle}{D_i} \qquad (i = 1,\dots,n_f)\,.
\label{fit:betaperp}
\ee
The deterministic part of the field is $\delta_{\rm det} = \sum_i\beta^\perp_iO^\perp_i$, and the
error field $\varepsilon = \delta_F - \delta_{\rm det}$ is, by construction, uncorrelated with all
operators in every bin. Its power is
\be
P_{\rm err}(k,\mu) = \langle\varepsilon\varepsilon^*\rangle = P_{FF} - \sum_i\big(\beta^\perp_i\big)^2D_i\,.
\label{fit:Perr}
\ee
Let's derive the variance of $\beta_a$, following the approach of~\cite{Schmittfull:2018yuk,Kokron:2026vgd}. Let us consider one bin with $N$ modes and write
$\delta_F = \sum_j\beta^\perp_jO^\perp_j + \varepsilon$. The estimator \eqref{fit:betaperp} is
\be
\hat\beta^\perp_i = \frac{1}{ND_i}\sum_{\vk\in{\rm bin}}\delta_F(\vk)\,O^{\perp*}_i(\vk)
= \beta^\perp_i + \frac{1}{ND_i}\sum_{\vk\in{\rm bin}}\varepsilon(\vk)\,O^{\perp*}_i(\vk)\,,
\ee
where the terms $j\neq i$ drop out exactly because the orthogonalization is done on the
same modes. We condition on the realization of the operator fields and treat $\varepsilon$ as
a Gaussian field with power $P_{\rm err}$, independent mode by mode and uncorrelated with the
operators, $\langle\varepsilon(\vk)\varepsilon^*(\vk')\rangle = P_{\rm err}\,\delta_{\vk\vk'}$, where $\delta_{\vk\vk'}$ is the Kronecker delta. Then
\be
\big\langle\Delta\hat\beta^\perp_i\,\Delta\hat\beta^\perp_j\big\rangle
= \frac{1}{N^2D_iD_j}\sum_{\vk,\vk'}\big\langle\varepsilon(\vk)\varepsilon^*(\vk')\big\rangle\,
O^{\perp*}_i(\vk)\,O^\perp_j(\vk')
= \frac{P_{\rm err}}{N^2D_iD_j}\sum_{\vk}O^{\perp*}_i(\vk)\,O^\perp_j(\vk)
= \frac{P_{\rm err}}{ND_i}\,\delta_{ij}\,,
\ee
implying
\be
{\rm Var}\big(\hat\beta^\perp_i\big) = \frac{P_{\rm err}(k,\mu)}{N(k,\mu)\,D_i(k,\mu)} \cs
{\rm Cov}\big(\hat\beta^\perp_i,\hat\beta^\perp_j\big) = 0\quad(i\neq j)\,.
\label{fit:var}
\ee
A transfer function is well determined if the error power is small compared with the
auto power of the corresponding orthogonalized field, and its error decreases with the
number of modes. 
The errors are diagonal only in the orthogonal basis: the same derivation
in the raw basis gives ${\rm Cov}(\hat\beta) = P_{\rm err}M^{-1}/N$, which is far from diagonal
and ill-conditioned for nearly degenerate operators. Eq.~\eqref{fit:var} neglects
(a) the scatter of $M$ because the transfer functions are constructed from the measured fields enabling the noise variance cancellation; 
(b) the non-Gaussianity of
$\varepsilon$ at high $k$ and the noise of $P_{\rm err}$, which is estimated from the data with
relative error $\sim\sqrt{2/N}$. When the three lines of sight of a box are averaged
($X$, $M$, $P_{\rm err}$ averaged before re-orthogonalizing), we keep the single-LOS $N$ in
\eqref{fit:var} because the three lines of sight share the density field, whose fluctuations do not reduce upon averaging different LOS.
This is a conservative choice.

\textbf{The fitting model.}
The model written in the raw (pre-orthogonalized) basis reads
\be
\delta_{\rm model}(\vk) = \sum_a\beta_a(k,\mu)\,O_a(\vk) \cs
\beta_a(\mu) = \sum_{j=1}^{22}C_{aj}(\mu)\,b_j \cs C_{aj}(\mu) = C^{(0)}_{aj} + \mu^2\,\Delta C_{aj}\,,
\label{fit:model}
\ee
where the $\beta_a$ are the dictionary \eqref{v2:beta}, linear in the 22 CLASS-PT parameters
\be
\begin{aligned}
b_j = \big(&b_1,\ b_\eta,\ b_2,\ b_{\GG_2},\ b_{\delta\eta},\ b_{\eta^2},\ b_{(KK)_\parallel},\ b_{\Pi^{[2]}_\parallel},\ b_{\Gamma_3},\
b_{\delta\Pi^{[2]}_\parallel},\ b_{\eta\Pi^{[2]}_\parallel},\ b_{(K\Pi^{[2]})_\parallel},\ b_{\Pi^{[3]}_\parallel},\\
&b_{\delta^3},\ b_{\delta\GG_2},\ b_{\GG_3},\ b_{\delta^2\eta},\ b_{\delta\eta^2},\ b_{\eta^3},\
b_{\delta(KK)_\parallel},\ b_{\eta(KK)_\parallel},\ b_{\eta\GG_2}\big)\,.
\end{aligned}
\ee
The
labels are the observed ones of \eqref{v2:Fs}, so there are no bias-independent constants.
This means that if the flux fluctuations go to zero, which is the case for some of the CAMELS boxes, all 
$b_j$ will go to zero too.

Each measured transfer function in addition receives a $k^2$ counterterm with the full
angular structure,
\be
\beta^\perp_i \supset k^2\big(c_{0,i} + c_{1,i}\,\mu^2 + c_{2,i}\,\mu^4\big)\,,
\label{fit:ct}
\ee
on all orthogonalized rows. The three coefficients on the $\td\delta_1^{\rm new}$ row
reach the power spectrum
where they describe the 
counterterm contribution $\delta_F\supset k^2(c_s + c_{s2}\mu^2 + c_{s4}\mu^4)\,\delta_1$. For other transfer functions $c_{n,i}$ are
treated as nuisance parameters that account for higher order effects. 

\textbf{Projection on the measured basis.}
The model \eqref{fit:model} is written on $22$ operators, the $n_f=21$ transfer fields and
$\Pi_2$ (for which $\beta_{\Pi_2}=b_{\Pi_2}$), while the data are the $n_f$ transfer
functions \eqref{fit:betaperp}, defined on the orthogonalized fields. To compare the two
we project $\delta_{\rm model}$ on the orthogonalized fields,
\be
\beta^{\perp,{\rm model}}_i = \frac{\langle\delta_{\rm model}\,O^{\perp*}_i\rangle}{D_i}
= \sum_{a=1}^{22}R_{ia}\,\beta_a \cs
R_{ia}\equiv\frac{\langle O_a\,O^{\perp*}_i\rangle}{D_i}\,.
\label{fit:rot}
\ee
The $21\times22$ matrix $R$ is known in every bin from the measured Gram matrix. For the
transfer fields $\langle O_aO^{\perp*}_i\rangle=(DL_t^T)_{ia}$, so the first $21$ columns
of $R$ are the Gram--Schmidt triangle $L_t^T$. The last column is the projection of
$\Pi_2$ on the orthogonalized fields, obtained from its cross spectra with the transfer
fields. 

The raw coefficients are fixed by the underlying bias
parameters $\beta_a=\sum_jC_{aj}b_j$, see \eqref{fit:model}. Inserting this in \eqref{fit:rot} gives the
transfer functions directly in terms of the bias parameters,
\be
\beta^{\perp,{\rm model}}_i = \sum_{j=1}^{22}\Big(\sum_{a=1}^{22}R_{ia}\,C_{aj}\Big)\,b_j\,.
\ee
Thus $C$ (a $22\times22$ matrix fixed by the dictionary \eqref{v2:beta}) converts the bias parameters into
raw operator coefficients, and $R$ converts the raw coefficients
into predicted transfer functions. In a
$(k,\mu)$ bin the $\mu^2$ in $C$ is replaced by its average over the modes of the bin,
\be
C_{{\rm bin},\,aj} = C^{(0)}_{aj}+\langle\mu^2\rangle\,\Delta C_{aj}\,,
\ee
which neglects the variation of the cross spectra with $\mu$ inside a wedge. Adding the
counterterms \eqref{fit:ct}, the model fitted to the data in a bin is
\be
\beta^{\perp,{\rm model}}_i(k,\mu) = \sum_{j=1}^{22}\big(R\,C_{\rm bin}\big)_{ij}\,b_j
+ k^2\big(c_{0,i}+c_{1,i}\langle\mu^2\rangle+c_{2,i}\langle\mu^4\rangle\big)\,,
\label{fit:tf}
\ee
with $R$, $C_{\rm bin}$ and the mode-weighted averages $\langle\mu^2\rangle$,
$\langle\mu^4\rangle$ evaluated in that bin. It is linear in the bias parameters $b_j$
and in the counterterm coefficients $c_{n,i}$.

\textbf{The $\chi^2$ fit.}
Eq.~\eqref{fit:tf} is linear in the unknowns $\theta=(b_j\,;\,c_{n,i})$: the $22$ bias
parameters, which are common to all bins, and the $3n_f=63$ counterterm coefficients. For
every bin in the window $k_{\min}\leq k\leq k_{\max}$ and every row $i$ we form the
residual weighted by the inverse error \eqref{fit:var},
\be
r_i(k,\mu) = w_i\,\Big[\beta^{\perp,{\rm model}}_i(k,\mu;\theta)-\beta^\perp_i(k,\mu)\Big] \cs
w_i\equiv\sqrt{\frac{N\,D_i}{P_{\rm err}}} \cs
\chi^2(\theta) = \sum_{k,\mu}\sum_{i=1}^{n_f}r_i^2(k,\mu)\,.
\label{fit:chi2}
\ee
Since the errors are diagonal in the orthogonal basis, this is the exact Gaussian
$\chi^2$. Stacking the residuals as $r=A\theta-y$, with rows labelled by $(k,\mu,i)$,
\be
A_{(k,\mu,i),\,j} = w_i\,\big(R\,C_{\rm bin}\big)_{ij} \cs
A_{(k,\mu,i),\,(n,i')} = w_i\,k^2\langle\mu^{2n}\rangle\,\delta_{ii'} \cs
y_{(k,\mu,i)} = w_i\,\beta^\perp_i\,,
\ee
the minimization is a linear least-squares problem with the closed-form solution
\be
\hat\theta = \big(A^TA\big)^{-1}A^Ty \cs {\rm Cov}(\hat\theta) = \big(A^TA\big)^{-1} \cs
\chi^2_{\min} = \big|A\hat\theta-y\big|^2\,,
\ee
and ${\rm dof}=n_{\rm rows}-n_\theta$. The bias parameters are the first $22$ components
of $\hat\theta$; their covariance is the corresponding block of $(A^TA)^{-1}$, which is
marginalized over the counterterms. The raw coefficients are not fitted separately but are transformed into $C_{aj}$ via \eqref{fit:model}. For the Sherwood box and
$0.15\leq k\leq1\,h\,{\rm Mpc}^{-1}$ the system has $1428$ rows ($17$ $k$ bins, $4$
wedges, $21$ transfer functions) for $85$ unknowns.

A single bin provides $21$ transfer functions for $22$ bias parameters, so it does not
determine them: at fixed $\mu$ a change of $b_{\Pi^{[2]}_\parallel}$ is absorbed by the other
parameters, which is the degeneracy discussed above. The parameter $b_{\Pi^{[2]}_\parallel}$ is determined only through the $\mu$ dependence of the transfer functions, i.e.\
by fitting several wedges together with bias parameters common to all of them.

Bins with a non-positive-definite Gram matrix or non-finite transfer functions are
skipped. The window is chosen by repeating the fit on a grid of $k_{\max}$ and monitoring
the drift $\max_p|b_p(k_{\max})-b_p(k_{\rm ref})|/\sigma_p(k_{\rm ref})$ of the
best-determined parameters; the largest $k_{\max}$ for which no parameter has shifted by
more than $3\sigma$ is retained. The fitted $b_j$, together with the counterterms $(c_s,c_{s2},c_{s4})$, are then passed to the
one-loop $P_{FF}$ model of~\cite{Ivanov:2023yla}.

\section{CAMELS Transfer Functions}\label{app:CAMELS_TF}

\begin{figure*}[t]
    \centering
    \includegraphics[width=\textwidth]{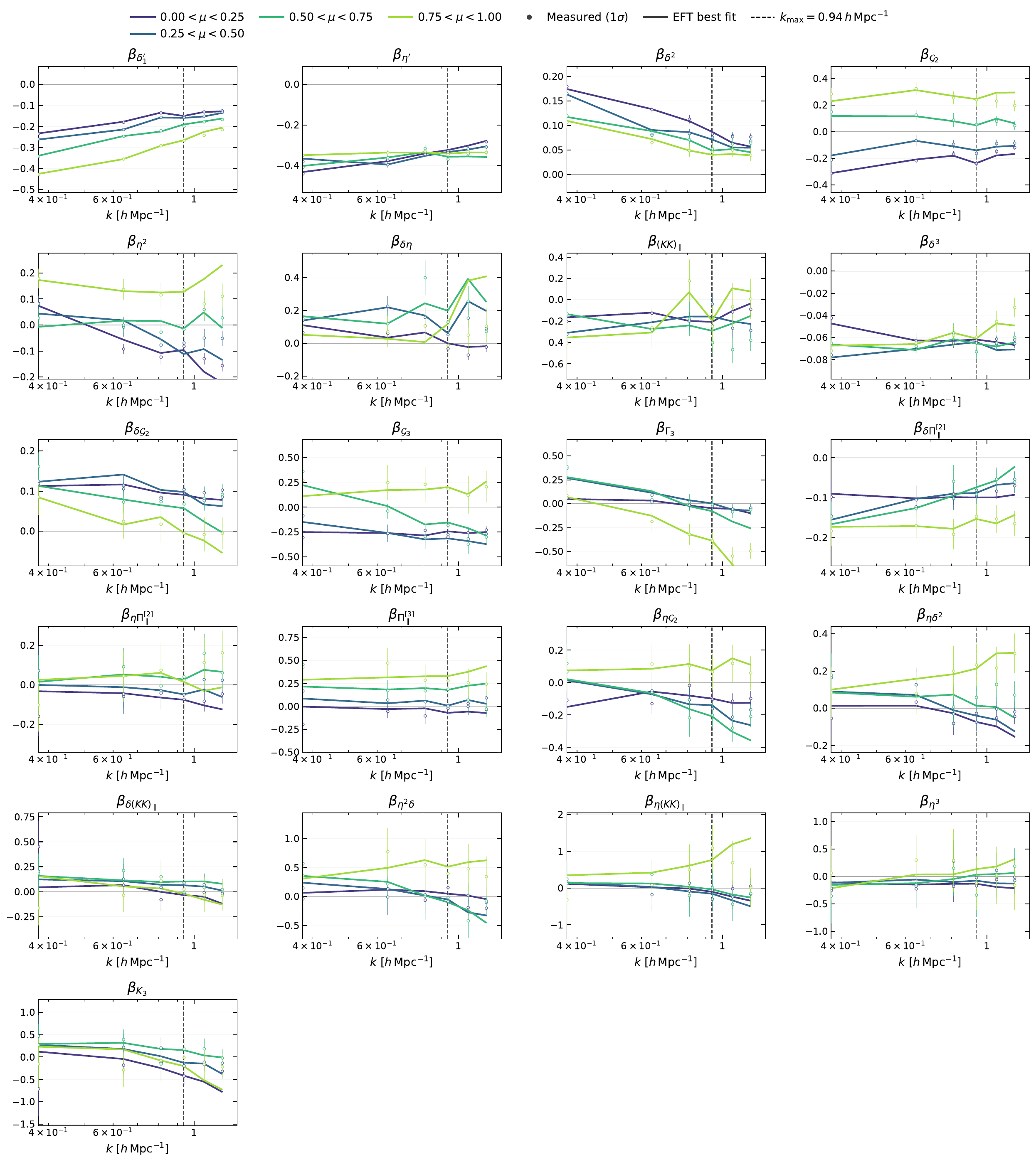}
    \caption{
    \textbf{Same as Fig.~\ref{fig:transfer_functions}, but for the near-Planck CAMELS simulation SB35-637.} The joint EFT fit yields $\chi^2/\mathrm{dof} = 0.448$.
    The CAMELS transfer functions show larger fluctuations than their Sherwood counterparts, primarily because of the substantially smaller simulation volume ($L=50\,h^{-1}\mathrm{Mpc}$ compared to $160\,h^{-1}\mathrm{Mpc}$ for Sherwood), which provides fewer independent Fourier modes, particularly on large scales.
    }
    \label{fig:camels_transfer_functions}
\end{figure*}

Here we show the complete set of cubic-order \lya forest transfer functions measured in the near-Planck CAMELS-SB35 simulation 637 at $z=2$, complementing the Sherwood results shown in Fig. \ref{fig:transfer_functions}. As shown in Fig. \ref{fig:camels_transfer_functions}, the EFT model provides a consistent description of the measured transfer functions across all four LOS bins, despite the larger statistical fluctuations associated with the smaller CAMELS simulation volume.

\section{Best-fit EFT bias parameters and contact terms}
\label{app:eft_parameters}
Here we report the best-fit bias coefficients,
counterterms, and contact terms used in our analysis. We present results for the Sherwood simulation at $z=2.8$
and the CAMELS-SB35 simulation 637 at $z=2.0$. Table \ref{tab:eft_bias_parameters} summarizes the complete
set of 22 Eulerian EFT bias coefficients through cubic order.
The first block contains the linear and quadratic bias
coefficients, while the second block contains the cubic
density, velocity-gradient, and line-of-sight selection
bias coefficients. The third block reports the
field-level counterterms entering the three-dimensional
power spectrum, together with the stochastic parameters.
Finally, the fourth block contains the contact terms
$C_0$, $C_2$, and $C_4$ obtained by fitting the
one-dimensional flux power spectrum.

For completeness, we additionally report the EFT bias parameters
extracted from the Sherwood
$L=80\,h^{-1}{\rm Mpc}$ hydrodynamical
simulation at four redshifts,
$z=2.0,\,2.4,\,2.8,\,3.2$. The resulting coefficients are presented in
Table~\ref{tab:sherwood_L80_bias_parameters}.

\begin{table*}[t]
\centering

\setlength{\tabcolsep}{12pt}
\renewcommand{\arraystretch}{1.08}

\begin{tabular}{lcccc}
\hline\hline
& \textbf{Sherwood}
& \multicolumn{3}{c}{\textbf{CAMELS-SB35 (sim. 637)}} \\
\cline{3-5}
$z$ & $2.8$ & $2.0$ & $3.0$ & $4.0$ \\
$L$ [$h^{-1}{\rm Mpc}$] & 160 & 50 & 50 & 50 \\
$k_{\max}$ [$h\,{\rm Mpc}^{-1}$] & 1.0 & 1.0 & 1.5 & 2.0 \\
\hline

\multicolumn{5}{c}{\textit{Linear and quadratic biases}} \\
\hline

$b_1$
& $-0.234\pm0.001$
& $-0.258\pm0.005$
& $-0.606\pm0.006$
& $-1.291\pm0.012$ \\

$b_{\eta}$
& $\phantom{-}0.355\pm0.003$
& $\phantom{-}0.295\pm0.013$
& $\phantom{-}0.447\pm0.014$
& $\phantom{-}0.545\pm0.029$ \\

$b_2$
& $\phantom{-}0.312\pm0.006$
& $\phantom{-}0.306\pm0.023$
& $\phantom{-}1.008\pm0.032$
& $\phantom{-}3.113\pm0.083$ \\

$b_{\mathcal G_2}$
& $-0.003\pm0.003$
& $\phantom{-}0.036\pm0.013$
& $-0.042\pm0.019$
& $-0.289\pm0.050$ \\

$b_{\delta\eta}$
& $\phantom{-}0.035\pm0.010$
& $-0.042\pm0.042$
& $-0.458\pm0.057$
& $-1.707\pm0.141$ \\

$b_{\eta^2}$
& $-0.325\pm0.012$
& $-0.228\pm0.048$
& $-0.290\pm0.065$
& $-0.018\pm0.164$ \\

$b_{(KK)_\parallel}$
& $-0.019\pm0.011$
& $\phantom{-}0.096\pm0.051$
& $\phantom{-}0.297\pm0.069$
& $\phantom{-}0.831\pm0.171$ \\

$b_{\Pi_\parallel^{[2]}}$
& $-0.132\pm0.008$
& $-0.348\pm0.036$
& $-0.555\pm0.045$
& $-1.025\pm0.101$ \\

\hline
\multicolumn{5}{c}{\textit{Cubic biases}} \\
\hline

$b_{\delta^3}$
& $-0.518\pm0.008$
& $-0.536\pm0.031$
& $-1.578\pm0.055$
& $-4.435\pm0.167$ \\

$b_{\delta\mathcal G_2}$
& $-0.018\pm0.008$
& $-0.048\pm0.031$
& $\phantom{-}0.060\pm0.056$
& $\phantom{-}0.548\pm0.173$ \\

$b_{\mathcal G_3}$
& $-0.082\pm0.020$
& $-0.118\pm0.079$
& $\phantom{-}0.017\pm0.130$
& $-0.453\pm0.384$ \\

$b_{\Gamma_3}$
& $\phantom{-}0.032\pm0.017$
& $-0.069\pm0.067$
& $\phantom{-}0.424\pm0.108$
& $\phantom{-}1.772\pm0.323$ \\

$b_{\delta^2\eta}$
& $\phantom{-}0.670\pm0.038$
& $\phantom{-}0.683\pm0.154$
& $\phantom{-}2.217\pm0.268$
& $\phantom{-}7.881\pm0.827$ \\

$b_{\delta\eta^2}$
& $\phantom{-}0.051\pm0.062$
& $\phantom{-}0.221\pm0.265$
& $\phantom{-}0.045\pm0.462$
& $\phantom{-}1.299\pm1.412$ \\

$b_{\eta^3}$
& $-0.042\pm0.048$
& $\phantom{-}0.034\pm0.191$
& $\phantom{-}0.241\pm0.339$
& $\phantom{-}3.116\pm1.038$ \\

$b_{\delta(KK)_\parallel}$
& $-0.121\pm0.040$
& $-0.181\pm0.172$
& $-0.373\pm0.302$
& $-3.212\pm0.941$ \\

$b_{\eta(KK)_\parallel}$
& $-0.114\pm0.072$
& $-0.125\pm0.262$
& $-0.120\pm0.493$
& $-3.156\pm1.553$ \\

$b_{\delta\Pi_\parallel^{[2]}}$
& $-0.135\pm0.024$
& $-0.254\pm0.098$
& $-0.511\pm0.163$
& $-0.029\pm0.487$ \\

$b_{\eta\Pi_\parallel^{[2]}}$
& $-0.401\pm0.039$
& $\phantom{-}0.524\pm0.173$
& $\phantom{-}0.451\pm0.290$
& $\phantom{-}0.436\pm0.875$ \\

$b_{\eta\mathcal G_2}$
& $\phantom{-}0.129\pm0.026$
& $\phantom{-}0.051\pm0.094$
& $\phantom{-}0.280\pm0.174$
& $\phantom{-}1.038\pm0.546$ \\

$b_{(K\Pi^{[2]})_\parallel}$
& $\phantom{-}0.119\pm0.039$
& $-0.304\pm0.172$
& $-0.596\pm0.277$
& $\phantom{-}0.555\pm0.824$ \\

$b_{\Pi_\parallel^{[3]}}$
& $\phantom{-}0.260\pm0.017$
& $\phantom{-}0.473\pm0.067$
& $\phantom{-}0.495\pm0.107$
& $-0.638\pm0.311$ \\

\hline
\multicolumn{5}{c}{\textit{3D counterterms}} \\
\hline
$c_s$
& $\phantom{-}0.004\pm0.002$ & $-0.011\pm0.007$ & $\phantom{-}0.004\pm0.005$ & $\phantom{-}0.032\pm0.007$ \\
$c_{s2}$
& $\phantom{-}0.072\pm0.008$ & $\phantom{-}0.086\pm0.031$ & $\phantom{-}0.053\pm0.020$ & $\phantom{-}0.023\pm0.029$ \\
$c_{s4}$
& $\phantom{-}0.026\pm0.009$ & $\phantom{-}0.016\pm0.038$ & $\phantom{-}0.015\pm0.023$ & $\phantom{-}0.017\pm0.032$ \\

\hline
\multicolumn{5}{c}{\textit{1D contact terms}} \\
\hline
$C_0$
& $\phantom{-}0.072\pm0.005$
& $-0.116\pm0.029$
& $-0.384\pm0.086$
& $-0.182\pm0.279$ \\
$C_2$
& $-0.155\pm0.019$
& $-0.267\pm0.122$
& $\phantom{-}0.775\pm0.361$
& $\phantom{-}3.260\pm1.188$ \\
$C_4$
& $\phantom{-}0.124\pm0.016$
& $\phantom{-}0.283\pm0.117$
& $-0.530\pm0.343$
& $-2.157\pm1.139$ \\

\hline\hline
\end{tabular}
\caption{
\textbf{Best-fit EFT coefficients from
field-level transfer-function fits.}
The Sherwood ($z=2.8$) and CAMELS-SB35
(simulation 637, $z=2.0,3.0,4.0$) columns show the
Eulerian EFT bias coefficients through cubic order,
the three-dimensional counterterms and stochastic
parameters, and the one-dimensional contact terms.
The field-level fits use
$k_{\max}=1\,h\,{\rm Mpc}^{-1}$ for Sherwood
and $k_{\max}=(1,1.5,2)\,h\,{\rm Mpc}^{-1}$
for CAMELS at $z=(2,3,4)$, respectively.
}
\label{tab:eft_bias_parameters}
\end{table*}

\begin{table*}[t]
\centering
\renewcommand{\arraystretch}{1.12}
\setlength{\tabcolsep}{10pt}
\begin{tabular}{lcccc}
\hline\hline
& \multicolumn{4}{c}{
\textbf{Sherwood}
($L=80\,h^{-1}{\rm Mpc}$)} \\
\cline{2-5}
$z$ & $2.0$ & $2.4$ & $2.8$ & $3.2$ \\
$L$ [$h^{-1}{\rm Mpc}$]
& 80 & 80 & 80 & 80 \\
$k_{\max}$ [$h\,{\rm Mpc}^{-1}$]
& 1.0 & 1.0 & 1.0 & 1.0 \\
\hline

\multicolumn{5}{c}{
\textit{Linear and quadratic biases}} \\
\hline
$b_1$
& $-0.124\pm0.001$ & $-0.173\pm0.002$ & $-0.241\pm0.002$ & $-0.335\pm0.003$ \\
$b_{\eta}$
& $\phantom{-}0.193\pm0.004$ & $\phantom{-}0.285\pm0.006$ & $\phantom{-}0.374\pm0.007$ & $\phantom{-}0.453\pm0.008$ \\
$b_2$
& $\phantom{-}0.110\pm0.005$ & $\phantom{-}0.186\pm0.008$ & $\phantom{-}0.303\pm0.011$ & $\phantom{-}0.467\pm0.015$ \\
$b_{\mathcal G_2}$
& $\phantom{-}0.013\pm0.003$ & $\phantom{-}0.020\pm0.005$ & $\phantom{-}0.026\pm0.007$ & $\phantom{-}0.026\pm0.009$ \\
$b_{\delta\eta}$
& $\phantom{-}0.019\pm0.011$ & $\phantom{-}0.019\pm0.016$ & $-0.022\pm0.022$ & $-0.149\pm0.029$ \\
$b_{\eta^2}$
& $-0.135\pm0.012$ & $-0.211\pm0.017$ & $-0.287\pm0.023$ & $-0.367\pm0.029$ \\
$b_{(KK)_\parallel}$
& $\phantom{-}0.053\pm0.012$ & $\phantom{-}0.041\pm0.017$ & $\phantom{-}0.032\pm0.024$ & $\phantom{-}0.078\pm0.032$ \\
$b_{\Pi_\parallel^{[2]}}$
& $-0.174\pm0.009$ & $-0.211\pm0.013$ & $-0.259\pm0.018$ & $-0.325\pm0.024$ \\

\hline
\multicolumn{5}{c}{\textit{Cubic biases}} \\
\hline
$b_{\delta^3}$
& $-0.229\pm0.006$ & $-0.352\pm0.010$ & $-0.532\pm0.015$ & $-0.781\pm0.020$ \\
$b_{\delta\mathcal G_2}$
& $-0.031\pm0.005$ & $-0.053\pm0.009$ & $-0.075\pm0.015$ & $-0.077\pm0.020$ \\
$b_{\mathcal G_3}$
& $-0.018\pm0.015$ & $-0.017\pm0.025$ & $-0.048\pm0.040$ & $-0.094\pm0.060$ \\
$b_{\Gamma_3}$
& $-0.009\pm0.013$ & $\phantom{-}0.014\pm0.021$ & $\phantom{-}0.022\pm0.033$ & $\phantom{-}0.031\pm0.049$ \\
$b_{\delta^2\eta}$
& $\phantom{-}0.338\pm0.027$ & $\phantom{-}0.472\pm0.045$ & $\phantom{-}0.730\pm0.071$ & $\phantom{-}1.182\pm0.104$ \\
$b_{\delta\eta^2}$
& $\phantom{-}0.045\pm0.044$ & $\phantom{-}0.022\pm0.072$ & $\phantom{-}0.005\pm0.115$ & $-0.082\pm0.171$ \\
$b_{\eta^3}$
& $\phantom{-}0.036\pm0.032$ & $\phantom{-}0.034\pm0.052$ & $\phantom{-}0.047\pm0.082$ & $-0.004\pm0.119$ \\
$b_{\delta(KK)_\parallel}$
& $-0.131\pm0.028$ & $-0.101\pm0.046$ & $-0.124\pm0.075$ & $-0.224\pm0.111$ \\
$b_{\eta(KK)_\parallel}$
& $-0.062\pm0.048$ & $\phantom{-}0.009\pm0.080$ & $\phantom{-}0.098\pm0.130$ & $\phantom{-}0.223\pm0.193$ \\
$b_{\delta\Pi_\parallel^{[2]}}$
& $-0.002\pm0.018$ & $-0.029\pm0.030$ & $-0.064\pm0.049$ & $-0.156\pm0.074$ \\
$b_{\eta\Pi_\parallel^{[2]}}$
& $-0.240\pm0.027$ & $-0.386\pm0.046$ & $-0.533\pm0.074$ & $-0.718\pm0.109$ \\
$b_{\eta\mathcal G_2}$
& $\phantom{-}0.059\pm0.018$ & $\phantom{-}0.046\pm0.030$ & $\phantom{-}0.036\pm0.048$ & $\phantom{-}0.067\pm0.071$ \\
$b_{(K\Pi^{[2]})_\parallel}$
& $\phantom{-}0.047\pm0.031$ & $\phantom{-}0.041\pm0.050$ & $\phantom{-}0.066\pm0.078$ & $\phantom{-}0.044\pm0.112$ \\
$b_{\Pi_\parallel^{[3]}}$
& $\phantom{-}0.220\pm0.015$ & $\phantom{-}0.280\pm0.023$ & $\phantom{-}0.355\pm0.035$ & $\phantom{-}0.480\pm0.051$ \\

\hline
\multicolumn{5}{c}{
\textit{3D counterterms}} \\
\hline
$c_s$
& $\phantom{-}0.025\pm0.002$ & $\phantom{-}0.027\pm0.003$ & $\phantom{-}0.027\pm0.004$ & $\phantom{-}0.032\pm0.005$ \\
$c_{s2}$
& $\phantom{-}0.135\pm0.011$ & $\phantom{-}0.101\pm0.014$ & $\phantom{-}0.073\pm0.018$ & $\phantom{-}0.045\pm0.023$ \\
$c_{s4}$
& $-0.002\pm0.013$ & $\phantom{-}0.033\pm0.018$ & $\phantom{-}0.055\pm0.023$ & $\phantom{-}0.076\pm0.028$ \\

\hline
\multicolumn{5}{c}{\textit{1D contact terms}} \\
\hline
$C_0$
& $-0.008\pm0.007$
& $\phantom{-}0.002\pm0.011$
& $\phantom{-}0.010\pm0.015$
& $\phantom{-}0.044\pm0.021$ \\
$C_2$
& $-0.054\pm0.027$
& $-0.058\pm0.041$
& $-0.008\pm0.059$
& $\phantom{-}0.099\pm0.082$ \\
$C_4$
& $\phantom{-}0.080\pm0.023$
& $\phantom{-}0.073\pm0.036$
& $\phantom{-}0.023\pm0.052$
& $-0.083\pm0.071$ \\

\hline\hline
\end{tabular}
\caption{Same as Tab.~\ref{tab:eft_bias_parameters} but the 
EFT parameters are measured from the Sherwood
$L=80\,h^{-1}{\rm Mpc}$ hydrodynamical simulation.
}
\label{tab:sherwood_L80_bias_parameters}
\end{table*}

\subsection{Robustness of the EFT approach}\label{sec:RobustResults}
\begin{figure*}
    \centering
    \includegraphics[width=\textwidth]{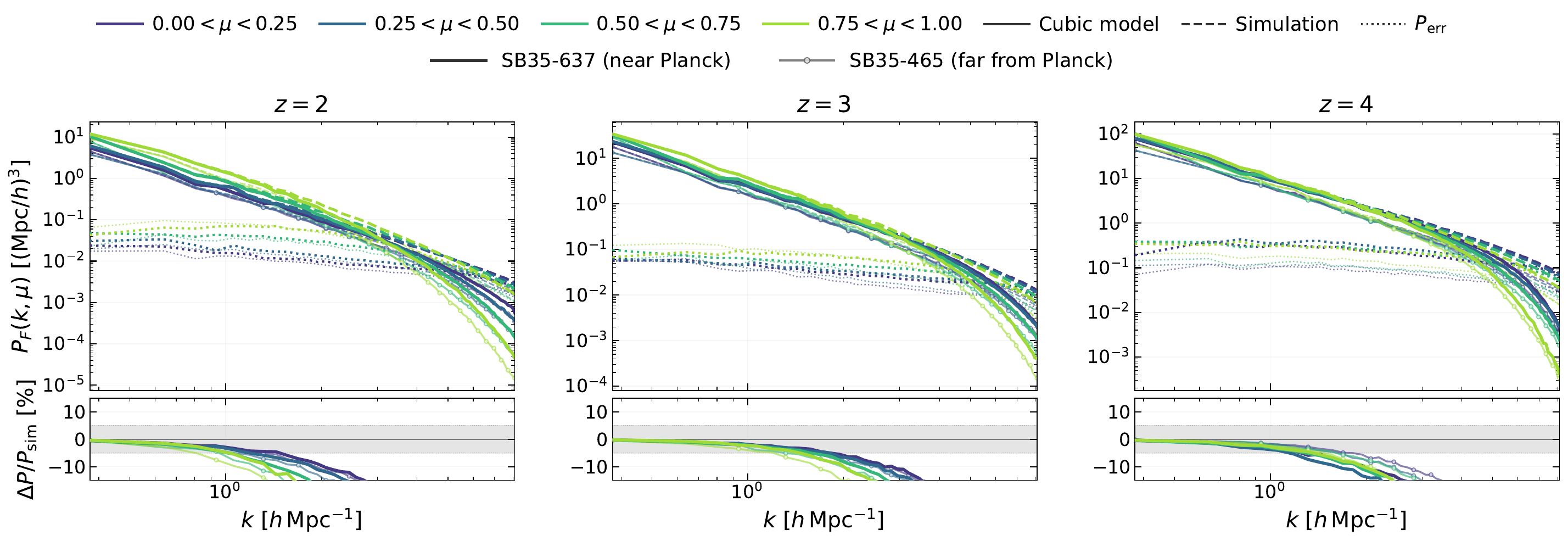}
    \caption{\textbf{Robustness test - Power spectrum fits} same as the third, cubic, column of Fig. \ref{fig:linear_quadratic_cubic_comparison} but comparing two representative simulations spanning different regions of the CAMELS cosmological parameter space to test the robustness of our EFT approach. Simulation SB35-637 shares Planck-like cosmology, while SB35-465 lies far from Planck. Thick, fully opaque lines denote SB35-637, while thinner, semi-transparent lines with circular markers denote SB35-465. The comparable performance across the cosmological parameter space demonstrates the robustness of our model.
    }
    \label{fig:robustness}
\end{figure*}

An important test of our EFT forward model is to test its performance across different regions of the parameter space. The broad cosmological and astrophysical parameter space provided by the CAMELS SB35 suite provides the breadth for testing our model's robustness. To illustrate this, we compare the performance of our EFT approach on two simulations occupying very different regions of the cosmological parameter space: SB35-637, our near-Planck simulation from earlier, and SB35-465, which lies far from Planck. Despite the substantial change in their underlying cosmology, Fig. \ref{fig:robustness} shows that our cubic model achieves comparable agreement for the two simulations. The fractional residuals are also of similar magnitude over the perturbative range for both angular wedges and across the three redshifts.

\begin{figure*}[t]
\centering
\includegraphics[width=0.49\textwidth]{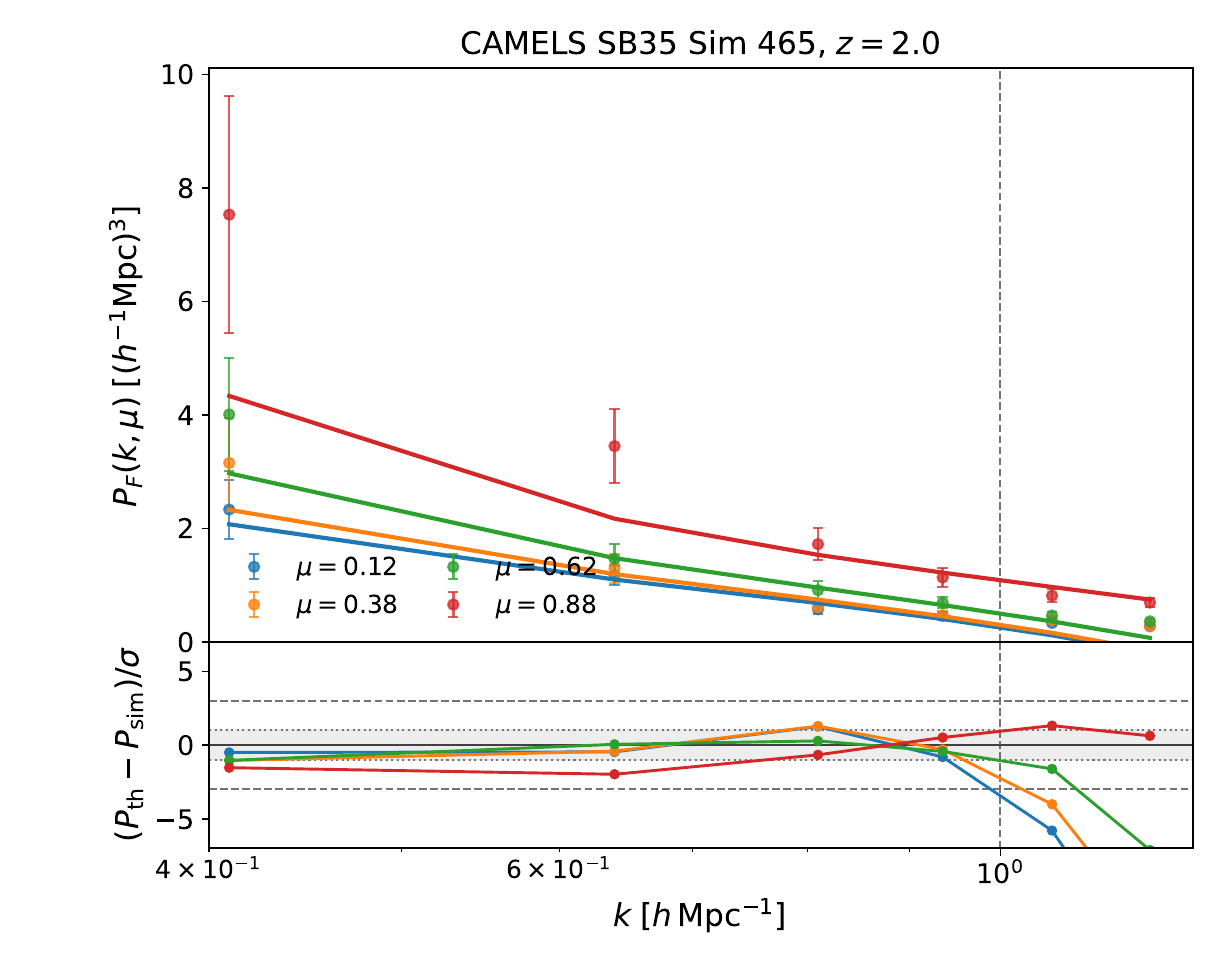}
\hfill
\includegraphics[width=0.49\textwidth]{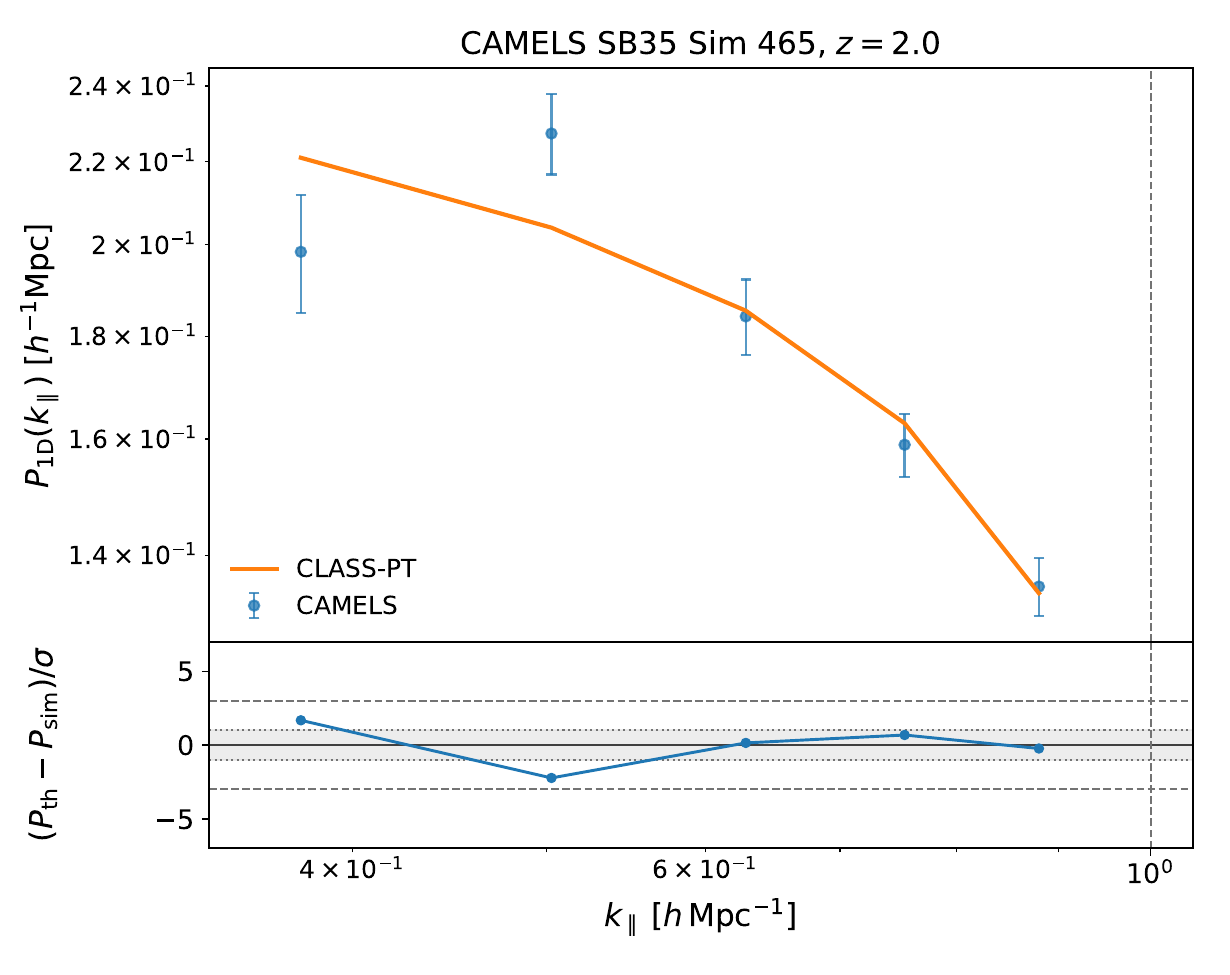}
\vspace{0.5em}
\includegraphics[width=0.49\textwidth]{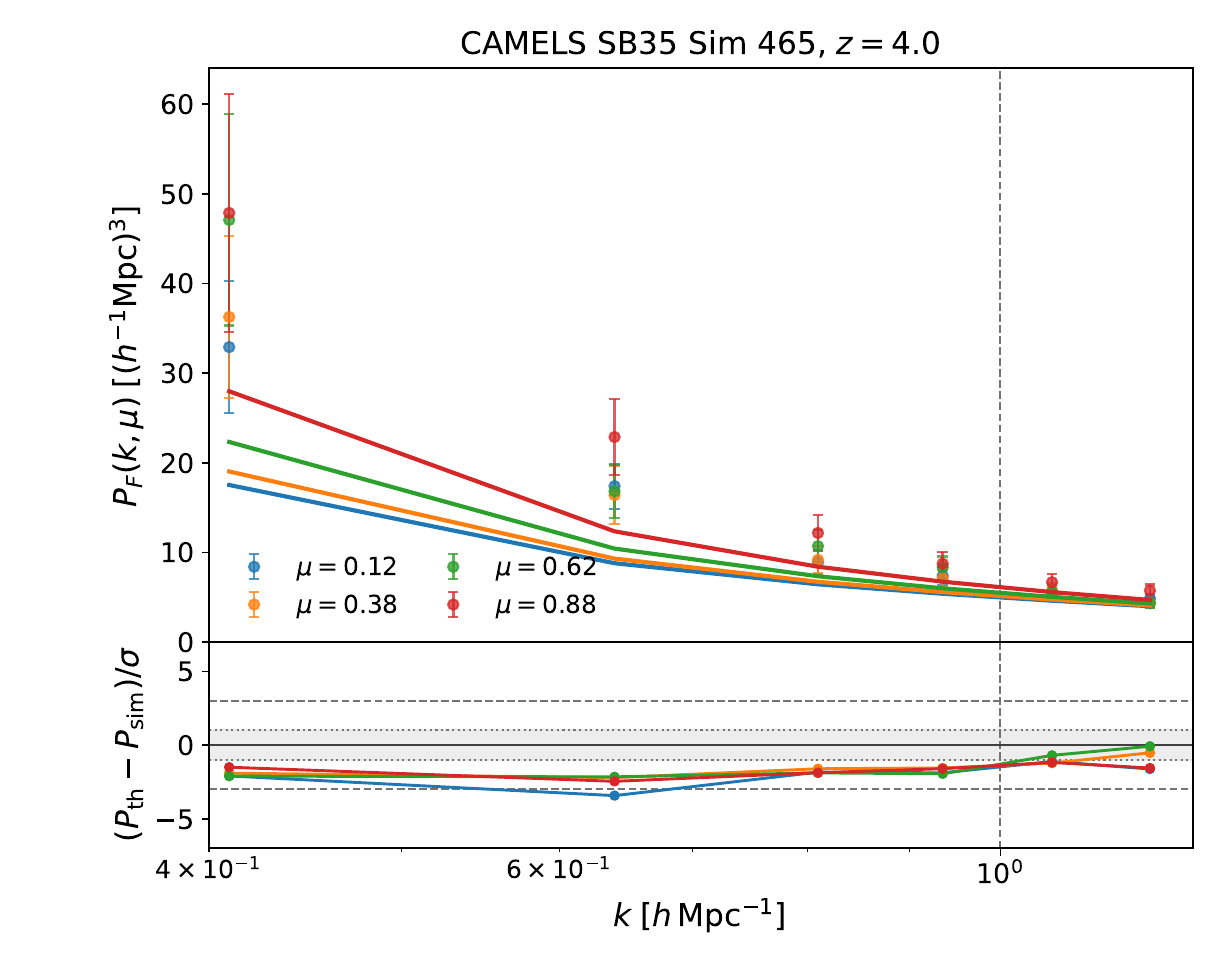}
\hfill
\includegraphics[width=0.49\textwidth]{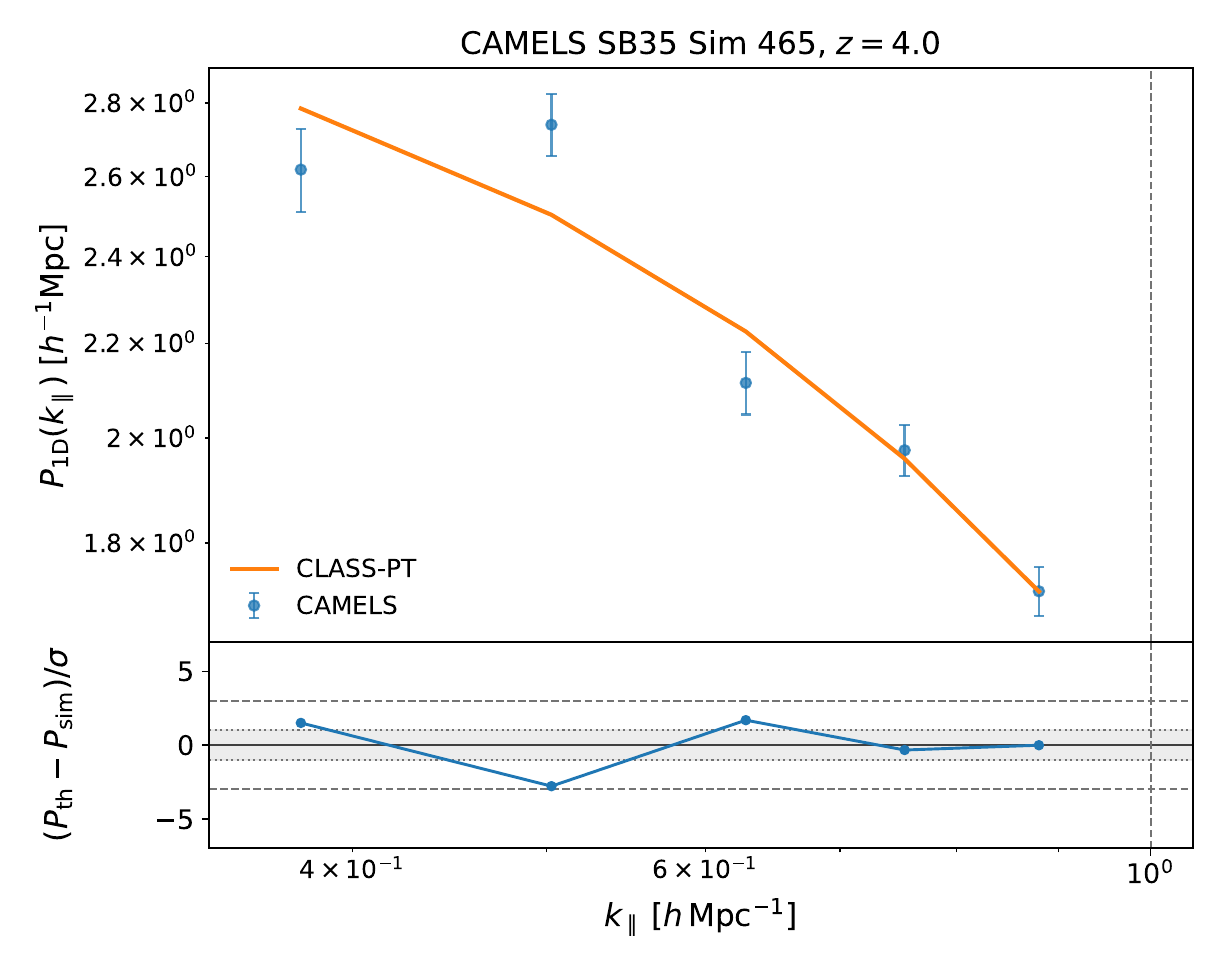}
\caption{\textbf{Robustness test - Power spectrum reconstruction} same as Fig. \ref{fig:CAMELS_classpt_pff}, but for the far-from-Planck CAMELS simulation SB35-465. The broad agreement across the angular wedges and redshifts provides an additional test that the perturbative two-point reconstruction remains robust away from the near-Planck region of the CAMELS parameter space.
}
\label{fig:CAMELS_classpt_pff_z2_z4_465}
\end{figure*}

We further test the robustness of our methods at the level of the power spectrum reconstruction. Figure \ref{fig:CAMELS_classpt_pff_z2_z4_465} repeats the CLASS-PT comparison of Fig. \ref{fig:CAMELS_classpt_pff}, now for the far-from-Planck simulation. In general, the reconstructed power follows the simulation at both $z=2$ and $z=4$ in the perturbative range. As in Fig. \ref{fig:CAMELS_classpt_pff}, the residuals are due to discreteness issues due to the small, low-resolution CAMELS boxes discussed in Sec. \ref{sec:PffReconstructionResults}. This comparison verifies that the accuracy of our forward model for the near-Planck simulation is representative of the broader CAMELS ensemble.

\bibliographystyle{aux_files/JHEP}
\bibliography{main.bib}

\end{document}